\documentclass[%
 reprint,
superscriptaddress,
bibnotes,
amsmath,amssymb,
aps,
prx,
floatfix,
showkeys
]{revtex4-2}

\usepackage{graphicx}%
\usepackage{subcaption}
\usepackage{adjustbox}
\usepackage{float}
\usepackage{dcolumn}
\usepackage{bm}
\usepackage{orcidlink}
\usepackage[mathlines]{lineno}
\usepackage{color}
\usepackage{lmodern}
\usepackage{booktabs}
\usepackage[separate-uncertainty=true, multi-part-units=single]{siunitx}
\usepackage[OT1]{fontenc}
\usepackage{placeins}
\usepackage{ragged2e}

\newcommand{\exciting}{{\usefont{T1}{lmtt}{b}{n}exciting}}
\newcommand{\larch}{{\usefont{T1}{lmtt}{b}{n}Larch}}
\DeclareSIUnit\angstrom{\text {Å}}
\DeclareSIUnit\bar{bar}
\begin{document}

\title{Ultrafast dynamic Coulomb screening of X-ray core excitons in photoexcited  semiconductors}

\author{Thomas C. Rossi\orcidlink{0000-0002-7448-8948}}
\thanks{Both authors contributed equally to this work.}
\affiliation{Helmholtz-Zentrum Berlin f\"ur Materialien und Energie GmbH, 14109 Berlin, Germany}
\email{thomas.rossi@helmholtz-berlin.de\\}
\email{renske.vanderveen@helmholtz-berlin.de}

\author{Lu Qiao\orcidlink{0009-0002-0344-1082}}
\thanks{Both authors contributed equally to this work.}
\affiliation{Department Physics and CSMB, Humboldt-Universit\"at zu Berlin, D-12489 Berlin, Germany}
\email{qiaolu@physik.hu-berlin.de}

\author{Conner P. Dykstra\orcidlink{0000-0001-5597-6914}}
\affiliation{Department of Chemistry, University of Illinois Urbana-Champaign, Urbana, Illinois 61801, United States}

\author{Ronaldo Rodrigues Pela\orcidlink{0000-0002-2413-7023}}
\affiliation{Supercomputing Department, Zuse Institute Berlin (ZIB), Takustraße 7, 14195 Berlin, Germany}

\author{Richard Gnewkow\orcidlink{0000-0002-6524-4909}}
\affiliation{Helmholtz-Zentrum Berlin f\"ur Materialien und Energie GmbH, 14109 Berlin, Germany}
\affiliation{Institute of Optics and Atomic Physics, Technische Universit\"at Berlin, 10623 Berlin, Germany}

\author{Rachel F. Wallick\orcidlink{0000-0002-7548-4850}}
\author{John H. Burke\orcidlink{0000-0001-9853-7292}}
\author{Erin Nicholas\orcidlink{0000-0001-6530-4097}}
\affiliation{Department of Chemistry, University of Illinois Urbana-Champaign, Urbana, Illinois 61801, United States}

\author{Anne-Marie March\orcidlink{0000-0003-2961-1246}}
\author{Gilles Doumy\orcidlink{0000-0001-8672-4138}}
\affiliation{Chemical Sciences and Engineering Division, Argonne National Laboratory, Lemont, Illinois 60439, United States}

\author{D. Bruce Buchholz\orcidlink{0009-0004-2747-1897}}
\affiliation{Department of Materials Science and Engineering, Northwestern University, Evanston, Illinois 60208, United States}

\author{Christiane Deparis\orcidlink{0000-0002-4578-8917}}
\affiliation{Universit\'e C\^ote d'Azur, CNRS, CRHEA, rue Bernard Gregory, Sophia Antipolis, 06560 Valbonne, France}

\author{Jesus Zuñiga-P\'erez\orcidlink{0000-0002-7154-641X}}
\affiliation{Universit\'e C\^ote d'Azur, CNRS, CRHEA, rue Bernard Gregory, Sophia Antipolis, 06560 Valbonne, France}
\affiliation{Majulab, International Research Laboratory IRL 3654, CNRS, Universit\'e C\^ote d'Azur, Sorbonne Universit\'e, National University of Singapore, Nanyang Technological University, Singapore, Singapore}

\author{Michael Weise}
\author{Klaus Ellmer\orcidlink{0000-0002-1874-8741}}
\affiliation{Optotransmitter-Umweltschutz-Technologie (OUT) e.V., K\"openicker Strasse 325, Haus 201, 12555 Berlin, Germany}

\author{Mattis Fondell\orcidlink{0000-0003-2689-8878}}
\affiliation{Helmholtz-Zentrum Berlin f\"ur Materialien und Energie, Institute for Methods and Instrumentation for Synchrotron Radiation Research, Albert-Einstein-Strasse 15, D-12489 Berlin, Germany}

\author{Claudia Draxl\orcidlink{0000-0003-3523-6657}}
\affiliation{Department Physics and CSMB, Humboldt-Universit\"at zu Berlin, D-12489 Berlin, Germany}

\author{Renske M. van der Veen\orcidlink{0000-0003-0584-4045}}
\affiliation{Helmholtz-Zentrum Berlin f\"ur Materialien und Energie GmbH, 14109 Berlin, Germany}
\affiliation{Department of Chemistry, University of Illinois Urbana-Champaign, Urbana, Illinois 61801, United States}
\affiliation{Institute of Optics and Atomic Physics, Technische Universit\"at Berlin, 10623 Berlin, Germany}

\date{\today}

\begin{abstract}

Ultrafast X-ray spectroscopy has been revolutionized in recent years due to the advent of fourth-generation X-ray facilities. In solid-state materials, core excitons determine the energy and line shape of absorption features in core-level spectroscopies such as X-ray absorption spectroscopy. The screening of core excitons is an inherent many-body process that can reveal insight into charge-transfer excitations and electronic correlations. Under non-equilibrium conditions such as after photoexcitation, however, core-exciton screening is still not fully understood. Here we demonstrate the dynamic Coulomb screening of core excitons induced by photoexcited carriers by employing X-ray transient absorption (XTA) spectroscopy with picosecond time resolution. Our interpretation is supported by state-of-the-art \emph{ab initio} theory, combining constrained and real-time time-dependent density functional theory with many-body perturbation theory. Using ZnO as an archetypal wide band-gap semiconductor, we show that the Coulomb screening by photoexcited carriers at the Zn K-edge leads to a decrease in the core-exciton binding energy, which depends nonlinearly on both the excitation density and the distribution of photoexcited carriers in reciprocal space. The effect of Coulomb screening dominates over Pauli blocking in the XTA spectra. We show that dynamic core-exciton screening is also observed at other X-ray absorption edges and theoretically predict the effect of core-exciton screening on the femtosecond time scale for the case of ZnO, a major step towards hard X-ray excitonics. The results have implications for the interpretation of ultrafast X-ray spectra in general and their use in tracking charge carrier dynamics in complex materials on atomic length scales.

\end{abstract}


\maketitle


\section{Introduction}

In photoexcited semiconductors, the electron dynamics is dominated by the interplay between carrier-carrier and carrier-phonon interactions \cite{Huber:2001ke}, which weakens with increasing excitation density due to the many-body (collective) screening of the Coulomb potential \cite{Haug:431405}. The effect of Coulomb screening can lead to dramatic changes in the electronic properties of semiconductors, preventing, for example, excitons to form (Mott transition) \cite{Kappei2005} or leading to insulator-to-metal transitions in materials with strong carrier correlations \cite{Jager:2017pn}. The onset of these light-induced collective perturbations usually occurs on femtosecond timescales \cite{Huber:2001ke,Zimin2023}, which offers great potential for building next-generation ultrafast optoelectronic devices.

X-ray absorption spectroscopy (XAS) probes the electronic and lattice structure of materials by promoting electrons between a core level and the unoccupied density of states (DOS) below and above the ionization threshold of resonant chemical elements \cite{Schnohr:2014cs}. Near the vacuum level, the energy of the transition depends on the strength of the Coulomb attraction between the core hole and the valence electrons in the final state, leading to a strong renormalization of the unoccupied DOS \cite{Shirley:1998pl}. In metal-oxide semiconductors, the screening of the core-hole potential is weak and limited to the electronic density in the vicinity of the absorbing atom \cite{Buczko2000:60211}, which leads to the formation of multiple core exciton peaks below the continuum of unbound states in the near-edge region of the XAS spectrum (dubbed \emph{core excitons}) \cite{Shirley2006}. In contrast, the large delocalized electronic density in noble metals efficiently screens the formation of core excitons \cite{Veal1985}, leading to XAS spectra without prominent resonances near the absorption edge \cite{Greaves1981}, although core excitons are still present \cite{Kitamura2020:266175}. Modeling of the core-hole screening in a material at equilibrium can yield information about the energy of interatomic charge-transfer excitations \cite{Kleiman1995} or the magnitude of electronic correlations and their anisotropy \cite{Koitzsch:2002cu,Juhin:2010dh}. 

Ultrafast X-ray transient absorption (XTA) spectroscopy on semiconductor materials is used to investigate charge density dynamics \cite{Rein2021} or local structural relaxation upon carrier trapping \cite{Penfold:2018ie} or polaron formation \cite{Cannelli:2021jc}. The presence of photoexcited charge carriers yields additional many-body interactions consisting of dynamic Coulomb screening and Pauli blocking. The effect of Coulomb screening on energy levels and optical excitons in photoexcited semiconductors is well established and used to investigate interfacial carrier injection \cite{Baldini:2017hx} or to control exciton transport \cite{Hao2020:84179}. However, it is unclear how non-equilibrium carriers in photoexcited semiconductors contribute to the Coulomb screening and Pauli blocking of core excitons in the excited-state XAS spectrum. Yet, characterizing core-exciton changes in photoexcited materials is crucial for understanding their perturbed electronic structure, similar to how the screening of optical excitons provides insights into band-gap dynamics \cite{Calati2023:134629}. Moreover, through XAS, core excitons reveal not only electronic changes but also chemical and structural information at atomic length scales, which makes XAS a powerful tool for probing materials with enhanced sensitivity.

On the theoretical side, several methods exist to calculate XAS spectra of materials. Real-time time-dependent density functional theory (RT-TDDFT), in particular, can handle complex excited states in photoexcited materials by providing a dynamic description of transition dipole moments \cite{ullrich2011tddft,Pemmaraju_2018}. However, it often simplifies the treatment of many-body effects by approximate exchange-correlation functionals that do not fully capture the intricate electron-hole interactions present in correlated systems \cite{ullrich2011tddft}. The Bethe-Salpeter equation (BSE) of many-body perturbation theory (MBPT) \cite{onida2002,Rohlfing_2000}, on the other hand, provides a precise description of electronic excitations by including electron-hole interactions \cite{Albrecht:1998si}. BSE has been successful in
describing the equilibrium absorption spectra of a large variety of systems \cite{Olovsson2009,Laskowski2006:58146,Yip2005:196093,Thygesen_2017,Garcia-Lastra2011:248053,Pela2024:14140}, including ZnO \cite{Schleife:2011hv,Leon2024:206327}, showing notable agreement with experiments. Nevertheless, it is rarely used in calculations for non-equilibrium states of photoexcited materials. Extensions to BSE \cite{Perfetto2015:206813} have been used to investigate the influence of non-equilibrium carriers on the absorption and luminescence spectra of laser-excited quantum wells \cite{noneq_QW} and semiconductors \cite{noneq_semiconductor}. However, no parameter-free computation has yet focused on the effect of non-equilibrium carriers on XAS spectra including core-exciton features.

In this work, we employ picosecond-resolved K- and L-edge XTA spectroscopy, constrained density functional theory (cDFT) and RT-TDDFT plus non-equilibrium BSE calculations to gain unprecedented insights into the many-body interactions between photoexcited carriers and core excitons in wide band-gap semiconductors. It constitutes the first fully \emph{ab initio} investigation of the effect of photoexcited carriers on core-level transitions, which requires the separation of thermal and non-thermal contributions to the experimental XTA transient spectrum to compare with theory. We apply our method to XTA spectra of a prototype II-VI wide band-gap semiconductor, ZnO, which is known for its photovoltaic \cite{Leschkies:2007ik,Tiwana:2011dt} and (photo)piezoelectronic \cite{Zhang:2012ee,Pan:2019cr} properties. The results show the sensitivity of the Coulomb screening of core excitons to the distribution and density of carriers in reciprocal space and the negligible Pauli blocking contribution from photoexcited carriers to excited-state XAS spectra at K- and L-edges on picosecond timescales. The calculations predict that dynamical Coulomb screening at the K-edge also dominates XTA spectra on femtosecond timescales, leading to changes an order of magnitude larger than at the picosecond timescale. These results help establish ultrafast XAS at new-generation light sources as a ubiquitous tool to track the dynamics of charge carriers and pave the way for the manipulation of hard X-ray excitons.  


\section{Experimental details\label{sec:experimental_details}}

X-ray absorption spectra (XAS) and X-ray transient absorption (XTA) spectra at the Zn K-edge of ZnO thin films were acquired at the Advanced Photon Source (Argonne National Laboratory). The experimental setup has been extensively described in the Supporting Information (SI) of reference\ \cite{Rossi2021}. The X-ray beam was $p$-polarized with $\sim\SI{70}{\pico\second}$ pulses at an incidence angle of \SI{45\pm2}{\degree} on the sample surface. The pump excitation was performed with the third harmonic (\SI{355}{\nano\meter}, \SI{3.49}{\electronvolt}, $p$-polarization) of a Nd:YVO$_4$ Duetto laser (Time-Bandwidth products), which delivered $\sim\SI{10}{\pico\second}$ pulses (FWHM) at \SI{100.266}{\kilo\hertz}. The time delay between the laser excitation and the X-ray probe pulse was set to \SI{100\pm10}{\pico\second}, which corresponds to the maximum of the XTA signal amplitude after excitation and is close to the time resolution of the instrument. During the XAS and XTA measurements, a nitrogen flow was applied to the sample to prevent adsorption and diffusion of carbon impurities and water inside the material. The nitrogen flow also provided active cooling, preventing static heating caused by the laser excitation. Temperature-dependent XAS spectra were recorded with the sample on a heating stage (Linkam THMSG-600) between room temperature (\SI{24\pm2}{\degree}C) and \SI{190\pm2}{\degree}C.

XAS and XTA spectra at the Zn L$_3$-edge of ZnO thin films were acquired in transmission at the UE52-SGM beamline of BESSY II \cite{Miedema2016}. The setup is described in reference\ \cite{Fondell:2017sd}. The nmTransmission NEXAFS chamber was modified with a sample tip to allow for measurements of thin film samples in transmission. The X-ray beam was $p$-polarized, impinging at normal incidence on the sample surface. The pump laser consisted of \SI{350}{\femto\second} pulses at the third harmonic (\SI{343}{\nano\meter}, \SI{3.61}{\electronvolt}, $p$-polarization) of a Yb-doped hybrid fibre/crystal laser system (Tangerine, Amplitude Syst\`emes) with \SI{10}{\kilo\hertz} repetition rate. The relative angle between the laser and the X-rays was \SI{45\pm3}{\degree}. Additional technical details about the measurements at the Zn K- and L-edge are provided in SI \ref{secSI:XTA_setup} \footnote{Supporting Information contains a full description of the XTA setups, the data processing of XAS and XTA spectra, time traces, the protocols for the synthesis of the samples and their characterizations, the effect of the temperature on XAS spectra and the procedure to separate XTA spectra into their thermal and non-thermal contributions, additional fluence dependence measurements, the calculations of the initial excitation density and at \SI{100}{\pico\second}, a description of the theoretical background for the calculations, and the computation details\label{SI}}.

XTA spectra measured in fluorescence mode at the Zn K-edge are the difference between XAS spectra measured in the non-equilibrium (pumped, excited) and equilibrium (unpumped, unexcited) states, normalized by the incident X-ray intensity recorded with an ion chamber \cite{Lima:2011dya}. XTA measured in transmission at the Zn L$_3$-edge instead is computed from the logarithmic ratio between the pumped and unpumped spectra divided by the incident X-ray intensity. XAS and XTA spectra are normalized with respect to the edge jump for a straightforward comparison of amplitudes between experiment and theory using the Python implementation of \larch\ \cite{Newville:2013go}. The XAS amplitude is indicated by a normalized absorption coefficient $\alpha$ in the figures. Additional information about data processing are provided in SI \ref{secSI:data_processing}. Kinetic decays of XTA signals are not discussed and provided in SI \ref{secSI:kinetics}. 

The ZnO thin films measured at the Zn K-edge have a thickness of \SI{283\pm2}{\nano\meter} and are grown by pulse laser deposition on \textbf{c}-sapphire substrates, while the films measured at the Zn L$_3$-edge have a thickness of \SI{400\pm5}{\nano\meter} and are grown either by molecular beam epitaxy or radiofrequency sputtering on polycrystalline silicon nitride membranes (\SI{100}{\nano\meter} thickness). The (0001) orientation of the thin films was checked by X-ray diffraction and the film thickness and optical constants were determined by spectroscopic ellipsometry. The band gap of ZnO is $\SI{3.32\pm0.01}{\electronvolt}$ determined by spectroscopic ellipsometry, hence the pump excitation (\SI{3.49}{\electronvolt} at the Zn K-edge and \SI{3.61}{\electronvolt} at the Zn L$_3$-edge) generates electron-hole pairs above the band gap. Sample methods and characterizations are provided in SI \ref{secSI:sample_synthesis_charac}.


\section{Theory\label{sec:theory_details}}

Our approach combines cDFT/RT-TDDFT with the Bethe-Salpether Equation (BSE). This allows for studying the spectroscopic response of core excitations to optical pumps and offers a theoretical basis for interpreting XTA spectra at picosecond/femtosecond time delays. We employ the all-electron, full-potential package \exciting\ \cite{Gulans:2014_exciting}, which implements the linear augmented plane wave plus local orbital (LAPW+LO) method, considered the gold standard for representing Kohn-Sham (KS) wavefunctions \cite{Gulans2018:135486}. Being an all-electron code, \exciting\ can describe optical and core excitations on the same footing by solving the BSE \cite{Vorwerk_2019,Vorwerk:2017gs,Draxl2020:176607,onida2002}, the state-of-the-art approach for computing neutral excitations in solids \cite{onida2002,Martin_Reining_Ceperley_2016}.
 
To investigate pump-induced changes in spectroscopic properties, this work extends the current implementation of the BSE in \exciting\ to account for non-equilibrium states. First, the BSE is solved to compute core excitations in equilibrium XAS spectra. Then, for XAS of non-equilibrium (photoexcited) states, the BSE is solved with constrained occupations of excited electrons and holes created by the pump pulse. This non-equilibrium carrier distribution over KS states and \textbf{k}-points is evaluated in two different ways: i) at \SI{100}{\pico\second} time delay (sections \ref{sec:many_body_interactions} and \ref{sec:L_edge}), by employing cDFT, excited electron (hole) populations are modeled by carriers following a Fermi-Dirac distribution at the conduction-band minimum (CBM) and valence-band maximum (VBM), with a temperature equal to the lattice temperature given by the analysis of the excited-state EXAFS spectrum (see SI \ref{secSI:XTA_thermal_non_thermal}); ii) at femtosecond time delays (section \ref{sec:simulation_fs_XAS}), we track the occupations of excited states by performing RT-TDDFT calculations, providing the occupations of the excited states after a pump pulse, where excited electrons (holes) populate multiple KS states across the Brillouin zone.

Having obtained XAS for the equilibrium and non-equilibrium states, the spectra are normalized by a constant factor such that the calculated absorption maximum of the equilibrium XAS spectrum matches the experimental one (at $\sim\SI{9.669}{\kilo\electronvolt}$). Finally, XTA spectra are calculated as the difference between the XAS spectra of non-equilibrium and equilibrium states.
 
Our extension to the existing BSE implementation also permits disentangling electronic effects, by decomposing many-body interactions into two contributions, namely i) Pauli blocking and ii) Coulomb screening. The former calculates oscillator strengths solely based on excited-state occupations, reflecting changes in transition probability due to the pump excitation. The latter uses excited-state occupations only in the calculation of screened Coulomb interaction, representing additional screening effects from photoexcited carriers.
 

\section{Non-thermal X-ray transient absorption spectra}

\begin{figure*}
    \centering
    \includegraphics[width=\linewidth]{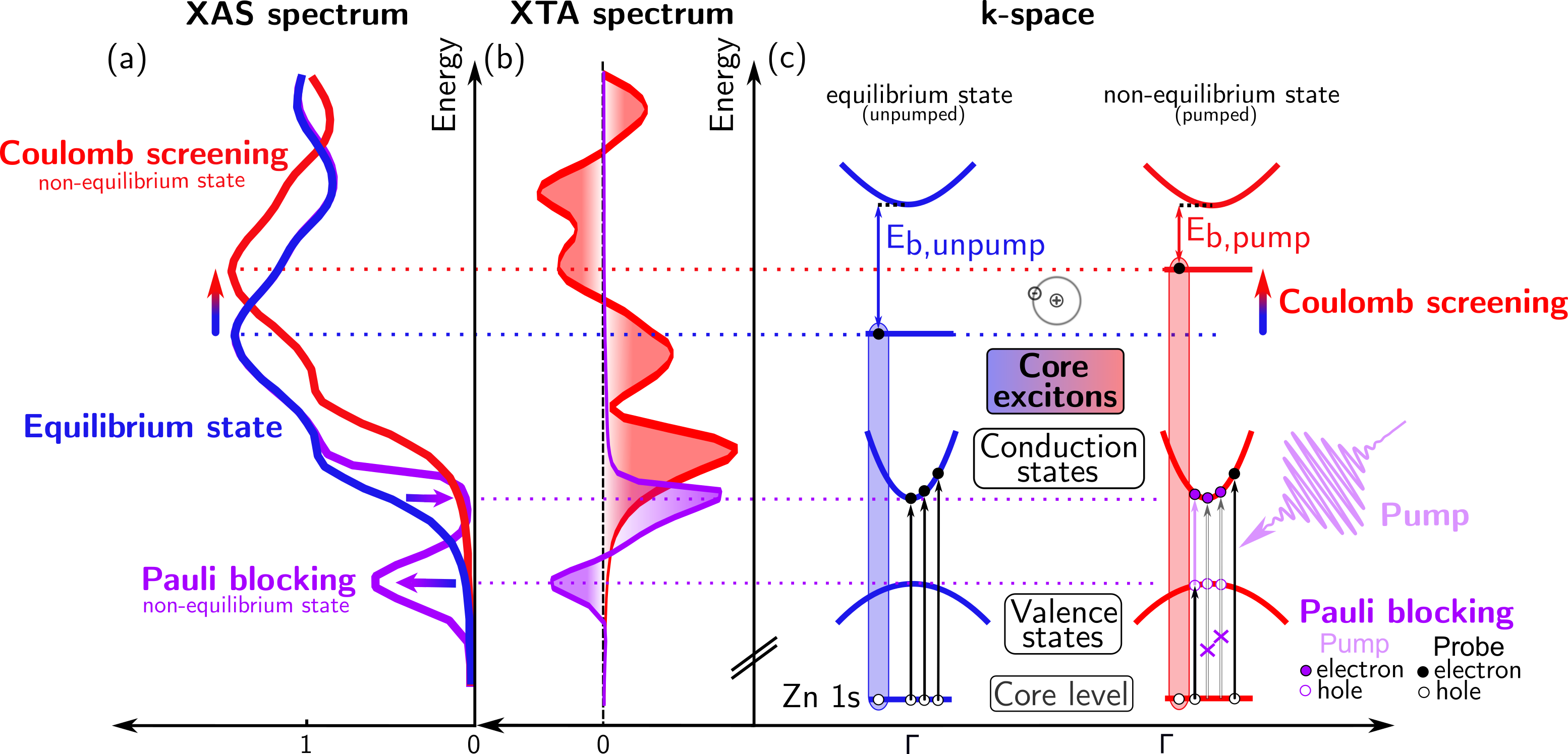}
    \caption{\justifying \textbf{Schematic presentation of the effect of the pump pulse on the (a) XAS spectrum, (b) XTA spectrum, and (c) electronic structure of ZnO at the Zn K-edge.} (a) Blue shift and absorption change of the XAS spectra between the equilibrium/unpumped (blue curve) and non-equilibrium/pumped states resulting from core-exciton Coulomb screening (red curve) and Pauli blocking (purple curve). (b) Corresponding XTA spectra obtained by the difference between the non-equilibrium and equilibrium XAS spectra. Positive (negative) regions, indicate higher (lower) absorption in the non-equilibrium state. (c) Illustration of the electronic structure of ZnO in the equilibrium (blue color) and non-equilibrium (red and purple color) states. In the non-equilibrium state, the pump pulse promotes electrons from the valence band (VB) to the conduction band (CB) (vertical purple arrow). Core excitons form between the core hole in the Zn 1s orbital and electron states (shaded oval areas), which correspond to resonance peaks in the XAS spectrum (blue and red horizontal dotted lines). The binding energy of the core exciton after the laser excitation ($\mathrm{E_{b,pump}}$) is smaller than before the excitation ($\mathrm{E_{b,unpump}}$) due to Coulomb screening by photoexcited carriers in the CB. Pauli blocking occurs in the pre-edge due to newly emptied (hole) states in the VB (black vertical arrow to the VB) and occupied states by photoexcited electrons in the CB (purple-crossed arrows to the CB). A spherical core exciton made of a hole with a large effective mass is illustrated in a semi-classical picture (black circle with positive and negative signs). Transitions to the continuum and thermal effects induced by the laser are omitted for clarity.}
    \label{fig:experiment_schematics}   
\end{figure*}


XAS spectra at the Zn K-edge involves the excitation of an electron from the Zn 1s core orbital to empty valence or continuum states by the concurrent absorption of a photon. The XTA spectrum is the difference between a XAS spectrum in the non-equilibrium (pumped) and the equilibrium (unpumped) state, as shown in Figure \ref{fig:experiment_schematics}a,b. The ZnO thin film is excited with above band-gap \SI{3.49}{\electronvolt} laser pulses, generating electrons and holes in the conduction band and the valence band, respectively (purple circles in Figure \ref{fig:experiment_schematics}c), affecting the XAS spectrum in different ways as illustrated by the black and gray arrows in Figure \ref{fig:experiment_schematics}c. The time delay between the pump and probe pulses is \SI{100\pm10}{\pico\second}. On this time scale, the excess energy of photoexcited carriers has been transferred to the lattice by electron-phonon coupling \cite{Zhukov:2010ex}, heating up the material to a temperature higher than room temperature \cite{Scorticati:2016aa}. Since XAS is sensitive to the incoherent thermal motion of atoms in the ground and excited state \cite{Hayes:2016bz,Rossi2021}, subtraction of the thermal contribution to the XTA signal is a prerequisite to gain access to the non-thermal electronic part that is of interest here. This approach was demonstrated previously for ZnO nanorods \cite{Rossi2021} and applied here as described in SI \ref{secSI:XTA_thermal_effects}. We note that analysis of the high-energy region of the spectrum shows the absence of charge trapping and/or polaron formation in the present experiment on ZnO thin films (see SI \ref{secSI:XTA_thermal_non_thermal}).


\begin{figure*}
	\centering	
    \includegraphics[height=0.5\linewidth]{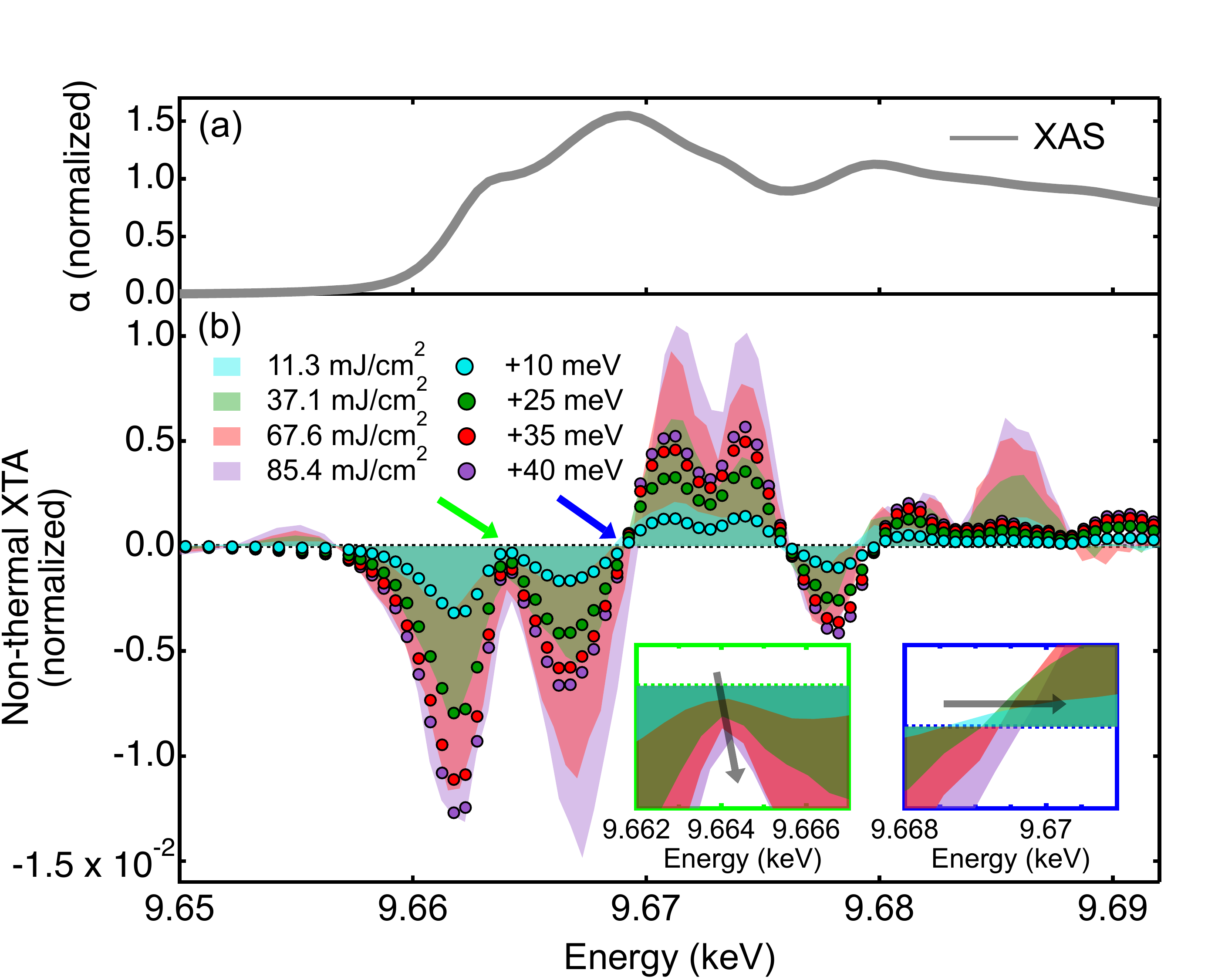}
    \includegraphics[height=0.5\linewidth]{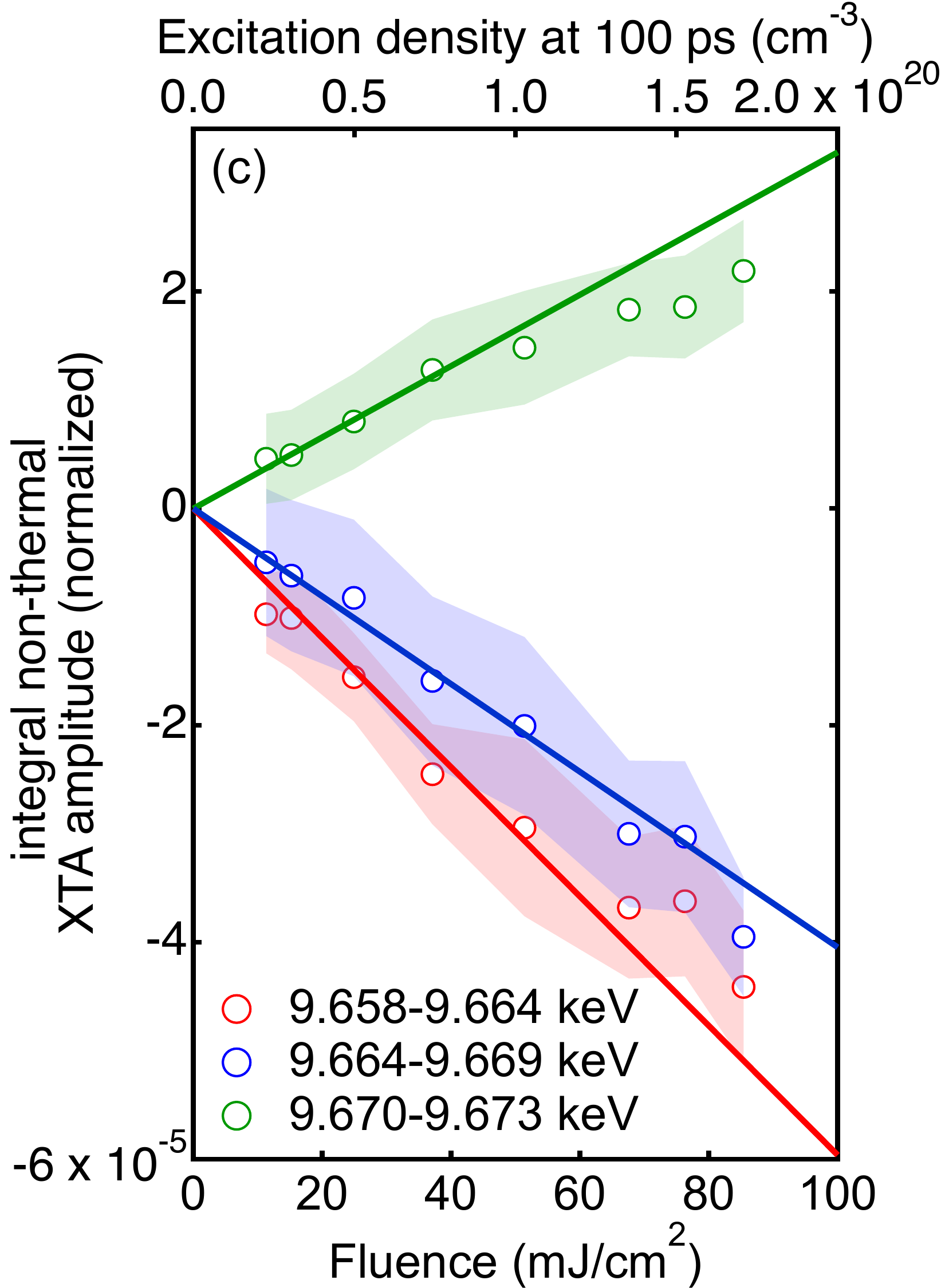}
	\caption{\justifying \textbf{Blue shifts of non-thermal excited-state XAS spectra with excitation fluence.} (a) Normalized equilibrium XAS spectrum at the Zn K-edge of ZnO. (b) Evolution of the normalized non-thermal contribution to the XTA spectrum in the XANES at the Zn K-edge of ZnO at \SI{100}{\pico\second} with excitation fluences (shaded colored areas). Insets zoom into the regions marked by the green and blue arrows. Black arrows indicate a blue shift of the local minimum at $\SI{9.664}{\kilo\electronvolt}$ and of the zero-crossing point in the XTA spectrum at $\SI{9.669}{\kilo\electronvolt}$. Simulated XTA spectra upon blue shifts of the equilibrium XAS spectrum between \SI{10}{} and \SI{40}{\milli\electronvolt} are shown based on a qualitative agreement with the experimental XTA amplitude at $\SI{9.661}{\kilo\electronvolt}$ (colored circles). (c) Evolution of the integrated non-thermal XTA amplitude as a function of excitation fluence over the energy ranges \SI{9.658}{}--\SI{9.664}{\kilo\electronvolt} (red circles), \SI{9.664}{}--\SI{9.669}{\kilo\electronvolt} (blue circles), and \SI{9.670}{}--\SI{9.673}{\kilo\electronvolt} (green circles). Shaded areas define confidence intervals based on the integration of the XTA amplitude $\pm$ one standard deviation. Linear fittings are weighted by the confidence intervals and constrained to the points at fluences $<\SI{50}{\milli\joule\per\centi\metre\squared}$ including the origin.}
    \label{fig:XTA_non_thermal_XANES} 
\end{figure*}


Figure \ref{fig:XTA_non_thermal_XANES}a shows the normalized equilibrium Zn K-edge XAS spectrum of an oriented ZnO (0001) thin film. Details about the background subtraction and edge-jump normalization procedures are given in SI \ref{secSI:data_processing}. The evolution of the \emph{non-thermal} XTA spectrum with increasing laser excitation fluence is given in Figure \ref{fig:XTA_non_thermal_XANES}b (shaded colored areas). It displays two negative features at $\SI{9.661}{\kilo\electronvolt}$ and $\SI{9.667}{\kilo\electronvolt}$ followed by a prominent double positive feature at $\SI{9.671}{\kilo\electronvolt}$ and $\SI{9.674}{\kilo\electronvolt}$. The relative amplitude between the two negative features changes as the fluence increases, which indicates that the transient amplitude scales differently with excitation fluence at different probe photon energies. This trend is corroborated by the lack of isosbestic points in the XTA spectra; the local maximum at $\SI{9.664}{\kilo\electronvolt}$ (green arrow and green-framed inset) and the zero-crossing point at $\SI{9.669}{\kilo\electronvolt}$ (blue arrow and blue-framed inset) shift to higher energies. The fluence dependence of the non-thermal XTA amplitude at different energies is shown in Figure \ref{fig:XTA_non_thermal_XANES}c. It reveals a slightly sublinear variation for the features centered at $\SI{9.661}{\kilo\electronvolt}$ (red circles) and $\SI{9.671}{\kilo\electronvolt}$ (green circles), illustrated by a linear fit constrained to the data points below $\SI{50}{\milli\joule\per\square\centi\metre}$, corresponding to the fluence above which the \emph{total} XTA amplitude (including thermal contributions) and its derivative deviate from linear behavior (SI \ref{secSI:XTA_fluence_dependence}, Figure \ref{figSI:XTA_amplitude_fluence_dependence}). In contrast, the amplitude of the negative feature at $\SI{9.667}{\kilo\electronvolt}$ (blue circles) evolves linearly within the error bars.

We model the non-thermal XTA spectra by assuming that the non-equilibrium XAS spectrum is equal to the equilibrium XAS spectrum but shifted to higher energies (qualitatively represented by the blue and red spectra in Figure \ref{fig:experiment_schematics}a). The results obtained by matching the amplitude of the first negative feature at $\SI{9.661}{\kilo\electronvolt}$ are displayed as colored circles in Figure \ref{fig:XTA_non_thermal_XANES}b. The main features of the experimental non-thermal XTA spectra are qualitatively reproduced by applying positive shifts between \SI{10}{} and \SI{40}{\milli\electronvolt}. While the blue shift accounts for the overall shape of the transient spectra, indicating an increase in the energy of core-level transitions upon photoexcitation, clearly, more subtle amplitude changes cannot be simulated in this way. We now turn to a deeper analysis of the transient XAS spectrum on a theoretical basis accounting for many-body interactions in the photoexcited state.


\section{Comparison between calculated and experimental XAS\label{sec:comp_exp_theo}}


\begin{figure*}
    \includegraphics[width=\linewidth]{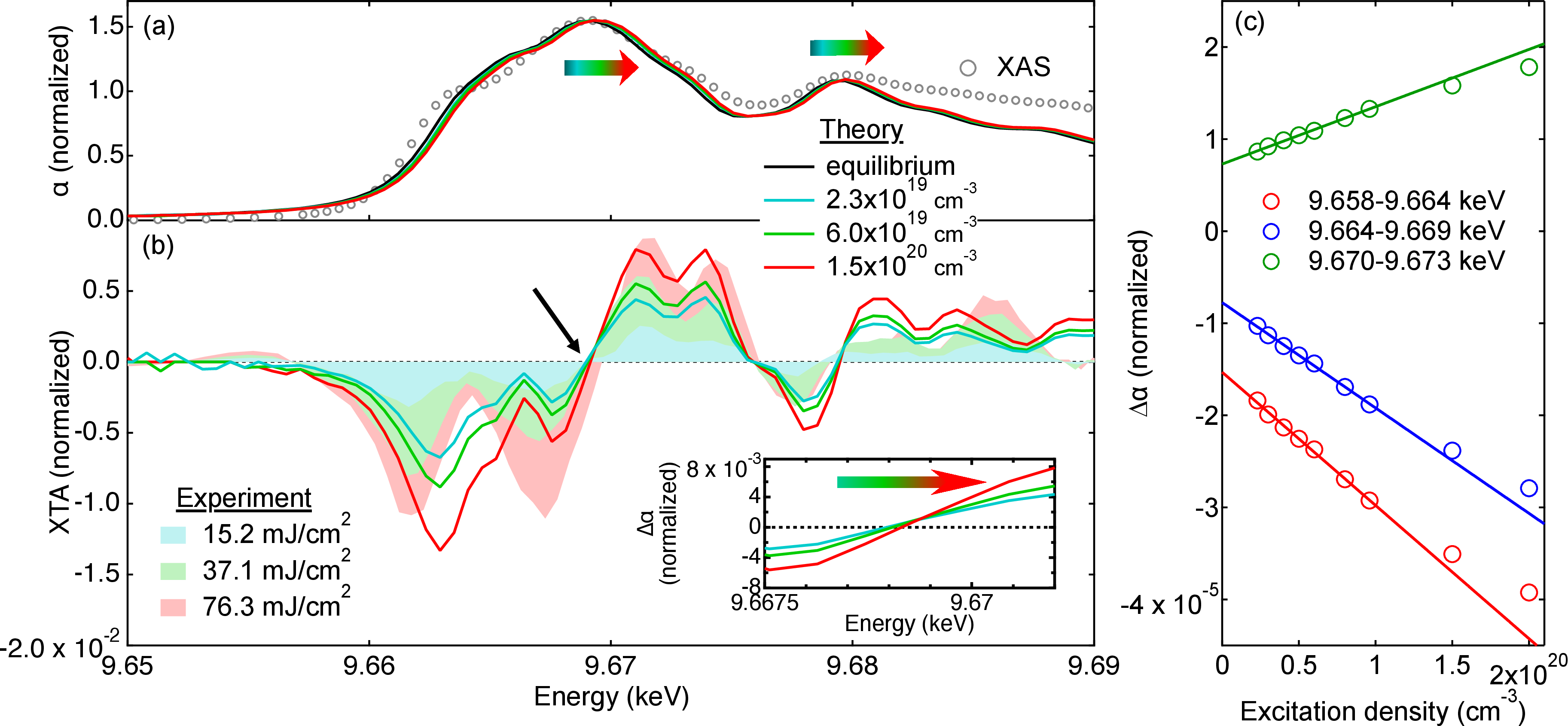}
	\caption{\justifying \textbf{\emph{Ab initio} calculations supporting the blue shift of excited-state XAS spectra.} (a) Computed normalized XAS spectra in the XANES at the Zn K-edge of ZnO with excitation densities ranging from \SI{2.3e19}{\per\cubic\centi\metre} to \SI{1.5e20}{\per\cubic\centi\metre} (colored curves). The computed (black curve) and experimental (gray circles) normalized equilibrium XAS spectra are shown for reference. (b) Calculated (colored curves) and experimental (shaded colored areas) XTA spectra at comparable excitation densities (same color coding). A vertical scaling factor of 0.125 is applied to the calculated XTA spectra. Inset: Magnification of the calculated normalized change of the absorption coefficient ($\Delta\alpha$) at $\sim\SI{9.670}{\kilo\electronvolt}$ (region marked by a black arrow). (c) Computed integral of the normalized XTA amplitude as a function of excitation density over the energy range \SI{9.658}{}--\SI{9.664}{} (red circles), \SI{9.664}{}--\SI{9.669}{} (blue circles), and \SI{9.670}{}--\SI{9.673}{\kilo\electronvolt} (green circles). Lines are linear fits constrained to excitation densities $<\SI{1e20}{\per\cubic\centi\metre}$ to show the sub-linearity or linearity of the calculated XTA amplitude.}
	\label{fig:non_thermal_XTA_theory}
\end{figure*}

\begin{figure*}
    \includegraphics[width=\linewidth]{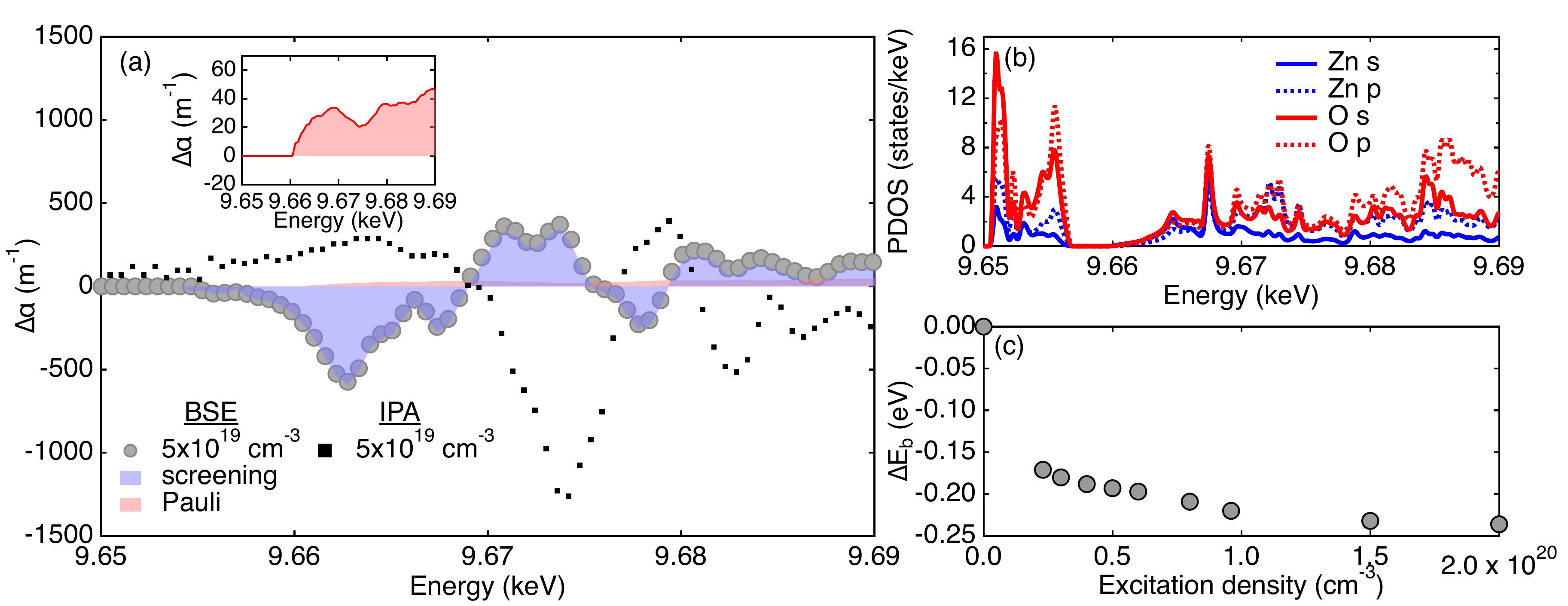}
    \caption{\justifying \textbf{XTA spectra originate from the Coulomb screening of core excitons.} (a) Calculated contributions of the screening modification (shaded blue area) and Pauli blocking (shaded red area) to the XTA spectrum (gray circles) calculated with the BSE for a thermalized excitation density of \SI{5e19}{\per\cubic\centi\metre} at a time delay of \SI{100}{\pico\second}. The calculated XTA spectrum in the IPA is shown for reference (black squares). Insets: zoom of the Pauli blocking contribution into the edge region. (b) Projected DOS in equilibrium. The VBM mainly comprises oxygen s and p orbitals, while the CBM comes from a combination of s and p orbitals from zinc and oxygen. (c) Computed change of the core-exciton binding energy ($\Delta E_b$) as a function of excitation density with respect to the equilibrium exciton binding energy (\SI{734}{\milli\electronvolt}).}
    \label{fig:BSE_IPA_exciton_binding_energy}
\end{figure*}

Figure \ref{fig:non_thermal_XTA_theory}a displays XAS spectra calculated with cDFT+BSE (solid lines), normalized to the peak amplitude of the experimental equilibrium XAS spectrum (gray circles). Computational details are given in the SI \ref{sec:computational_details}. The calculated equilibrium XAS spectrum (black curve) reproduces the main features of the experimental counterpart. The increasing mismatch on the high-energy side can be attributed to the limited number of unoccupied states considered in the calculations, however, we focus here on the lower-energy part of the spectrum. Increasing photoexcited carrier densities results in an increasing blue shift of the calculated XAS spectra (indicated by colored arrows) without a significant change in amplitude. For the shown curves, the shift amounts to $\sim\SI{170}{}-\SI{240}{\milli\electronvolt}$. Calculations of the XAS (SI Figure \ref{figSI:XAS_BSE_VS_IPA}) and XTA spectra (black dotted curve in Figure \ref{fig:BSE_IPA_exciton_binding_energy}a)  in the independent particle approximation (IPA), which disregard electron-hole interactions, instead fail to reproduce the experimental spectra, in particular at the absorption edge. This highlights the importance of including bound states to accurately describe the equilibrium XAS spectrum of ZnO at the Zn K-edge.


Figure \ref{fig:non_thermal_XTA_theory}b highlights the nonlinearity of the XTA spectra with excitation fluence. It compares calculated (colored curves) and experimental non-thermal (shaded colored curves) XTA spectra at similar excitation densities at \SI{100}{\pico\second} (details about the calculation of the excitation density in the SI \ref{secSI:calculation_excitation_density}). A scaling factor of \SI{0.125}{} is applied to the calculated XTA spectra to align them with the experimental ones and to compensate for the larger spectral shifts in the calculations ($\sim\SI{170}{}-\SI{240}{\milli\electronvolt}$) compared with experiment ($\sim\SI{10}{}-\SI{40}{\milli\electronvolt}$). This discrepancy can stem from (i) approximations in determining the laser excitation densities (\emph{e.g.}\ measurement of the laser power or beam waist), (ii) inaccuracies in the estimate of the excitation density at \SI{100}{\pico\second} used in the calculations, which is based on previously reported carrier recombination and trapping time constants (see SI \ref{secSI:calculation_excitation_density}), and/or (iii) the approximate description of the thermalized Fermi-Dirac distribution in the calculations due to the limited \textbf{k}-sampling in the cDFT method.

The computed XTA spectra using the BSE formalism present good qualitative agreement with the experimental spectra, reproducing the two negative signals at $\SI{9.661}{\kilo\electronvolt}$ and $\SI{9.667}{\kilo\electronvolt}$, along with the double positive signal at $\SI{9.672}{\kilo\electronvolt}$. The zero-crossing point around \SI{9.670}{\kilo\electronvolt} blue shifts with the excitation density (inset of Figure \ref{fig:non_thermal_XTA_theory}b), in the same fashion as in the experimental spectra (blue-framed inset of Figure \ref{fig:XTA_non_thermal_XANES}b). In the energy region between $\sim\SI{9.680}{\kilo\electronvolt}$ and $\sim\SI{9.690}{\kilo\electronvolt}$, the calculated XTA spectra display oscillations above zero, similar to experiment.


Figure \ref{fig:non_thermal_XTA_theory}c depicts the evolution of the calculated XTA amplitude as a function of excitation density integrated over the same energy range as for the experimental data in Figure \ref{fig:XTA_non_thermal_XANES}c (same color coding). The calculated XTA amplitude shows a sublinear increase at excitation densities $>\SI{1e20}{\per\cubic\centi\metre}$ at all energy points, similar to experiment. 


\section{Many-body interactions in XTA spectra \label{sec:many_body_interactions}}

In the absence of charge trapping/transfer, polaron formation, and heating effects (see discussion in SI \ref{subsecSI:XTA_contributions}), the non-thermal XTA spectrum is dominated by changes in many-body interactions such as Pauli blocking and long-range Coulomb screening (sketched in Figure \ref{fig:experiment_schematics}). Pauli blocking can be further categorized into phase-space filling and short-range exchange interaction contributions \cite{je2002separation}. Following pump excitation, phase-space filling prohibits X-ray absorption from the Zn 1s core level to the occupied CB (crossed-out arrows in Figure \ref{fig:experiment_schematics}c), leading to a reduction of the XAS amplitude in the excited state. Similarly, it enhances the X-ray absorption to the unoccupied VB states that are emptied by photoexcitation. Besides, electron-hole exchange interactions prevent two particles (either electrons or holes) with the same spin from occupying the same quantum state. After excitation, the delocalized nature of the photoexcited carriers means that they do not significantly affect the short-range exchange interaction, which relies on the close spatial proximity between particles. The attractive long-range electron-hole Coulomb interaction instead is a prerequisite for the formation of bound excitons. After photoexcitation, the excited carriers screen and thus decrease the electron-hole Coulomb interaction. Here, we refer to the combined effects of phase-space filling and exchange interaction modifications as Pauli blocking, while the screening of the Coulomb interaction is referred to as Coulomb screening. To distinguish these different interactions and their effects on the XTA spectra, we perform non-equilibrium BSE calculations with constrained excited-carrier distributions and decompose the many-body effects in the two contributions (details in Section \ref{sec:theory_details}).

Figure \ref{fig:BSE_IPA_exciton_binding_energy}a separates the contributions of Pauli blocking and Coulomb screening to the non-thermal XTA spectra. The calculated spectra (gray circles) are dominated by the effect of the modified core-exciton screening (shaded blue area). The effect of Pauli blocking (shaded red area), mainly originating from the modification of the exchange interaction due to photoexcited carriers (SI \ref{secSI:supp_decomposition_XTA}, Figure \ref{figS:BSE_IP}), is nearly an order of magnitude weaker. The negligible contribution of phase-space filling is related to the localized distribution of excited electrons in the Brillouin zone (BZ) after $\SI{100}{\pico\second}$, which only blocks a small fraction of the states contributing to the X-ray absorption transitions at the K-edge (carrier distribution displayed in SI \ref{secSI:supp_decomposition_XTA}, Figure \ref{figS:ps_fs_occupation}a).


The increase of the XTA signal (Figure \ref{fig:non_thermal_XTA_theory}a,b,c) with increasing excitation density is related to the charge transfer occurring between oxygen and zinc atoms upon photoexcitation above the band gap due to the orbital composition at the VB maximum and the CB minimum (see the projected DOS in Figure \ref{fig:BSE_IPA_exciton_binding_energy}b). The proximity of the photoexcited electron density to the Zn 1s core hole enhances the core-exciton screening, which is expected to increase in magnitude with photoexcitation density, corresponding to an increase of the XTA amplitude (Figure \ref{fig:non_thermal_XTA_theory}b,c). 

The Coulomb screening enhancement renormalizes the exciton binding strength. Figure \ref{fig:BSE_IPA_exciton_binding_energy}c shows the exciton binding energy of the transition from the Zn 1s core level to the CBM+1 state at $\Gamma$ as a function of excitation density. At equilibrium, the exciton binding energy is \SI{734}{\milli\electronvolt}, which is reduced by \SI{171}{} and by \SI{236}{\milli\electronvolt} for excitation densities of \SI{2e19}{} and \SI{2e20}{\per\cubic\centi\metre}, respectively, saturating at around \SI{240}{\milli\electronvolt} at high excitation densities $\gtrsim\SI{1e20}{\per\cubic\centi\metre}$. Figure \ref{figS:IP_renormalization} in the SI shows that even at the highest excitation density the renormalization of the core-level transition energy in the IPA is $\sim\SI{4}{\milli\electronvolt}$ only. Hence, the blue shift of the non-equilibrium XAS spectrum (Figure \ref{fig:non_thermal_XTA_theory}a) is almost uniquely due to the reduction of the exciton binding energy induced by the enhancement of the core-exciton screening by photoexcited carriers. The current results represent a substantial improvement over previous DFT-based calculations of XTA spectra \cite{Rossi2021} which were calculated without explicitly considering electron-hole interaction and relied on a pseudopotential approach. In Ref. \cite{Rossi2021}, the contribution of core-hole screening was modeled by the difference between XAS spectra calculated with fully screened and unscreened core-hole potentials. Here instead, we employ the all-electron LAPW+LO method, which treats core excitons on an equal footing with valence excitons. By incorporating the electron-hole interaction within the BSE formalism, we are able to capture partial screening of the core hole, clearly resulting in better agreement with experimental XTA spectra.


\begin{figure}[ht!]
    \begin{subfigure}[t]{0.38\linewidth}
        \adjustbox{trim={-0.05\width} {0.01\height} {.00\width} {.00\height},clip}%
        {\includegraphics[width=\linewidth]{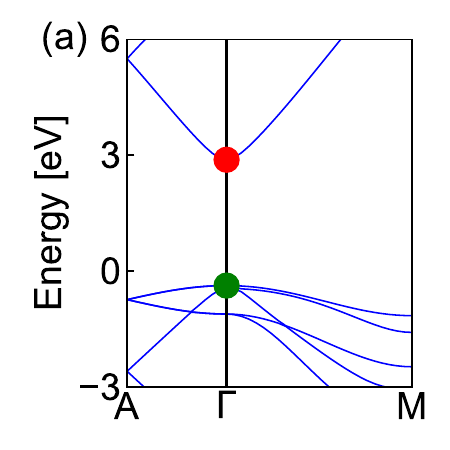}}
    \end{subfigure}
    \begin{subfigure}[t]{0.5\linewidth}
        \adjustbox{trim={-0.03\width} {.01\height} {0.0\width} {.00\height},clip}%
        {\includegraphics[width=\linewidth]{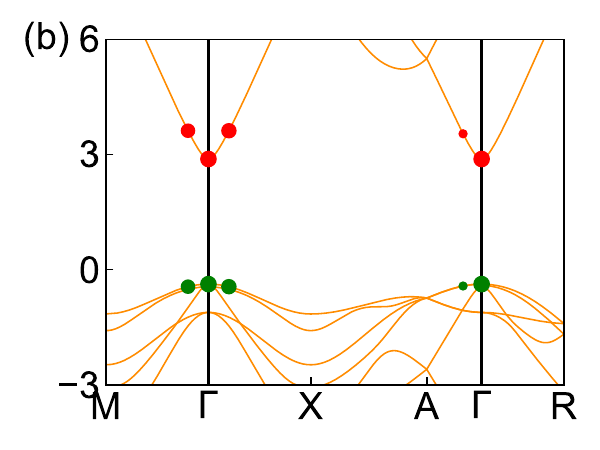}}
    \end{subfigure}
    \begin{subfigure}[t]{0.9\linewidth}
        \adjustbox{trim={-0.03\width} {.00\height} {-0.3\width} {.00\height},clip}%
        {\includegraphics[width=\linewidth]{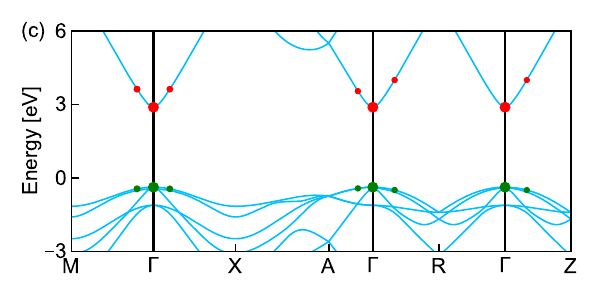}}
    \end{subfigure}
    \begin{subfigure}[t]{0.97\linewidth}
        \adjustbox{trim={0.0\width} {.00\height} {0.0\width} {.00\height},clip}%
        {\includegraphics[width=\linewidth]{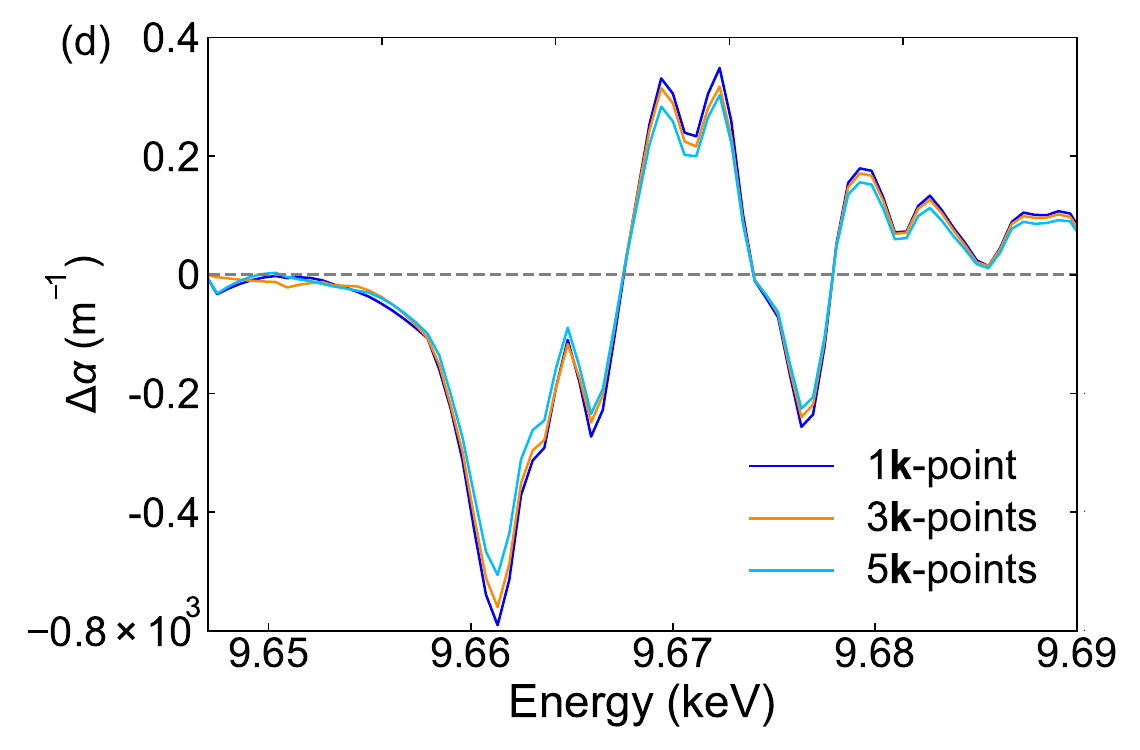}}
    \end{subfigure}
    \caption{\justifying \textbf{Reduced screening of core excitons by delocalized carrier distributions.} Distributions of photoexcited electrons (red circles) and holes (green circles) over (a) 1, (b) 3, and (c) 5 \textbf{k}-points at a fixed excitation density of $\SI{5.0e19}{\per\cubic\centi\metre}$. The area of the circles is proportional to the occupation at a given \textbf{k}-point. (d) Computed XTA spectra for the different distributions of excited carriers (same color coding as in the band diagrams).}
    \label{fig:local_delocal}
\end{figure}


To explain the nonlinear dependence of the XTA amplitude with excitation density, as shown in Figure \ref{fig:non_thermal_XTA_theory}c, we examine the effect of excited carrier occupations in the band diagram. Figure \ref{fig:local_delocal} presents three examples of occupations of the VB and the CB, all at the same excitation density of $\SI{5.0e19}{\per\cubic\centi\metre}$. In panel (a) only the $\Gamma$-point is populated, while in panels (b) and (c) only \SI{41.2}{\percent} and \SI{30}{\percent}, respectively, of the excited electrons are at the $\Gamma$-point, and the remaining electrons extend farther out in \textbf{k}-space following a Fermi-Dirac distribution. The corresponding XTA spectra in panel (d) show that increased delocalization of carriers leads to reduced amplitudes. Since Pauli blocking is negligible in all three cases, the differences are attributed to variations in the core-exciton screening. This reveals that photoexcited carriers away from the $\Gamma$-point contribute comparatively little to the modification of the core-exciton screening. Thermalized distributions at increasing excitation densities populate states further away from $\Gamma$ due to the phase-space filling at band extrema. These additional carriers are expected to modify less the core exciton screening, which explains the sublinear increase in the non-thermal XTA amplitude with increasing photoexcited carrier density as shown in Figure \ref{fig:non_thermal_XTA_theory}c, in agreement with the experimental observations in Figure \ref{fig:XTA_non_thermal_XANES}c. We conclude that the modification of the core-exciton screening depends on both the excitation density and the distribution of excited carriers in the BZ. We propose that these two parameters can be used to control core-exciton screening with tunable excitation pulses, as discussed in Section \ref{sec:simulation_fs_XAS}. Reciprocally, combining experiment and theory can be used to yield local carrier concentrations and distributions in the band diagram of photoexcited semiconductors or under operando conditions.

\section{Extension to L-edge spectroscopy\label{sec:L_edge}}


\begin{figure*}
    \centering
    \includegraphics[height=0.34\linewidth]{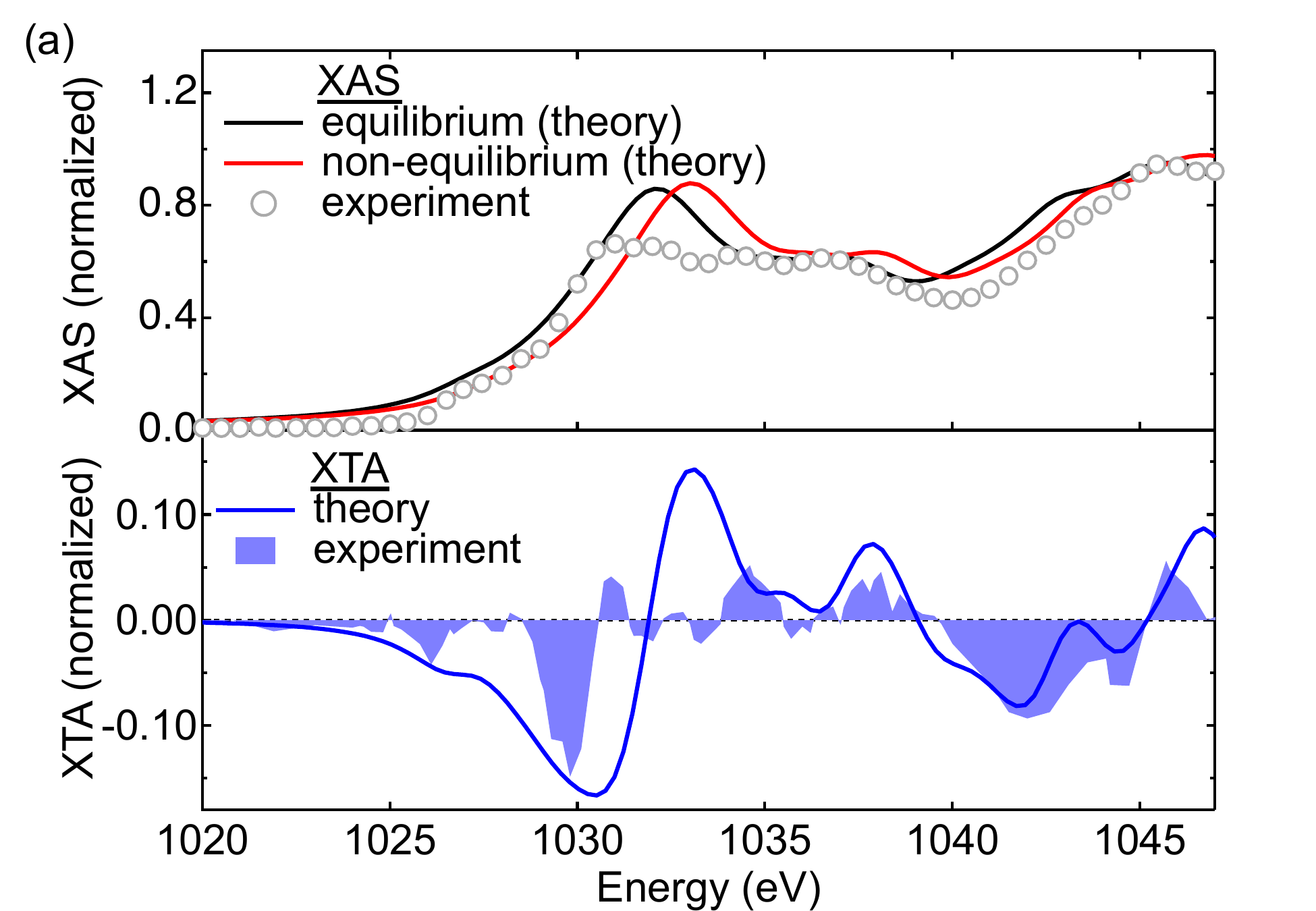}
    \includegraphics[height=0.34\linewidth]{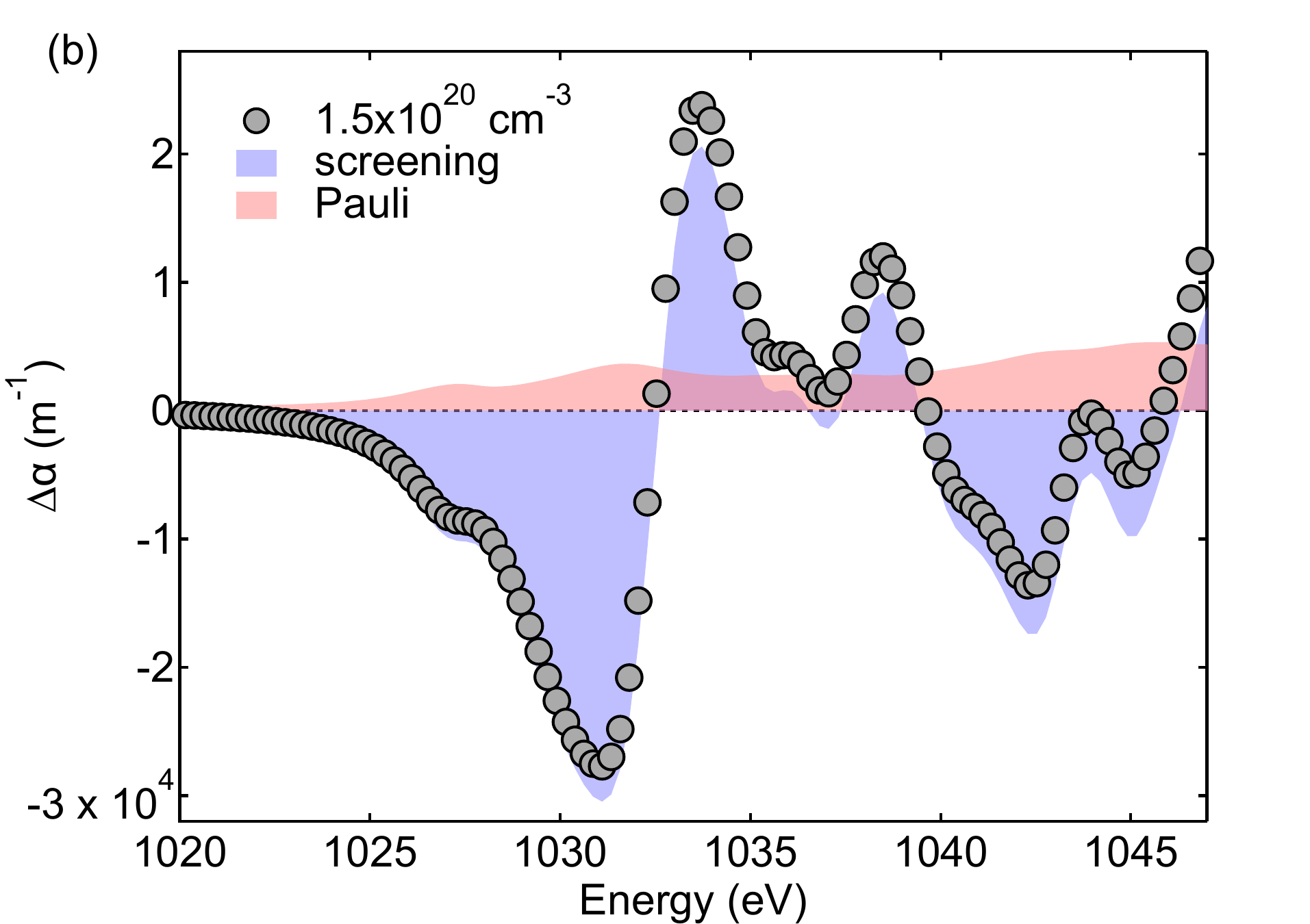}
    \caption{\justifying \textbf{Core-exciton screening also present at the Zn L$_3$-edge in ZnO.} (a) Experimental normalized non-thermal XTA spectrum at \SI{100}{\pico\second} (shaded blue area) and calculated XTA spectrum (blue curve, vertically scaled by a factor of 2.2, lower panel). The experimental XAS spectrum (gray circles) and the calculated XAS spectra in equilibrium (black curve) and non-equilibrium (red curve) are shown for reference (upper panel). (b) Calculated contributions of the Coulomb screening modification (shaded blue area) and Pauli blocking (shaded red area) to the XTA spectrum (black circles). The laser fluence is \SI{14}{\milli\joule\per\square\centi\metre}, the excitation wavelength is \SI{343}{\nano\meter} (\SI{3.61}{\electronvolt}). The calculations are performed for an excitation density of \SI{1.5e20}{\per\cubic\centi\metre}, comparable to the estimated experimental excitation density at \SI{100}{\pico\second}.}
    \label{fig:ZnL23_core_hole_screening}
\end{figure*}

The effect of the modified core-exciton screening is not limited to the case of the K-edge. To illustrate its larger scope, we have performed XTA measurements in transmission of ZnO thin films at the Zn L$_3$-edge. Figure \ref{fig:ZnL23_core_hole_screening}a shows a comparison between the non-thermal XTA spectrum at \SI{100}{\pico\second} and a calculation by cDFT+BSE. The calculated spectrum (continuous blue curve) reproduces the main features of the experimental counterpart (shaded blue curve, details of the procedure in SI \ref{secSI:XTA_thermal_effects}), but being weaker by a factor $\sim2.2$. The calculation achieves good agreement for the relative amplitudes of the different features, which are, qualitatively, a blue shift of the XAS spectrum in the excited state by $\sim\SI{750}{\milli\electronvolt}$ (black and red curves in Figure \ref{fig:ZnL23_core_hole_screening}a). This is $\sim$3 times larger than the calculated renormalization of the exciton binding energy at the Zn K-edge for the same excitation density of \SI{1.5e20}{\per\cubic\centi\metre} (\SI{232}{\milli\electronvolt}). The equilibrium exciton binding energy corresponding to transitions from the Zn L$_3$ core level (2p$_{3/2}$) to the CBM+1 is \SI{886}{\milli\electronvolt}, slightly larger than that at the Zn K-edge ($\SI{734}{\milli\electronvolt}$). Based on an early description of core-hole screening by Slater \cite{Slater1930:189278}, 1$s$ core holes are efficiently screened by a localized cloud of outer electrons, while more delocalized 2$p$ core holes are less efficiently screened due to deeper electrons not participating in the screening \cite{Mauchamp2009,Bunau2013}. However, in the case of XTA, where photoexcited electrons are more delocalized and highly polarizable, we expect a more efficient screening of the delocalized 2$p$ core hole at the L$_3$-edge than of the 1$s$ core hole at the K-edge, in line with experimental observations.

Figure \ref{fig:ZnL23_core_hole_screening}b displays the Coulomb screening and Pauli blocking contributions to the calculated L$_3$-edge XTA spectra, showing that the former dominates over the latter at the L$_3$-edge as well. This result highlights the importance of core-exciton screening modifications in core-level spectra of photoexcited materials, which should \emph{a priori} affect any core-level transition. In fact, in recent works employing RT-TDDFT, it is claimed that electronic screening of the core hole contributes to the XTA spectra at the M-edge of photoexcited transition metals \cite{Volkov2019,Cushing:2020sa} and low-band gap semiconductors \cite{Cushing:2019jl}.


\section{Core-exciton screening on the femtosecond time scale\label{sec:simulation_fs_XAS}}

\begin{figure*}[ht!]
    \includegraphics[width=1\linewidth]{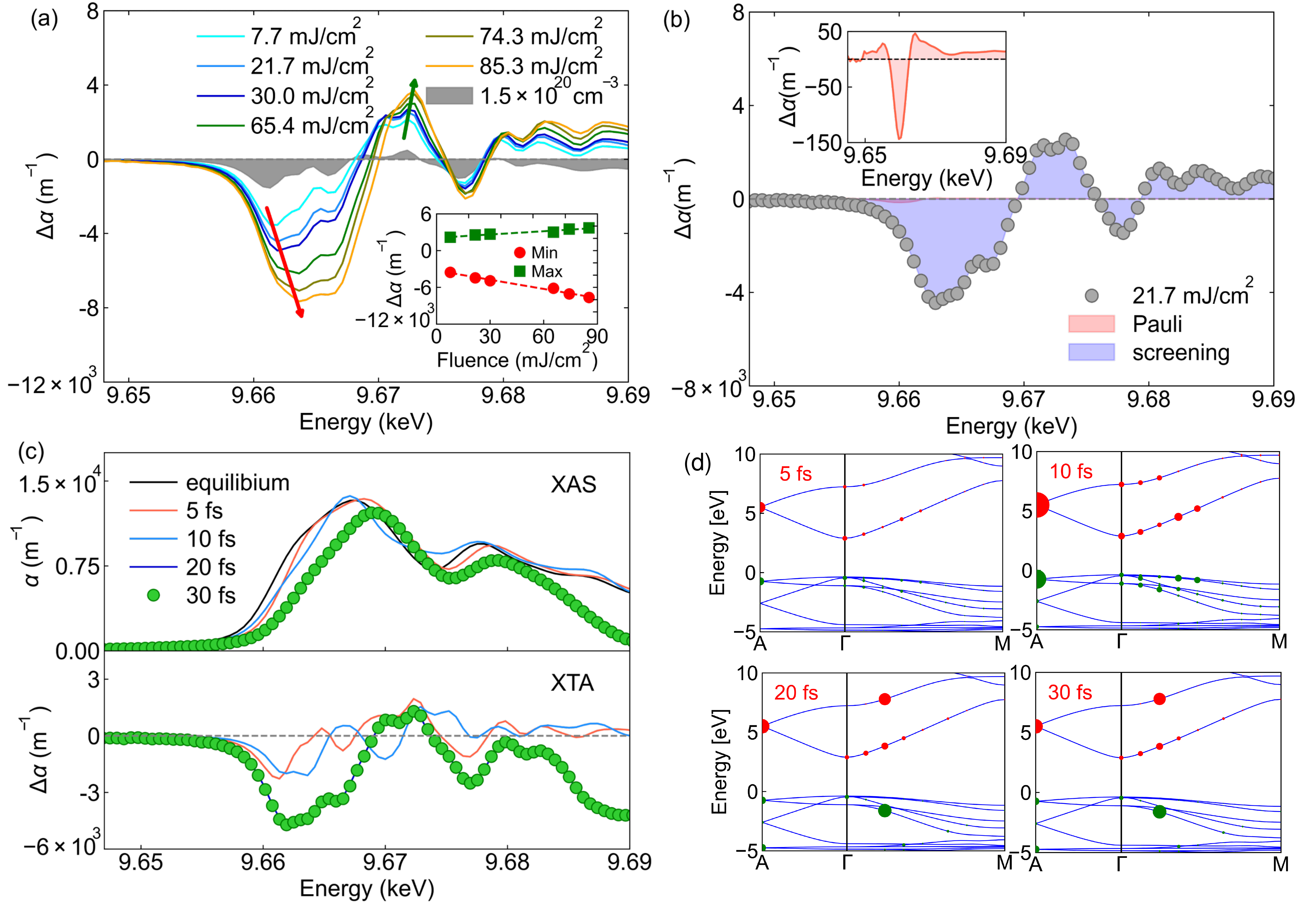}
    \caption{\justifying \textbf{Optical control of core excitons by Coulomb screening in calculated femtosecond XTA spectra at the Zn K-edge.} (a) Computed XTA spectra with pump fluences ranging from \SI{7.7}{\milli\joule\per\square\centi\metre} to \SI{85.3}{\milli\joule\per\square\centi\metre} (colored curves) at a time delay of \SI{20}{\femto\second}. The XTA spectrum at \SI{100
    }{\pico\second} time delay with an excitation density of \SI{1.5e20}{\per\cubic\centi\metre} is shown for reference (shaded grey area). Inset: Computed XTA amplitude at the minimum ($\SI{9.662}{\kilo\electronvolt}$, red circles) and maximum ($\SI{9.672}{\kilo\electronvolt}$, green squares) of the XTA spectra as a function of pump fluence. (b) Contributions of the screening modification (shaded blue curve) and Pauli blocking (shaded red area) to the XTA spectra (black circles) for an excitation fluence of \SI{21.7}{\milli\joule\per\centi\metre\squared} at a time delay of \SI{20}{\femto\second}. Inset: zoom into the edge region. (c) XTA spectra at time delays of \SI{5}{}, \SI{10}{}, \SI{20}{}, and \SI{30}{\femto\second} at a pump fluence of \SI{21.7}{\milli\joule\per\square\centi\metre} and a pulse duration of \SI{10}{\femto\second}. (d) Distribution of excited electrons (red circles) and holes (green circles) at \SI{5}, \SI{10}, \SI{20}, and \SI{30}{\femto\second} at a pump fluence of \SI{21.7}{\milli\joule\per\square\centi\meter}.}
    \label{fig:tddft}
\end{figure*}

In the above sections, we focused on picosecond XTA spectra, which cannot resolve the ultrafast dynamics that occurs immediately after short-pulse photoexcitation. Therefore, femtosecond XTA spectra are necessary to capture transient changes that evolve on sub-picosecond timescales, such as the spectral motion of the exciton manifold and ultrafast screening effects, which are essential for understanding the full picture of ultrafast phenomena in materials. Unlike the picosecond XTA spectra calculations, where the excited-state occupations are obtained using cDFT, the occupations for the femtosecond XTA spectra are derived from RT-TDDFT, followed by non-equilibrium BSE calculations to compute the spectra. This approach can capture the change of core excitations after a short pump pulse, which can be measured with X-ray free-electron lasers \cite{Asakura:2020jq}.

The temporal evolution of the number of excited electrons per unit cell obtained by RT-TDDFT simulations is shown in Figure \ref{figSI:density_vs_time} in the SI \ref{secSI:occ_fs}. The intensity of the excitation is characterized by the pump fluence instead of the excitation density, since the latter changes with time. The carrier occupations at \SI{20}{\femto\second} are used to compute the femtosecond XTA spectra shown in Figure \ref{fig:tddft}a (colored curves), which exhibit a lineshape similar to the XTA spectra at \SI{100}{\pico\second} (Figure \ref{fig:non_thermal_XTA_theory}b), with two negative signals at $\SI{9.662}{}$ and $\SI{9.667}{\kilo\electronvolt}$ and positive signals at $\SI{9.672}{\kilo\electronvolt}$. The spectra with larger amplitudes are associated with larger blue shifts of the non-equilibrium XAS spectra, which is attributed to the larger number of excited carriers at femtosecond time delays and thus a stronger core-exciton screening. The relative amplitude between the two negative low-energy features changes with increasing excitation density to a much larger extent than in the cDFT calculations (Figure \ref{fig:non_thermal_XTA_theory}b), which leads to a blue shift of the minimum (red arrow) and the maximum (green arrow) of the XTA spectra. The larger change in relative amplitude is attributed to excitonic transitions shifting in energy by different amounts and leading to a "shuffling" of the excitonic transitions. At time delays shorter than the characteristic cooling time ($\lesssim\SI{200}{\femto\second}$ at the carrier excess energy given by the pump \cite{Zhukov:2010ex}), this effect is particularly large, since variations of both the density of carriers and their distribution are simultaneous, leading to changes in both the XAS spectral shape and the energy shift (Figure \ref{figSI:XAS_fs} in the SI \ref{secSI:occ_fs}). We note that the relative change in the amplitude of the XTA spectral features is similar to the experimental trend at \SI{100}{\pico\second} in Figure \ref{fig:XTA_non_thermal_XANES}a, although no "shuffling" of excitonic transitions takes place at picosecond time delays (Figure \ref{fig:non_thermal_XTA_theory}b). This observation illustrates that cDFT does not account for dynamic carrier occupations on picosecond timescales (see discussion in \ref{sec:discussion}). The inset of Figure \ref{fig:tddft} shows the calculated evolution of the XTA amplitude at the minimum (red circles) and maximum (green squares) of the XTA spectrum, revealing a linear increase with the excitation fluence.

Figure \ref{fig:tddft}b decomposes the XTA spectra into contributions from modified Coulomb screening (shaded blue area) and Pauli blocking (shaded red curve). Similar to the picosecond spectra, the modified Coulomb screening dominates the femtosecond spectra. The contribution from Pauli blocking (shaded red area) is nearly two orders of magnitude weaker than the overall XTA spectrum. Pauli blocking (inset of Figure \ref{fig:tddft}b) exhibits a prominent negative absorption change around $\sim\SI{9.662}{\kilo\electronvolt}$, differing from the positive signal observed in the picosecond spectra (inset of Figure \ref{fig:BSE_IPA_exciton_binding_energy}a). At short time delays, compared to the electron-phonon scattering time \cite{Zhukov:2010ex}, excited electrons occupy a delocalized region across the BZ (SI \ref{secSI:supp_decomposition_XTA}, Figure \ref{figS:ps_fs_occupation}b), contributing more efficiently to blocking core excitations (phase-space filling), consistent with previous femtosecond XTA measurements in semiconductors \cite{Verkamp:2019tc,Cushing:2019jl,Park2022:nc}.

Figure \ref{fig:tddft}c shows the time evolution of calculated XAS and XTA spectra at a fluence of \SI{21.7}{\milli\joule\per\centi\metre\squared}. They exhibit large lineshape changes with the time delay due to highly dynamic carrier occupations (Figure \ref{fig:tddft}d) and variations of the excitation density (Figure \ref{figSI:density_vs_time} in Supporting information). The excitation density is \SI{5.8e20}{\per\cubic\centi\metre} and \SI{1.3e21}{\per\cubic\centi\metre} at time delays of \SI{5}{\femto\second} and \SI{10}{\femto\second}, and it remains at \SI{2.9e21}{\per\cubic\centi\metre} after \SI{20}{\femto\second}. The largest blue shift in the excited-state XAS spectra is achieved at \SI{20}{\femto\second} (blue curve), and remains at \SI{30}{\femto\second} (green circles). Although the carrier density at \SI{30}{\femto\second} is smaller than that at \SI{10}{\femto\second}, the blue shift is larger, which illustrates that the effect of a higher degree of localization of the electron distribution in \textbf{k}-space on the core-exciton screening at later time delays overwhelms the effect of a smaller excitation density. Hence, the time evolution of the excitation density and its distribution in the BZ leads to dynamical core-excitation screening on femtosecond timescales, which shows that core-exciton resonances can be tuned depending on photoexcited carrier densities and distributions on ultrafast timescales.


\section{Discussion\label{sec:discussion}}


The effect of dynamic core-exciton screening on time-resolved XAS spectra has not often been discussed in the literature. One of the possible reasons for this lack of previous examples is the fact that only a handful of studies adequately incorporate thermal effects in the analysis of photoexcited states of solids at picosecond time delays \cite{Hayes:2016bz, Rossi2021, Rein2021}. The accurate subtraction of thermal contributions requires temperature-dependent XAS and XTA spectra with comparable and sufficiently high statistics. Heating takes hundreds of femtoseconds to picoseconds to develop in solid-state materials. Therefore, on ultrafast ($<\SI{100}{}$--\SI{200}{\femto\second}) time scales, it is possible to measure transient spectra representative of electronic perturbations in photoexcited materials before significant lattice heating occurs. We thus expect to see more systematic studies on the modification of core-exciton screening on shorter timescales using lab-based fs X-ray sources \cite{Liu2021} and at X-ray free-electron lasers \cite{Asakura:2020jq}. The state-of-the-art computational method presented in this paper serves as a crucial approach for interpreting such ultrafast dynamics.

Femtosecond photoemission measurements have shown that core-hole spectral functions in WSe$_2$ are renormalized by photoexcited carriers \cite{Dendzik2020}. The renormalization observed at subpicosecond time delays is substantial ($\sim\SI{100}{\milli\electronvolt}$), contrasting with the significantly smaller single-particle energy shifts seen in this work ($<\SI{4}{\milli\electronvolt}$). In ref.\ \cite{Dendzik2020}, the magnitude of the spectral shift is intimately related to the temporal increase in the population of free charge carriers, which originates from the initial exciton population through a Mott transition. Given that the excitation densities in this work are comparable to or exceed the Mott density for optical excitons ($n_M\sim\SI{3.7e19}{\per\cubic\centi\metre}$ \cite{Ozgur:2005it}), photoexcited carriers in ZnO primarily exist as free charge carriers that screen core excitons. The findings from time-resolved photoemission are a strong indication that core excitons can enter a different screening regime in measurements performed below the Mott density, where optical excitons and free charge carriers coexist. Previous XTA measurements in this regime revealed significant signals attributed to phase-space filling and core-exciton screening \cite{Rossi2021}. This implies that the effect of core-exciton screening on the XTA spectrum becomes sufficiently weak to be comparable to phase-space filling. Future XTA measurements in this excitation regime and at short time delays represent an interesting extension to the current work, potentially unveiling alternative mechanisms to control the screening of core excitons.

Reported Wannier-Mott exciton radii at the optical gap of ZnO range from \SI{0.9} to \SI{2.34}{\nano\meter} \cite{Fonoberov:2004kh,Gil2002:79003,Dietz:1961hf,Wischmeier2006:276991,Thomas1960:219862,Park:1966jy,Senger2003:27149}, with the most recent works reporting a value of $\sim\SI{1.4}{\nano\meter}$. In contrast, the equilibrium core-exciton radius in the hydrogenoid model is \SI{2}{\angstrom} at the Zn K-edge, comparable to the Zn-O bond distance at room temperature ($\SI{1.97}{}-\SI{1.99}{\angstrom}$), which falls into the category of Frenkel excitons \cite{Frenkel1931:17034} (SI \ref{secSI:electron distribution}, Figure \ref{figSI:electron_distribution}a). It implies that core excitons are only sensitive to structural dynamics occurring within the first coordination shell. We explain the validity of the separation between thermal (atomic vibrations) and non-thermal (exciton) contributions in the near-edge region of the XTA spectrum to originate from the small dimension of core excitons. This is a fundamental hypothesis made in this work. For instance, since acoustic phonons cause mainly long-range crystal deformations, their coupling to localized core excitons is weak \cite{Cherukara2017}. The weak coupling between optical excitons and phonons in covalent semiconductors is a fundamental difference to ionic compounds, in which excitons with binding energies of $\sim\SI{1}{\electronvolt}$ exhibit a strong coupling to phonons \cite{Carson1987:24618}. Core excitons also have limited coupling with thermal vibrations, since the mean-square displacement of the atoms, quantified by the Debye-Waller factor, is of the order of \SI{2}{\percent} ($\sim\SI{40}{\milli\angstrom}$) of the equilibrium Zn-O bond distance at room temperature, which is negligible compared to the exciton radii. Due to the short lifetime of the core hole at transition metal K-edges, we exclude the effect of structural motion occurring between the absorption and emission of the X-ray photon on the core exciton spectrum \cite{Mahan:2012wd}. At the Zn L$_3$-edge, the exciton radius is 1.78 \AA\ (SI \ref{secSI:electron distribution}, Figure \ref{figSI:electron_distribution}b), which is also smaller than the first coordination shell. However, since the lattice temperature is expected to be significantly higher during the Zn L$_3$-edge experiment due to the vacuum environment and the low thermal contact conductance with the substrate, we expect the incoherent atomic motion to span a significant fraction of the interatomic bond distance, and thus the dynamics of core excitons and atoms cannot be considered as independent as at the K-edge. It likely explains the poorer agreement between the non-thermal XTA and the calculations shown in Figure \ref{fig:ZnL23_core_hole_screening}a. Hence, the thermal vibrations encountered here are expected to have a limited impact on the electronic structure of core excitons in covalent semiconductors.

The calculations show that the change of relative amplitude between the two negative low-energy features in the XTA spectrum is negligible in cDFT (Figure \ref{fig:non_thermal_XTA_theory}a), whereas it is substantial in RT-TDDFT (Figure \ref{fig:tddft}a). A fundamental approximation made in the cDFT calculations is that the excited-state distribution of carriers matches a Fermi-Dirac distribution with a defined temperature, which is an accurate description when averaging instantaneous distributions over timescales significantly longer than the electron-electron and electron-phonon scattering times. In reality, the distribution of carriers is dynamic, with more high-energy carriers present at a given time due to thermal fluctuations in the material as the temperature increases. Hence, we expect an increase in the lattice temperature with the excitation density to promote higher energy carriers and to increase the dynamic fluctuations of the carrier density. RT-TDDFT captures dynamic fluctuations in the carrier distribution due to the temporal variation of the pump field, and thus correctly captures the relative change in amplitude between the features in the XTA spectrum as more energy is generated at increasing excitation fluences. The current results show that improvements are needed that incorporate dynamic carrier fluctuations, even at long time delays, to capture the dynamic screening and reproduce the excited-state XAS spectrum of photoexcited semiconductors.


\section{Conclusions}

In conclusion, by combining experiment and theory, we have shown with the example of ZnO that electronic effects in X-ray transient absorption (XTA) spectra of photoexcited semiconductors result dominantly from the non-equilibrium carrier screening of core excitons with a negligible contribution from phase-space filling. These results apply to any semiconducting material when photoexcitation does not lead to the formation of localized carriers (polarons and/or trapped charges). Screening of the core hole reduces exciton binding energies, resulting in a blue shift of XAS spectra in the excited state. The evolution of the XTA spectral amplitude as a function of excitation density indicates that the core-exciton screening depends on both the excitation density and the degree of excited carrier delocalization in reciprocal space. The inclusion of many-body effects in the theoretical description of these phenomena is a prerequisite to accurately reproduce experimental results. The modification of the core-exciton screening could be used to quantify local densities of conduction carriers in materials, taking advantage of the chemical sensitivity of XAS to probe elements in specific materials.  For instance, it could be used to investigate the efficiency of carrier transfer at interfaces \cite{PelliCresi2021} or in materials under operando conditions \cite{Chen:2011jc}. The current work also shows that the electronic properties of core excitons with kiloelectronvolt resonances can be modulated by photoexcitation, which paves the way for excitonics beyond the soft X-ray range \cite{Moulet2017:122783}.


\section*{Data availability}

The experimental data that support the findings of this article are available upon reasonable request. All input and output files of the \emph{ab initio} calculations are available in the NOMAD data infrastructure \cite{nomad}.


\begin{acknowledgments}

This work made use of the Pulsed Laser Deposition Shared Facility at the Materials Research Center at Northwestern University supported by the National Science Foundation MRSEC program (DMR-2308691) and the Soft and Hybrid Nanotechnology Experimental (SHyNE) Resource (NSF ECCS-2025633). This work is supported by the U.S. Department of Energy, Office of Science, Office of Basic Energy Sciences under Award No.\ DE-SC0018904. RMV acknowledges funding from the Initiative and Networking Fund of the Helmholtz Association. CPD and RFW acknowledge funding from the Department of Energy Solar Photochemistry program with grant number DE-SC0021062. LQ acknowledges funding from the Alexander von Humboldt Foundation and computing time on the high-performance computer "Lise" at the NHR Center NHR@ZIB. C.Draxl appreciates funding from the DFG, projects 182087777 and 424709454. EN acknowledges DOE funding under grant DE-SC0018904 and from the Department of Defense SMART Scholarship Program. JHB acknowledges support from the National Science Foundation Graduate Research Fellowship Program under Grant No. DGE21-46756. GD and AMM were supported by the U.S. Department of Energy, Office of Science, Basic Energy Sciences, Geosciences, and Biosciences Division under contract No. DE-AC02-06CH11357. This research used resources at the Advanced Photon Source, a U.S. Department of Energy (DOE) Office Science User Facility operated by the DOE Office of Science by Argonne National Laboratory under Contract No. DE-AC02-06CH11357. Measurements were carried out at the UE52-SGM beamline at the BESSY II electron storage ring operated by the Helmholtz-Zentrum Berlin für Materialien und Energie.

We thank Christian Albinus and Rene Gr\"uneberger (Helmholtz Zentrum Berlin, HZB) for the 3D printing of the sample holder used in the Zn K-edge measurements, and for the design of the sample tip used in the Zn L$_3$-edge measurements, respectively. We thank Christopher Otolski and Donald Walko (Argonne National Laboratory) for their support during the measurement. We thank the X-ray Corelab team at HZB for their valuable instrumental and scientific support and Yves Joly for fruitful discussions. We thank Keith Gilmore for contributing to implementing the non-equilibrium Bethe–Salpeter Equation in \exciting.

\end{acknowledgments}


\section*{Author Contributions}

T.C.R and L.Q. contributed equally to this work. T.C.R. L.Q., C.Draxl, and R.M.V. conceptualized the work. T.C.R, C.P.D. R.F.W., J.H.B., E.N., A.-M.M., and G.D. conducted the experiment at the Zn K-edge. T.C.R., C.P.D., R.G. and M.F. conducted the experiment at the Zn L$_3$-edge. L.Q. performed the \emph{ab initio} calculations. R.R.P implemented the cDFT and RT-TDDFT methods. D.B.B., C.Deparis, J.Z.P., M.W. and K.E. synthesized the samples. T.C.R., L.Q., C.Draxl, and  R.M.V. wrote the manuscript with input contributions from every author.


\clearpage


\bibliographystyle{apsrev4-2}
\bibliography{apssamp}

\cleardoublepage
\onecolumngrid
\begin{center}
\textbf{\Large Ultrafast dynamic Coulomb screening of X-ray core excitons in photoexcited semiconductors\\ Supporting Information\\
(Dated: \today)}
\end{center}
\setcounter{section}{0}
\setcounter{subsection}{0}
\setcounter{equation}{0}
\setcounter{figure}{0}
\setcounter{table}{0}
\setcounter{page}{1}
\makeatletter
\renewcommand{\theequation}{S\arabic{equation}}
\renewcommand{\thefigure}{S\arabic{figure}}

\section{XTA experimental setup\label{secSI:XTA_setup}}

\subsection{Zn K-edge}

The experimental setup used in this work has been previously described in detail in the Supporting Information of reference \cite{Rossi2021}. In brief, X-ray absorption spectra (XAS) and X-ray transient absorption (XTA) spectra were acquired at the 7ID-D beamline of the Advanced Photon Source (APS, Argonne National Laboratories). The spectra were measured in total fluorescence yield with an avalanche photodiode detector (APD0008, FMB Oxford). The X-rays were monochromatized with a double silicon (111) crystal and focused with a pair of Kirkpatrick-Baez (KB) mirrors with the beam footprint imaged on a scintillator crystal in the focal plane of a microscope camera. The X-ray beam was a square-shaped focal spot of $\SI{15\pm1}{}\times\SI{39\pm1}{\micro\meter}$ at normal incidence used for both XAS and XTA measurements. The sample was glued with silver paint (Ted Pella, Pelco\textsuperscript{\textregistered}) to a 3D printed plastic sample holder that had a surface at \SI{45}{\degree} from normal incidence. The X-rays are $p$-polarized and are at \SI{45}{\degree} incidence angle at the sample surface (with equal projections along the ($\mathbf{a}$,$\mathbf{b}$) and $\mathbf{c}$ axes of ZnO). The temperature-dependent XAS spectra were recorded with a Linkam THMSG-600 oven (accuracy \SI{0.01}{\degree}C). A nitrogen flow was applied onto the sample during the measurement to prevent adsorption and diffusion of carbon impurities and water inside the material and to provide some active cooling to prevent static heating.

\begin{figure}
	\centering
	\includegraphics[width=0.4\linewidth]{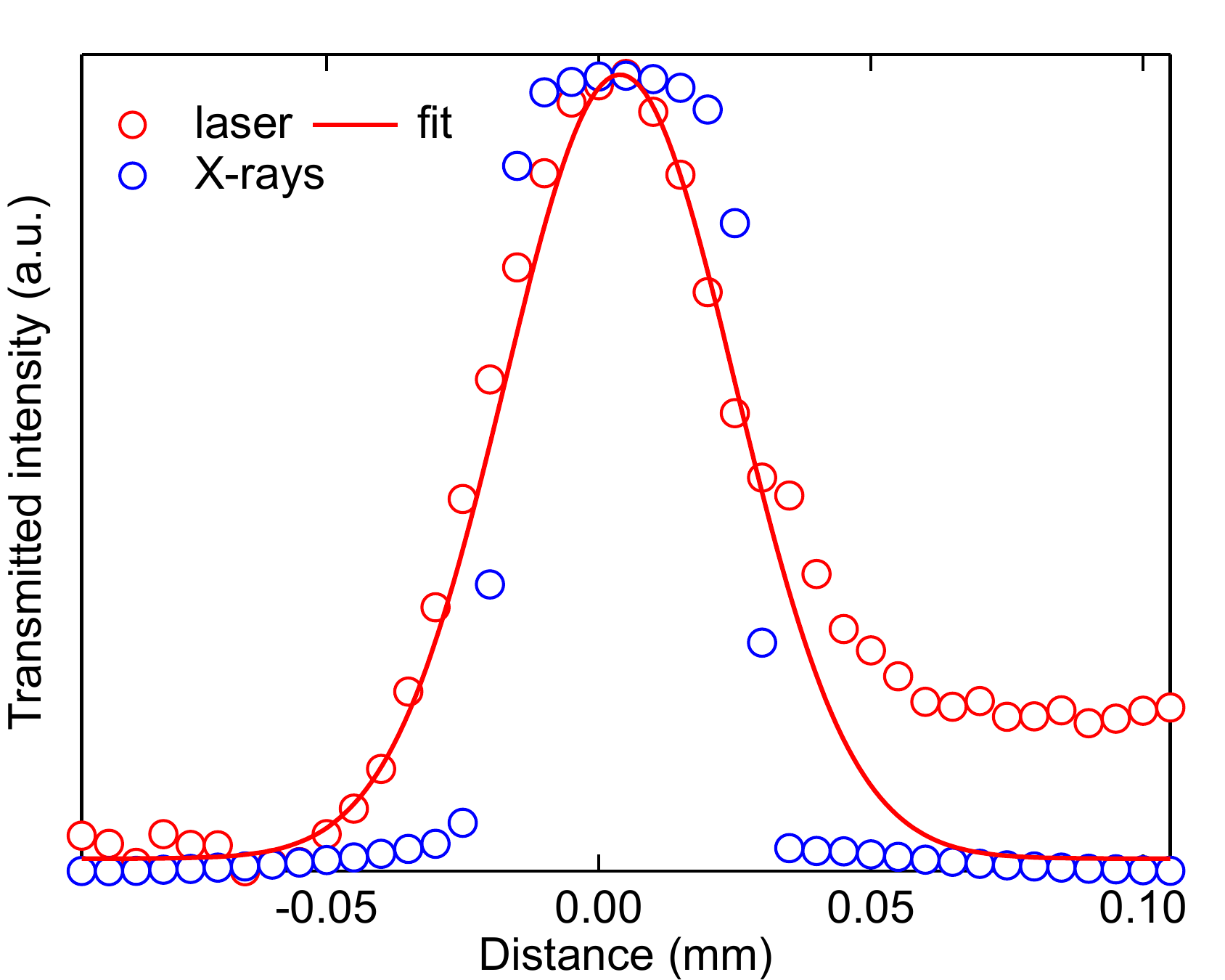}
	\caption{\textbf{Spatial overlap at the Zn K-edge.} Laser (red circles) and X-ray (blue circles) beam profiles measured through a \SI{50}{\micro\metre} diameter pinhole at normal incidence and at focus. A Gaussian fit to the laser profile (red curve) gives a laser spot size of $\SI{50\pm3}{\micro\metre}$ (FWHM) or $\SI{85\pm5}{\micro\metre}$ (1/e$^2$).}
	\label{figSI:pinhole_scan}
\end{figure}

The XTA measurements were performed in 24-bunch mode filling pattern with a ring current of \SI{102}{\milli\ampere}. The XTA setup has been fully described in reference \cite{March:2011dn}. The photon current detected by the APD is transferred to a boxcar averager (UHF-BOX from Zurich Instruments) controlled remotely with the LabOne user interface. The APD was positioned at \SI{90}{\degree} from the incident X-ray beam to minimize the contribution from elastic scattering. The pump excitation was performed with a Nd:YVO$_4$ Duetto laser (Time-Bandwidth products, \SI{1064}{\nano\metre} fundamental wavelength), which delivered $\sim\SI{10}{\pico\second}$ pulses (FWHM) at \SI{100.266}{\kilo\hertz}. The \SI{355}{\nano\metre} excitation wavelength was achieved after frequency doubling and sum frequency generation in two LBO crystals (placed in ovens at $\sim$\SI{100}{\degree}C for long-term stability and noncritical phase matching). The frequency converted pulses were $\sim\SI{10}{\pico\second}$ (FWHM) with a spectral bandwidth of \SI{0.1}{\nano\metre} (FWHM). The relative angle between the laser and the X-rays is $\sim\SI{7\pm1}{\degree}$. Spatial overlap between the laser and the X-rays was achieved through a \SI{50}{\micro\metre} diameter pinhole. The laser diameter is $\SI{85\pm5}{\micro\metre}$ (1/e$^2$) measured by a pinhole scan (Figure \ref{figSI:pinhole_scan}). The excitation fluences are calculated with the FWHM dimensions of the pump beam ($\SI{50\pm3}{\micro\metre}$). Fluences are reported with an experimental uncertainty calculated by propagating the uncertainties of the measurements of the laser power (5\% of the measured value with a thermal power sensor head) and the uncertainty in laser spot size (see Figure \ref{figSI:pinhole_scan}). It yields the following equation for the uncertainty over the fluence $F$ with the laser average power $P$ and the laser spot area $A$,
\begin{equation}
\Delta F=F\sqrt{\frac{\Delta P^2}{P^2}+\frac{\Delta A^2}{A^2}}.
\end{equation}
The laser is incident on the sample at \SI{52}{\degree} incidence angle and the incident laser spot size has a circular shape with radius $R=\SI{25\pm2}{\micro\meter}$ which gives,
\begin{equation}
\Delta F=F\sqrt{\frac{\Delta P^2}{P^2}+\frac{4\Delta R^2}{R^2}}.
\end{equation}
Temporal overlap of the laser and X-ray pulses was achieved with a fast metal-semiconductor-metal photodetector (Hamamatsu, model G4176-03) connected to a Lecroy digital oscilloscope. An electronic delay was introduced for the laser pulse to achieve a temporal overlap of $\sim\SI{10}{\pico\second}$ precision. The time delay was set to \SI{100\pm10}{\pico\second} between the laser and the X-ray pulses for energy scans. The XTA signal was computed from the difference between the X-ray fluorescence detected with and without the laser pulse divided by the incident X-ray intensity at a given photon energy. The XTA was then normalized to the XAS edge jump such that the amplitude of the transient is indicative of a normalized change in the XAS with respect to a unitary edge jump (more details in section \ref{secSI:data_processing}).

\begin{figure}
	\centering
	\includegraphics[width=0.47\linewidth]{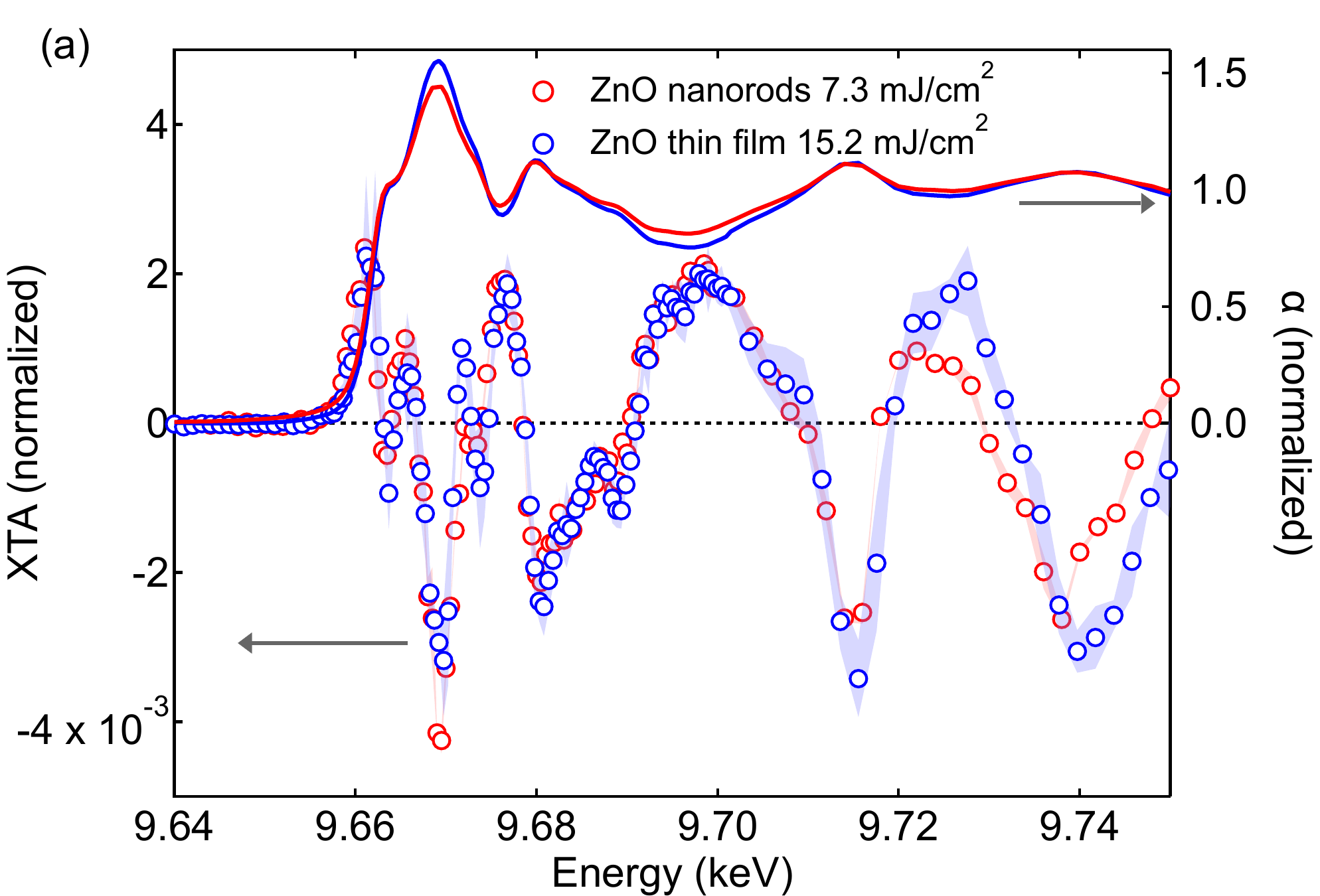}
	\includegraphics[width=0.47\linewidth]{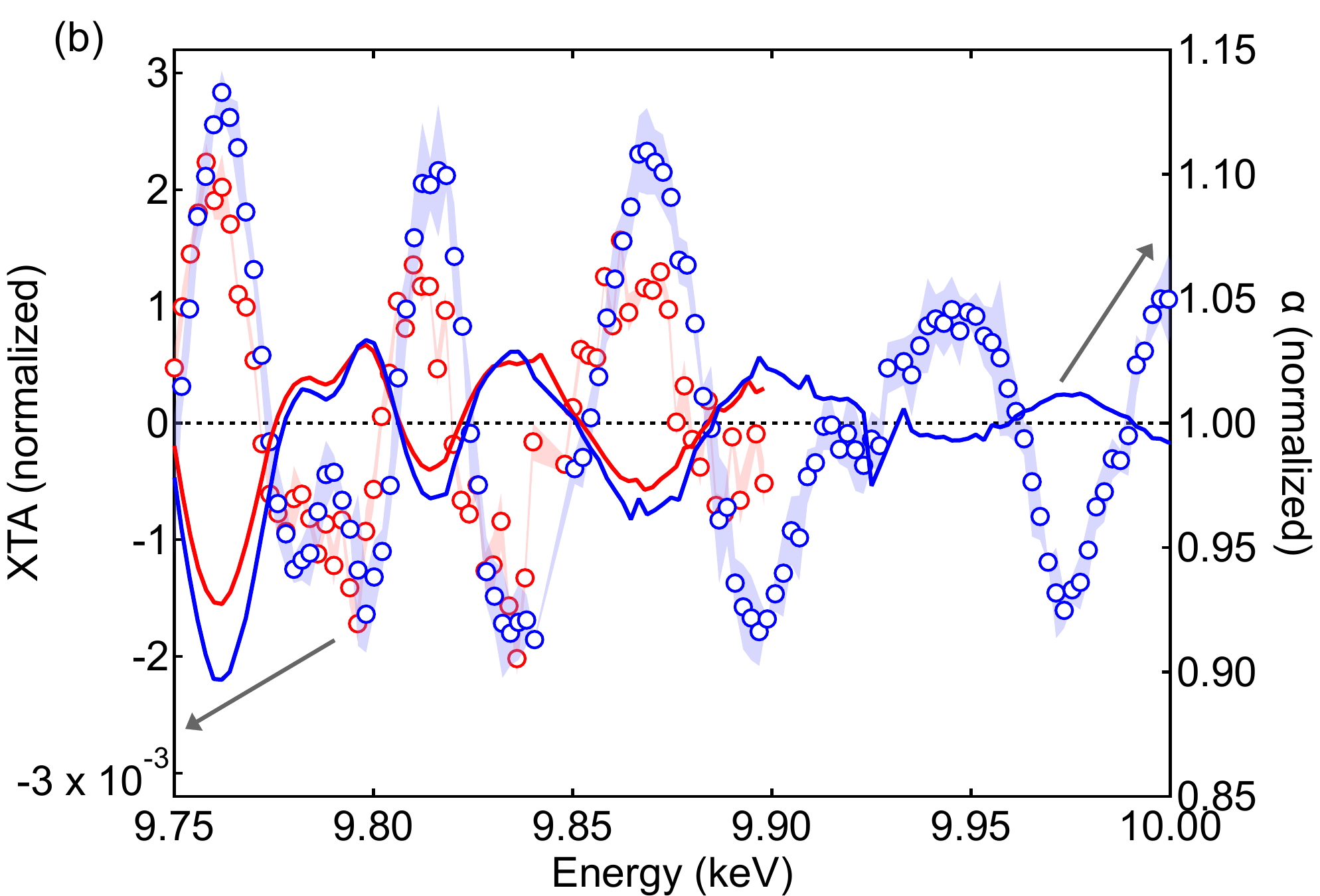}
	\caption{\textbf{XTA spectra at the Zn K-edge of ZnO thin films and nanorods.} Comparison between XTA measurements at the Zn K-edge of ZnO nanorods (red circles) and ZnO (0001) thin film (blue circles) at \SI{100}{\pico\second} upon excitation at \SI{355}{\nano\meter} in the (a) XANES and (b) EXAFS (left axis). Vertical error bars represent the standard deviation between individual measurements. XAS spectra are shown with continuous curves for reference (right axis). Measurements on ZnO nanorods are reproduced from \cite{Rossi2021}.}
	\label{figSI:comparison_XTA_film_nanorods}
\end{figure}

The X-ray energy calibration was performed with a zinc foil and compared with the EXAFS Materials reference spectra for metal foils. The energy axes of the XAS spectra also matched a previous work with the same calibration procedure \cite{Rossi2021}. Additionally, the energy scale of XTA measurements on ZnO thin films in this work and in a previous work on ZnO nanorods show that the energy scales are compatible (Figure \ref{figSI:comparison_XTA_film_nanorods}).

\begin{figure}
	\centering
	\includegraphics[width=0.5\linewidth]{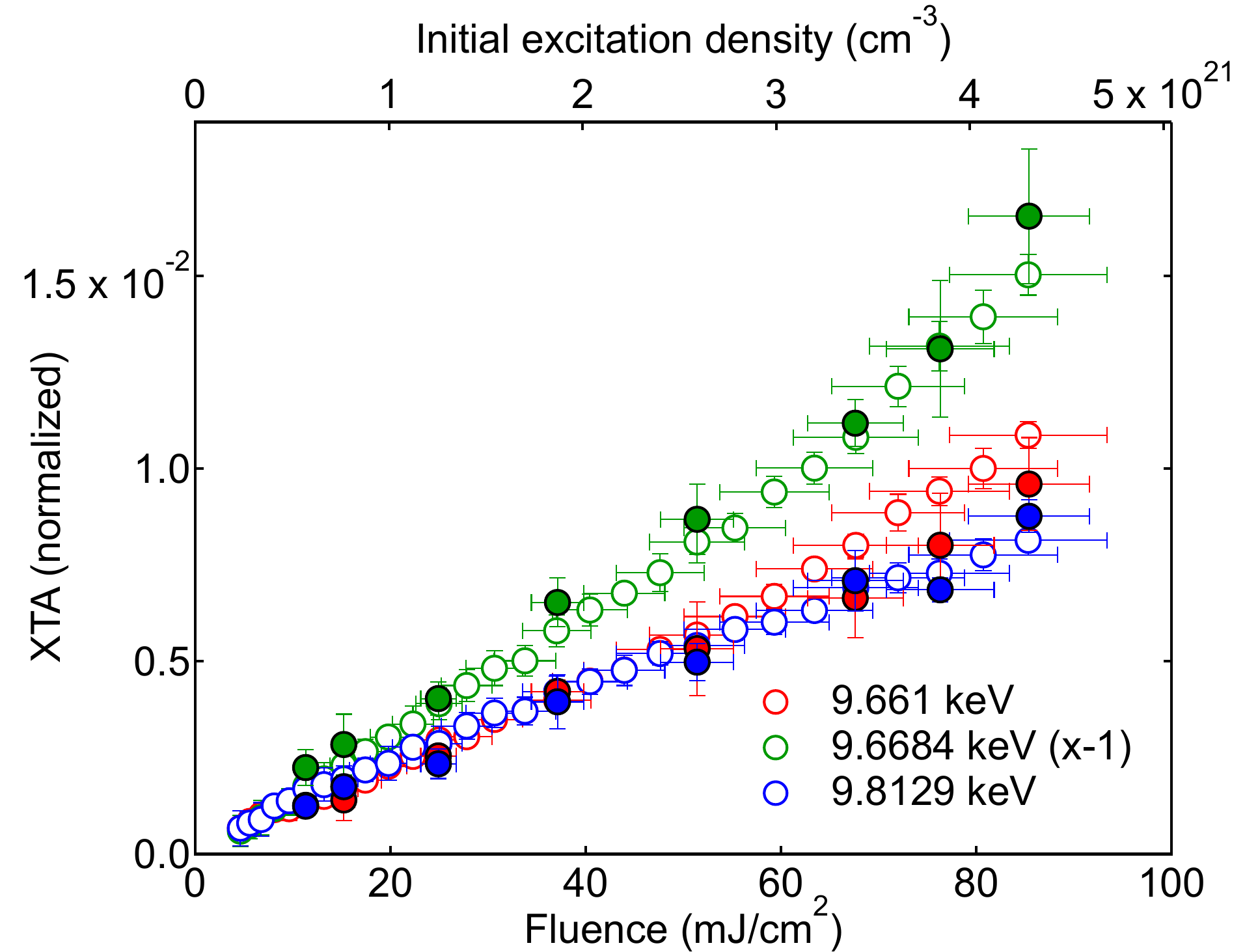}
	\caption{\textbf{Fluence dependence of the XTA amplitude at the Zn K-edge.} XTA amplitude at selected energy points with the excitation fluence for the dedicated fluence dependence measurements (white-filled circles) and full energy scans (colored-filled circles). The time delay is \SI{100}{\pico\second} and the excitation is at \SI{355}{\nano\metre}.}
	\label{figSI:XTA_fluence_dependence_with_mono_scans}
\end{figure}

Figure \ref{figSI:XTA_fluence_dependence_with_mono_scans} displays a comparison of XTA amplitude (including thermal effects) between dedicated fluence dependence measurements at selected energy points (open circles) and the amplitude of the XTA in measurements over the entire spectrum (filled circles). The results are compatible with the standard deviation of the XTA amplitude and the uncertainty over the excitation fluence.

\subsection{Zn \texorpdfstring{L$_3$}{L3}-edge}

\begin{figure}
    \centering
    \includegraphics[width=0.45\linewidth]{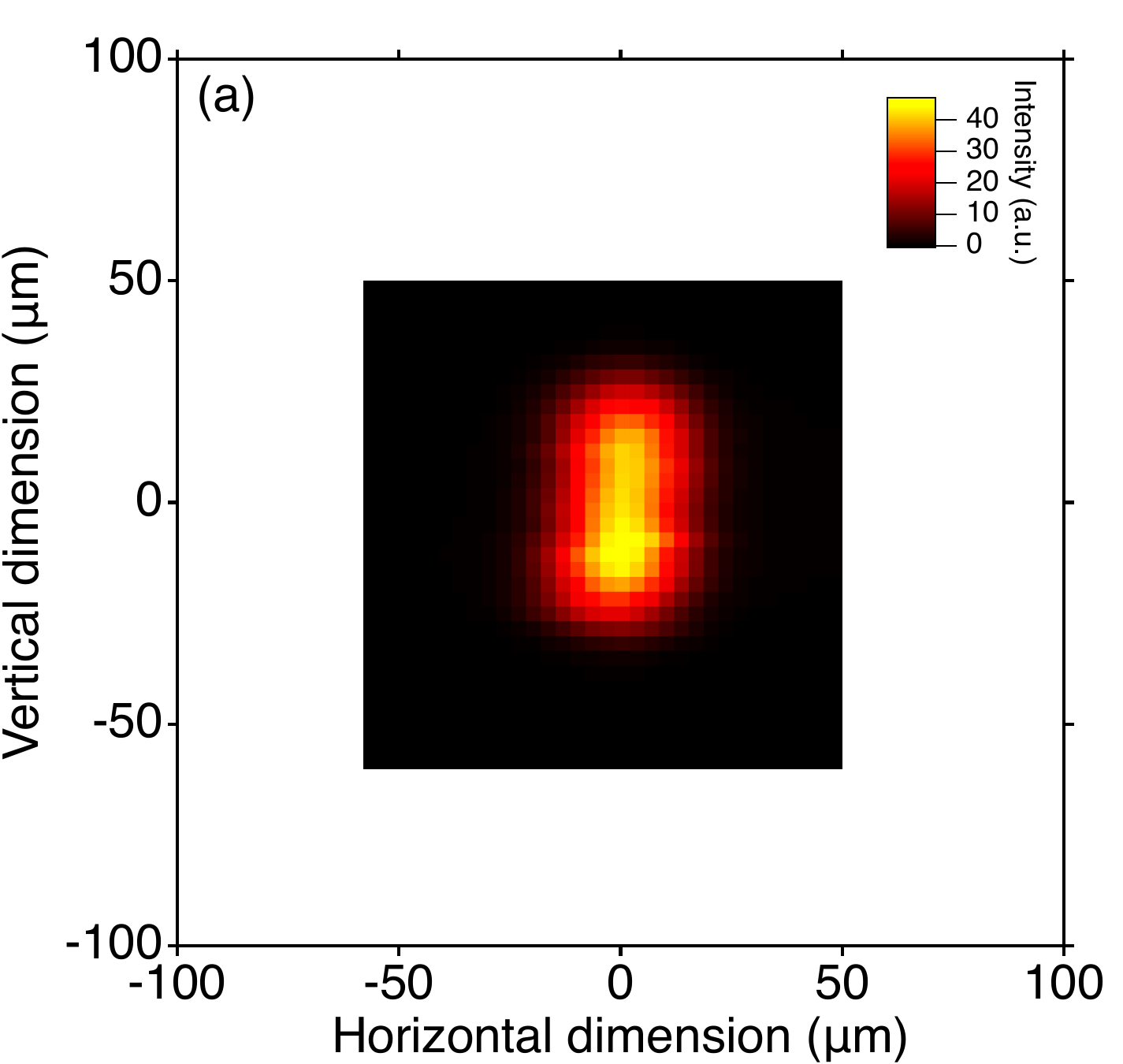}
    \includegraphics[width=0.45\linewidth]{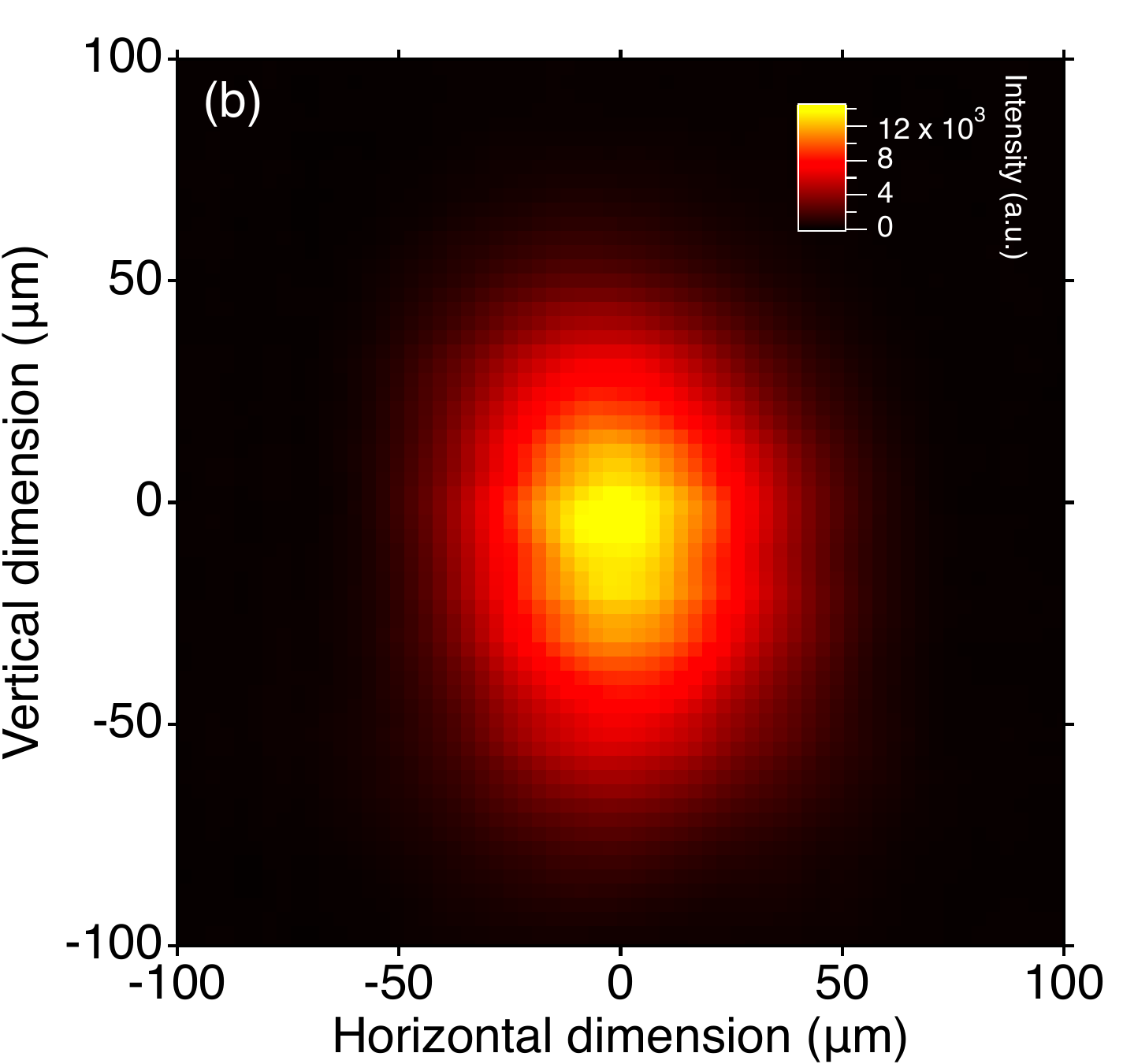}
    \caption{\textbf{Beam profiles at the Zn L$_3$-edge.} Profiles of (a) the X-rays at \SI{1040}{\electronvolt} measured through a \SI{20}{\micro\metre} pinhole (after spatial deconvolution), and (b) the \SI{343}{\nano\meter} pump laser measured on a beam profiler in a focal mirror plane of the sample.}
    \label{figSI:pinhole_scan_ZnL23_laser}
\end{figure}

\begin{figure}
    \centering
    \includegraphics[width=0.45\linewidth]{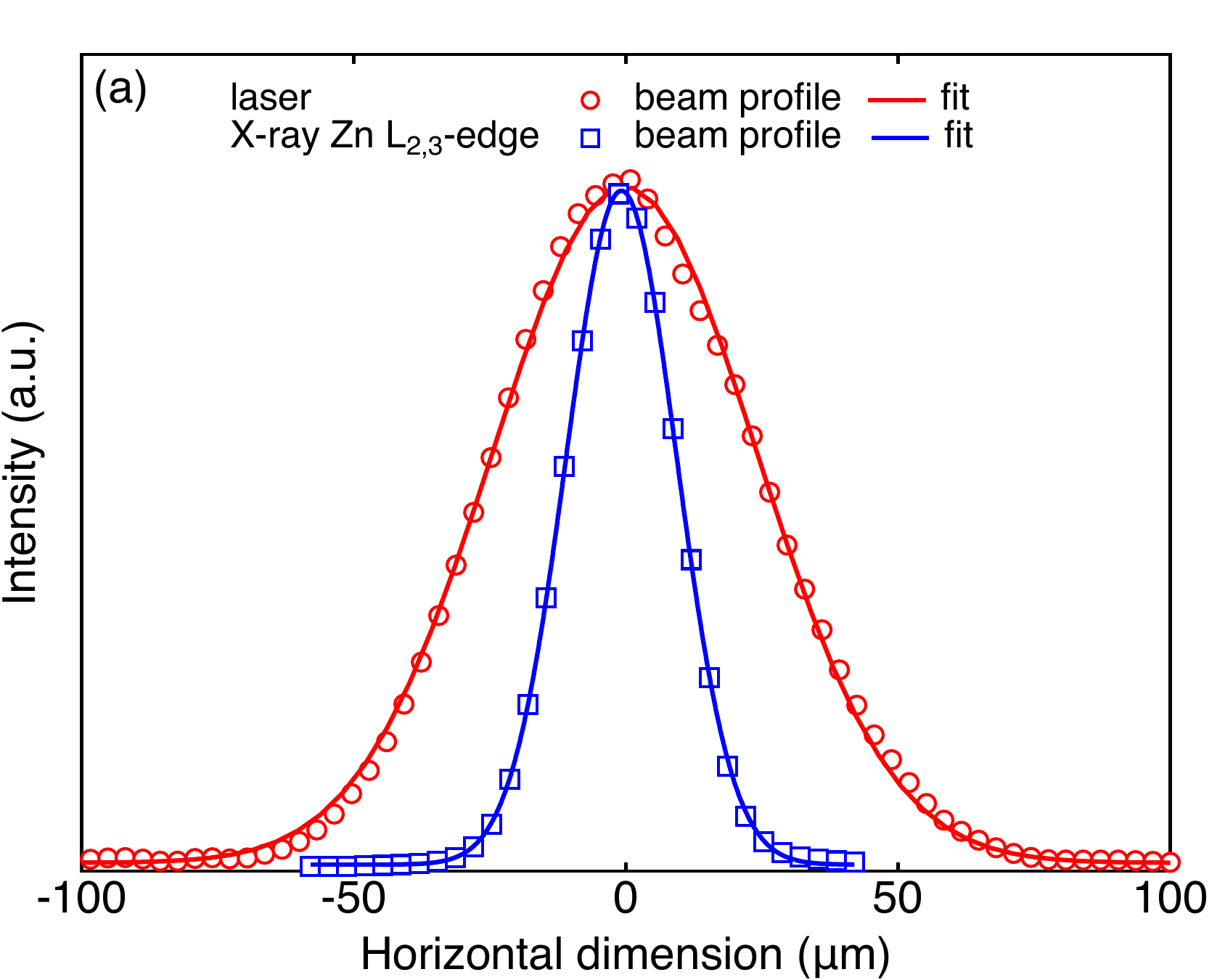}
    \includegraphics[width=0.45\linewidth]{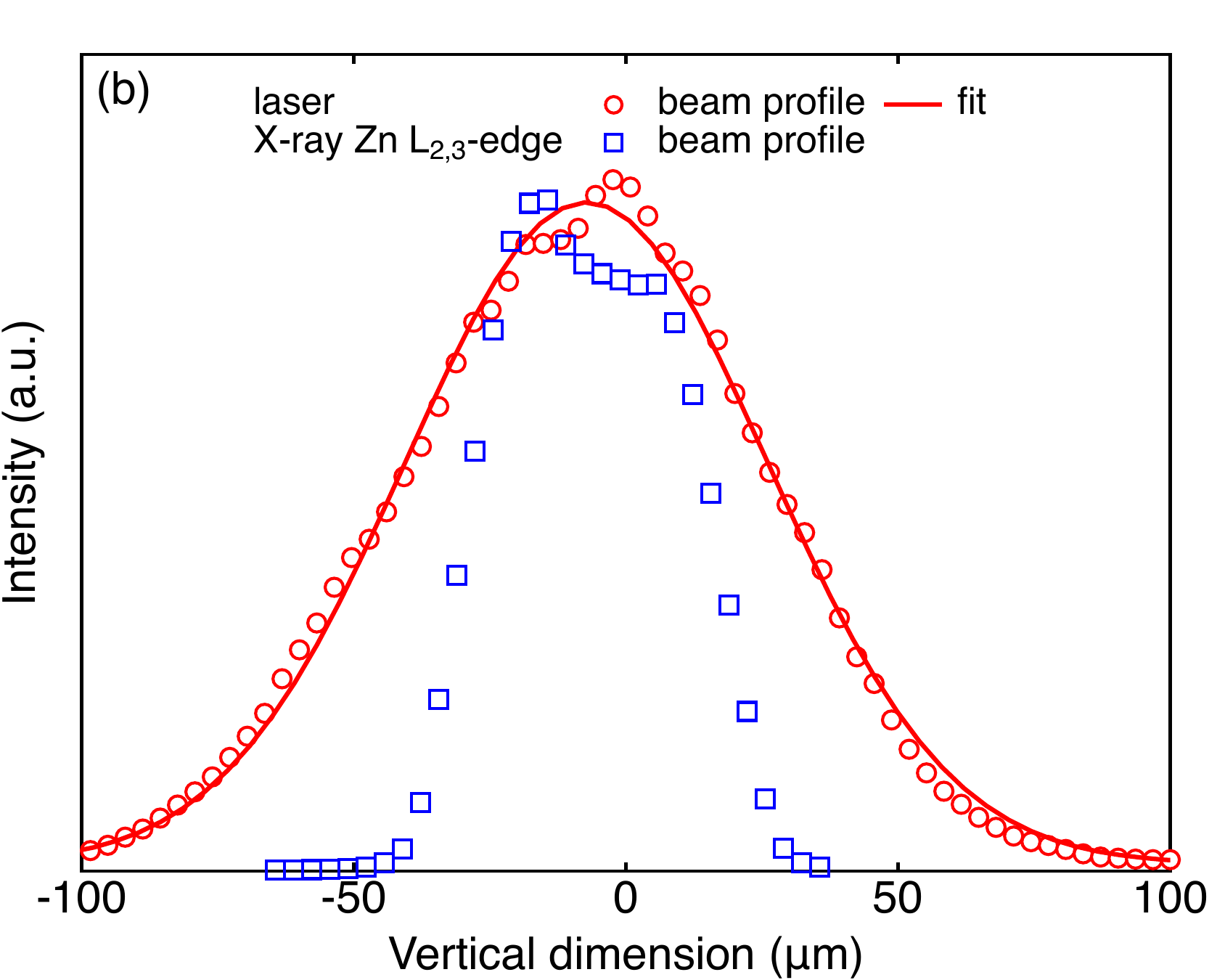}
    \caption{\textbf{Spatial overlap at the Zn L$_3$-edge.} (a) Horizontal and (b) vertical beam profiles at spatial overlap for the laser at \SI{343}{\nano\meter} (red circles) and the X-rays at \SI{1040}{\electronvolt} (blue squares). Gaussian fits the beam profiles show that the laser beam waist is \SI{57\pm2}{\micro\meter} FWHM (\SI{69\pm2}{\micro\meter} 1/e$^2$) horizontally and \SI{78\pm3}{\micro\meter} FWHM (\SI{94\pm4}{\micro\meter} 1/e$^2$) vertically while the X-ray beam is \SI{23\pm1}{\micro\meter} FWHM (\SI{30\pm1}{\micro\meter} 1/e$^2$) horizontally and $\sim\SI{46}{\micro\meter}$ FWHM ($\sim\SI{63}{\micro\meter}$ 1/e$^2$) vertically.}
    \label{figSI:beam_profile_spatial_overlap}
\end{figure}

XTA spectroscopy measurements at the Zn L$_3$-edge were conducted at the UE52-SGM beamline of Bessy II in the nmTransmission NEXAFS chamber \cite{Fondell:2017sd} modified with a sample tip to allow for measurements in transmission of thin film samples. The pump-probe experiment uses the camshaft from the hybrid filling pattern of Bessy II, which yields a temporal resolution of $\sim\SI{70}{}$--\SI{80}{\pico\second}. The polarization of the X-rays is horizontal, incident at the thin film surface at normal incidence with the polarization in the plane defined by the (\textbf{a},\textbf{b}) lattice vectors of the ZnO wurtzite unit cell. The operating vacuum pressure of the chamber was approximately \SI{1e-5}{\milli\bar}. Initial spatial overlap was determined by visualization of the laser and the X-ray spot on a Ce:YAG screen mounted in the same plane as the sample. The X-ray beam waist is $\SI{30\pm1}{}\times\SI{63}{\micro\meter\pm2}$ (1/e$^2$ width), measured by deconvoluting an intensity profile measured in transmission of a pinhole with \SI{20}{\micro\metre} diameter (Thorlabs P20HK, Figure \ref{figSI:pinhole_scan_ZnL23_laser}a). The laser beam waist is measured on a beam profiler (WinCamD-LCM from DataRay) in a focal mirror plane of the sample outside the vacuum chamber (Figure \ref{figSI:pinhole_scan_ZnL23_laser}b). The final spatial overlap is achieved on the sample by scanning the laser position to achieve the maximum amplitude of the transient signal. Figure \ref{figSI:beam_profile_spatial_overlap} shows the horizontal and vertical beam profiles of the laser beam (red circle) and the X-ray beam at the spatial overlap as well as the results of Gaussian fits.

The sample was pumped with \SI{350}{\femto\second} pulses at \SI{343}{\nano\meter} (third harmonic of the \SI{1030}{\nano\meter} laser fundamental) from a Tangerine laser (Amplitude Syst\`emes) triggered at \SI{10}{\kilo\hertz} by a frequency-divided signal from the synchrotron radiofrequency cavity ($p$-polarization). The relative angle between the laser and the X-rays is \SI{45\pm3}{\degree} with a beam waist of $\SI{69\pm2}{}\times\SI{94\pm4}{\micro\meter}$ (1/e$^2$ width) at the sample position. The incident laser fluence on the sample is calculated from the FWHM of the laser beam waist.

The transmitted X-ray pulses through the sample were collected on a silicon APD (SAR3000, Laser Components) connected to a boxcar integrator (UHF-BOX Zurich Instruments). Electronic gates are set on the camshaft the closest in time to the laser pulse (pumped pulse) and on the camshaft delivered from an earlier synchrotron period by \SI{800}{\nano\second} on the sample (unpumped camshaft). The transient XTA signal is computed by the difference between the integrated response of the pumped and unpumped X-ray pulses through the sample divided by the single bunch current. For the measurement of time traces, a relative time delay is introduced between the laser and X-ray pulses by delaying the laser trigger.

Equilibrium XAS spectra were measured without the laser and with the full intensity of the X-ray beam (given by the total ring current) on a GaAs detector in transmission. The transmitted intensity is normalized by the ring current, which is proportional to the incident X-ray intensity on the sample. The measurement of static heating was performed on the same detector with the laser impinging on the sample at \SI{10}{\kilo\hertz} meaning one laser pulse for every 125 synchrotron periods.

\begin{figure}
    \centering
    \includegraphics[height=0.35\linewidth]{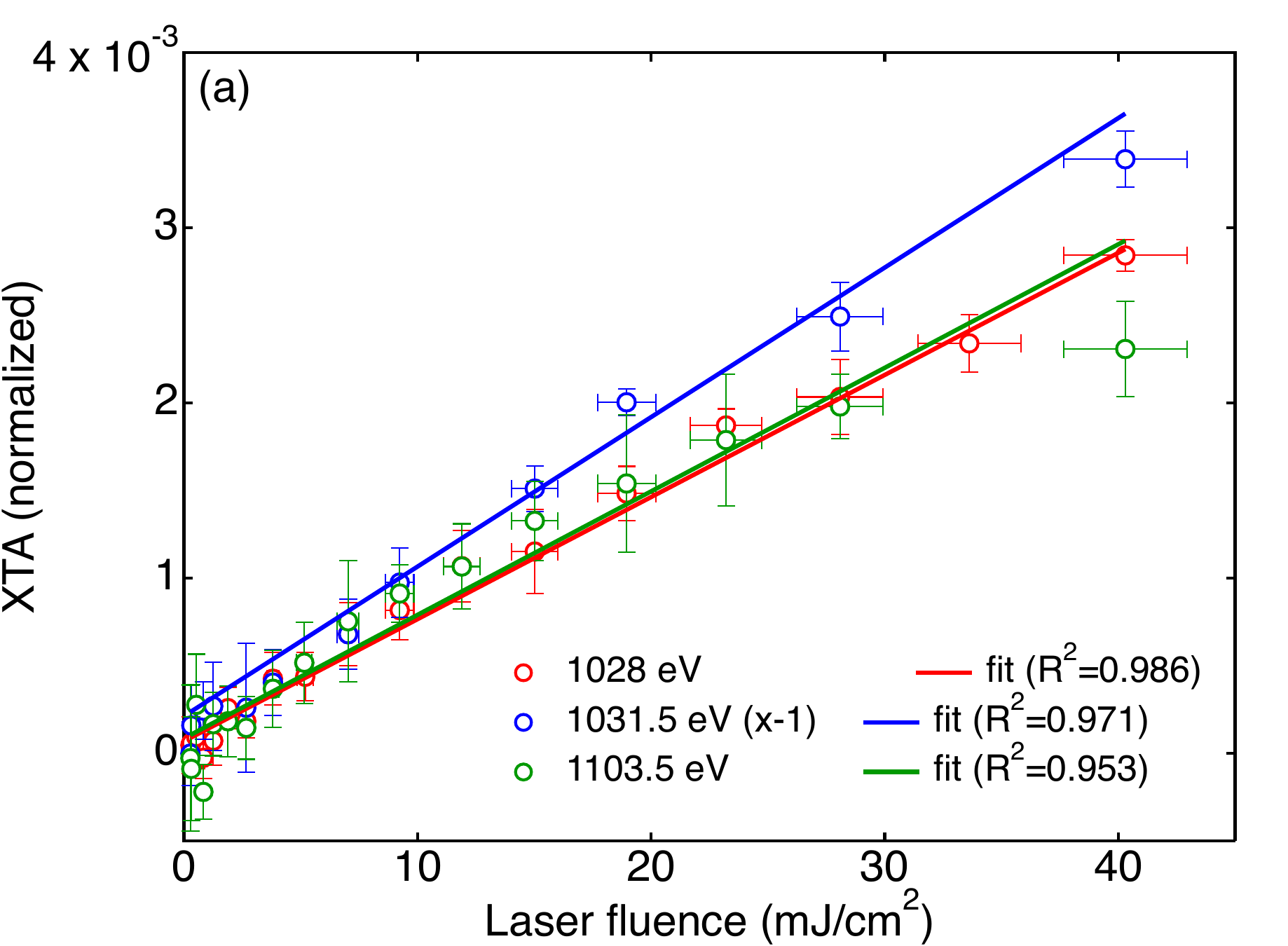}
    \includegraphics[height=0.35\linewidth]{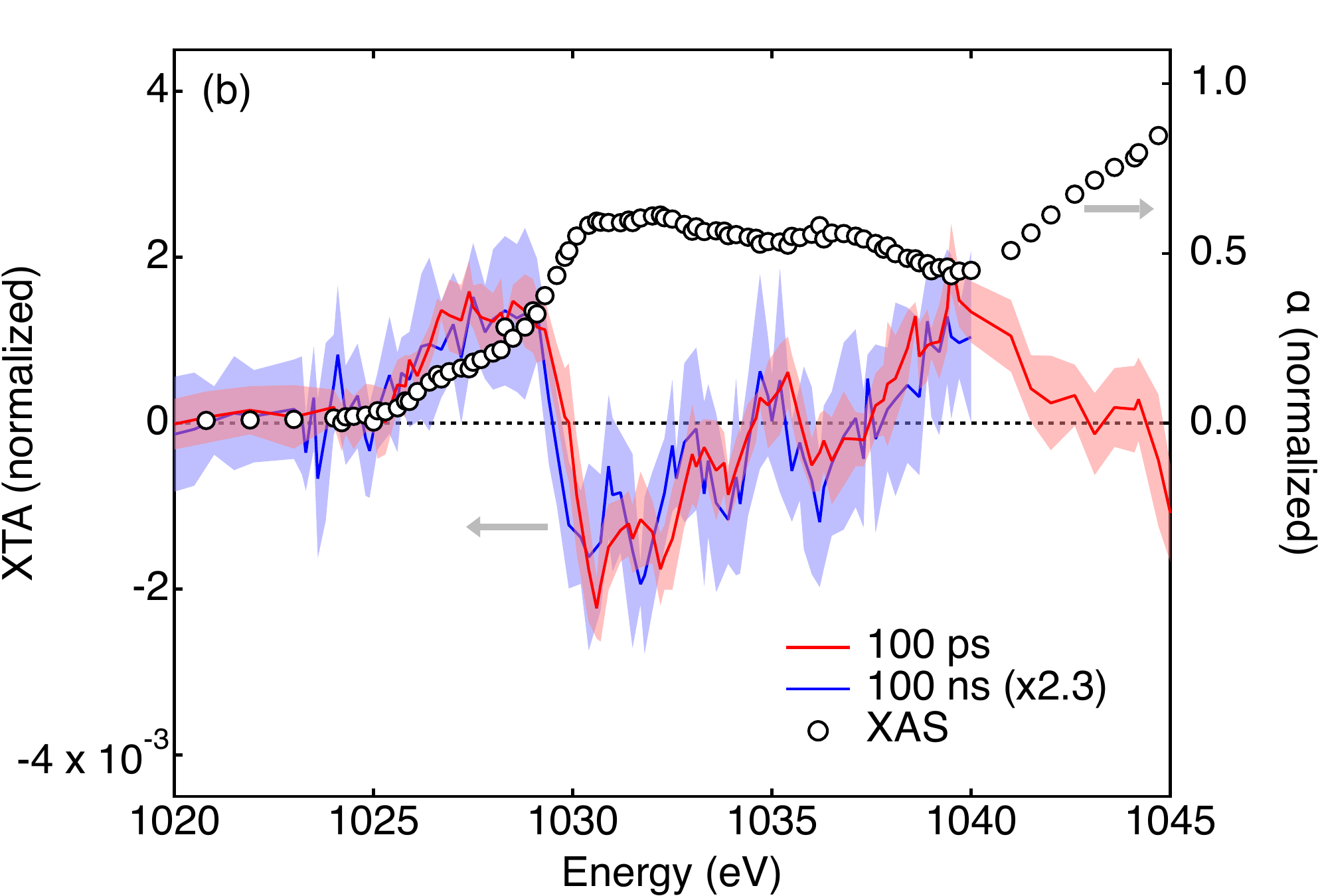}
    \caption{\textbf{Fluence dependence and measurements at later time delays at the Zn L$_3$-edge.} (a) Fluence dependence at the Zn L$_3$-edge of ZnO (colored circles, \SI{100}{\pico\second} time delay). Selected energy points are \SI{1028}{\electronvolt} (red circles), \SI{1031.5}{\electronvolt} (blue circles) and \SI{1103.5}{\electronvolt} (green circles). The vertical error bars represent standard deviations between individual measurements and the horizontal error bars represent the uncertainty in the calculation of the laser fluence. Linear fits are weighted by the standard deviation of the measurement. (b) Comparison between normalized XTA spectra at \SI{100}{\pico\second} (red curve) and \SI{100}{\nano\second} (blue curve, rescaled by a factor 2.3). Shaded areas represent standard deviations between individual measurements. The XAS is shown with black circles for reference. The laser excitation fluence is \SI{14}{\milli\joule\per\square\centi\metre}.}
    \label{figSI:ZnL23_fluence_dependence}
\end{figure}

\begin{figure}
    \centering
    \includegraphics[width=0.5\linewidth]{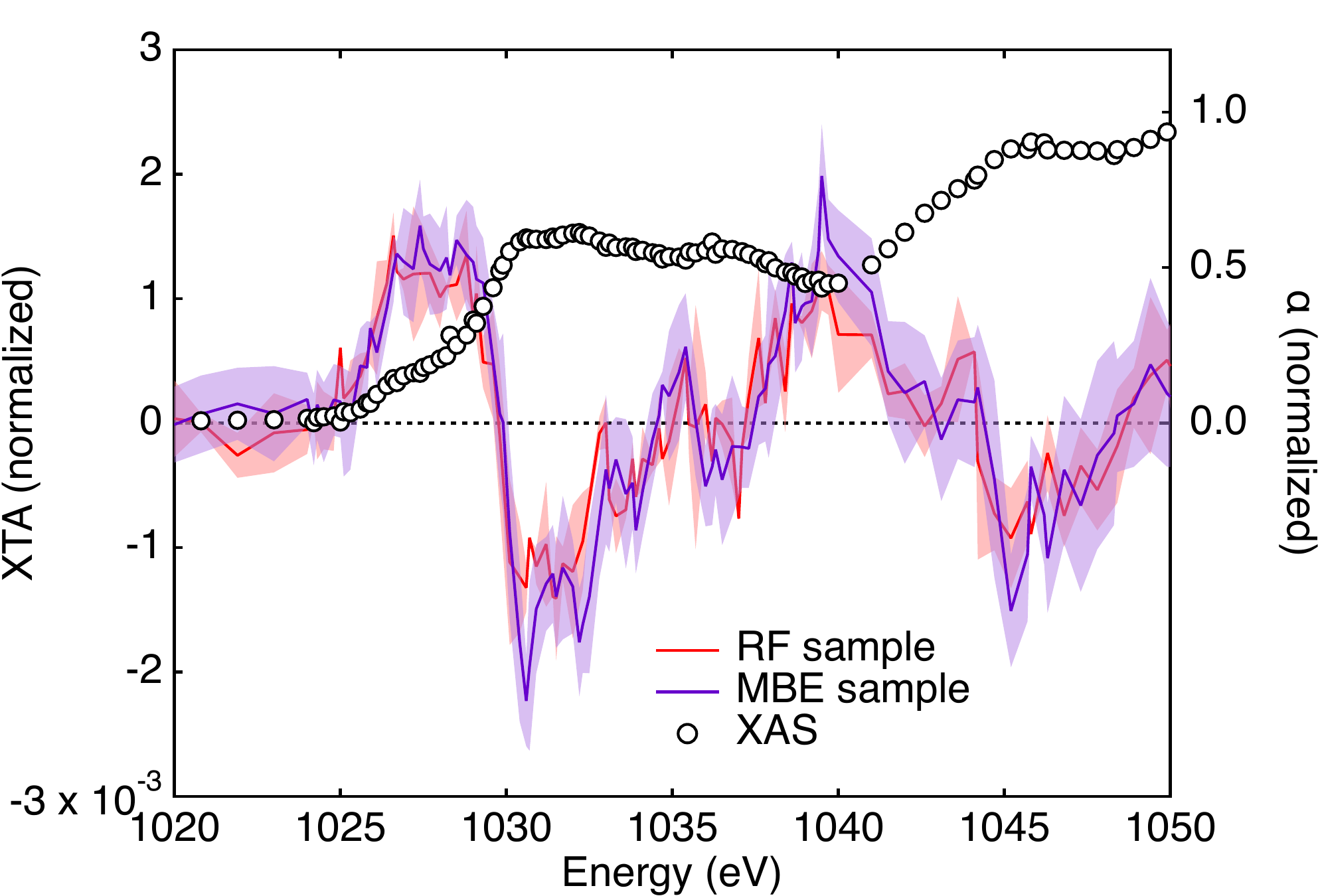}
    \caption{\textbf{Effect of sample morphology on XTA spectra at the Zn L$_3$-edge.} Effect of the sample growth method on the XTA spectrum at \SI{100}{\pico\second} under \SI{355}{\nano\metre} excitation: radiofrequency sputtering (red shaded curve) and molecular beam epitaxy (purple shaded curve). The XAS spectrum of the MBE sample is shown with black circles for reference. Shaded areas in the XTA spectra represent standard deviations between individual measurements. The laser excitation fluence is \SI{14}{\milli\joule\per\square\centi\metre}.}
    \label{figSI:ZnL23_XTA_sample_morphology}
\end{figure}

Under vacuum, the ZnO thin film undergoes a large static heat load, which does not recover between consecutive laser pulses. Because of the difference operation between the pumped and the unpumped response to compute the XTA signal, this contribution is removed. The amplitude of the XTA signal did not show any sign of non-linearity within the accessible pump fluences (Figure \ref{figSI:ZnL23_fluence_dependence}a) with similar lineshapes between \SI{100}{\pico\second} and \SI{100}{\nano\second} time delay (Figure \ref{figSI:ZnL23_fluence_dependence}b). Samples grown by molecular beam epitaxy (MBE) or radio frequency sputtering (RF) did not exhibit XTA spectra with significant differences in the near-edge region. However, larger differences appear above the edge, which are assigned to different laser-induced heat loads on the two samples (Figure \ref{figSI:ZnL23_XTA_sample_morphology}).

\cleardoublepage


\section{Data processing\label{secSI:data_processing}}

\subsection{XAS spectra data processing}

\subsubsection{Zn K-edge}

The background subtraction, edge jump normalization, and flattening of XAS spectra in the post-edge were performed with the \texttt{xraylarch} python package (version 0.9.71) \cite{Newville:2013go}. The ionization potential was set to \SI{9673.5}{\electronvolt} and the pre-edge energy range and post-edge energy ranges (relative to the ionization potential) were set between \SI{-35}{} and \SI{-25}{\electronvolt} and between \SI{35}{} and \SI{330}{\electronvolt}, respectively. The pre-edge and post-edge backgrounds were both subtracted with a linear function. Deglitching was applied to a maximum of 3 points per spectra, which originate from diffraction peaks apparent in XAS spectra at specific energy points. Self-absorption effects were not considered in the data analysis since the attenuation length of the most intense K$\alpha_{1,2}$ emission lines (\SI{8615.8}{} and \SI{8638.9}{\electronvolt} \cite{Bearden1967}) is \SI{35\pm1}{\micro\metre} (1/e) \cite{Henke:1993eda}, which is more than orders of magnitude longer than the thickness of the ZnO thin film (\SI{283\pm2}{\nano\metre}, see SI \ref{subsec:spectroscopic_ellipsometry}). 

\subsubsection{Zn \texorpdfstring{L$_3$}{L3}-edge}

Edge jump normalization of XAS spectra is performed with our own python script. The pre-edge background is subtracted by fitting a second-order polynomial function between \SI{1000}{} and \SI{1010}{\electronvolt} while the post-edge normalization and flattening are performed by normalizing the edge jump with respect to a linear fit between \SI{1040}{} and \SI{1055}{\electronvolt}.

\subsection{XTA spectra data processing}

After the normalization of individual XTA spectra to the edge jump, statistical outliers 4 standard deviations away from the mean value were removed before averaging and recomputing the standard deviation (performed only once).

\cleardoublepage


\section{Kinetics\label{secSI:kinetics}}

\subsection{Zn K-edge}

\begin{figure}[!ht]
	\begin{center}
		\includegraphics[width=0.325\linewidth]{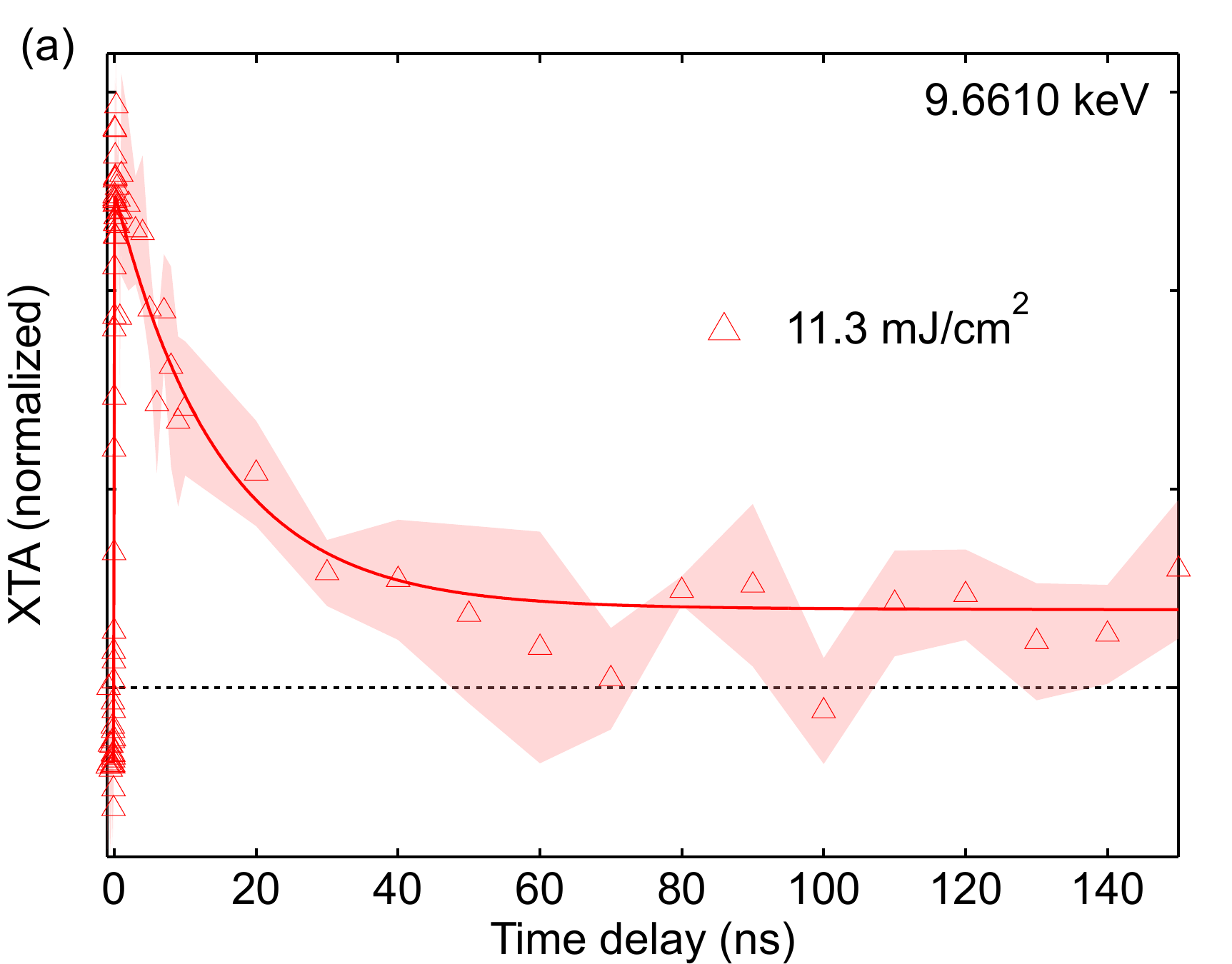}
		\includegraphics[width=0.325\linewidth]{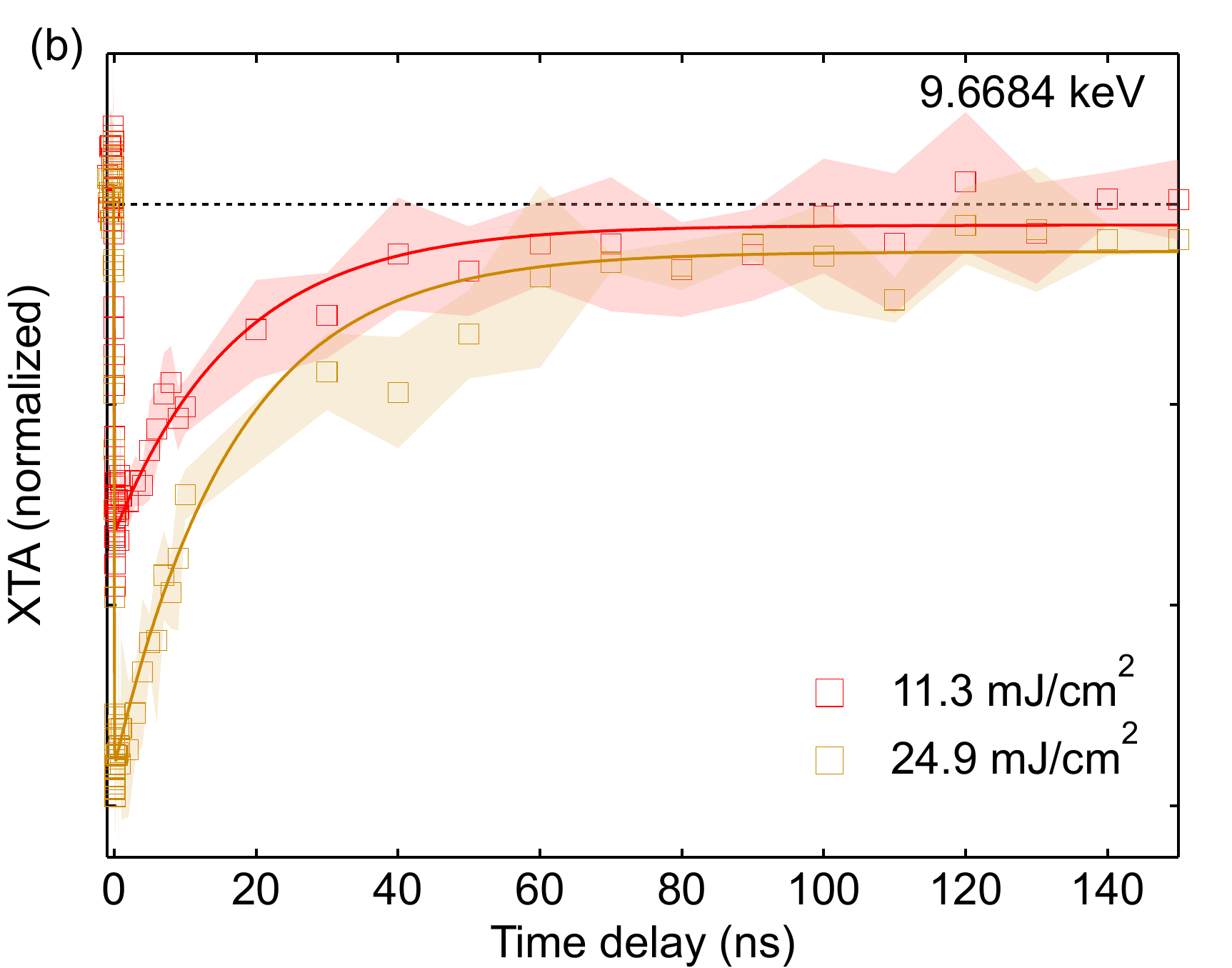}
		\includegraphics[width=0.325\linewidth]{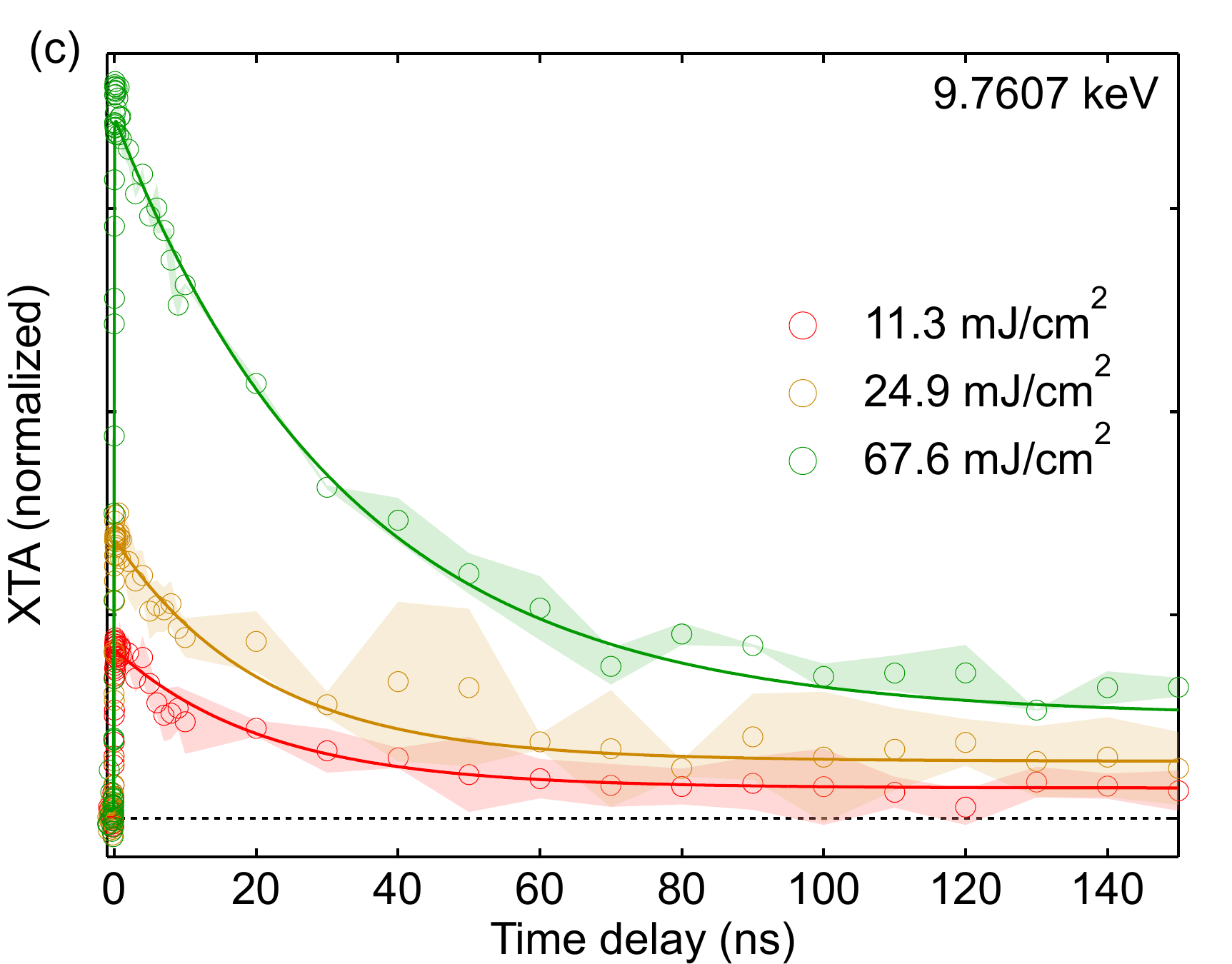}
	\end{center}
	\caption{\textbf{XTA kinetics at the Zn K-edge.} Evolution of the kinetics at (a) \SI{9.661}{\kilo\electronvolt}, (b) \SI{9.6684}{\kilo\electronvolt}, and (c) \SI{9.7607}{\kilo\electronvolt} at various fluences with \SI{3.49}{\electronvolt} pump photon energy (colored markers). Exponential fits are shown with continuous curves. Shaded areas represented standard deviations between individual measurements.}
    \label{figSI:kinetics}
\end{figure}

Time traces at the Zn K-edge at different X-ray energies and excitation fluences are shown in Figure \ref{figSI:kinetics}. Both in the XANES (Figure \ref{figSI:kinetics}a,b) and in the EXAFS (Figure \ref{figSI:kinetics}c), the time traces show that $\sim\SI{80}{\percent}$ of the XTA signal decays with a time constant $\tau_1\sim\SI{15}{}$--\SI{20}{\nano\second} for fluences $<\SI{25}{\milli\joule\per\square\centi\metre}$. This time constant is similar to the long time constant from a previous measurement at the Zn K-edge of ZnO nanorods \cite{Rossi2021}. Since the transient EXAFS is uniquely described by lattice heating, $\tau_1$ is assigned to heat diffusion in the ZnO thin film. This time constant is in excellent agreement with the short timescale of heat diffusion of $\sim\SI{10}{\nano\second}$ calculated upon pulse excitation of a ZnO thin film on a glass substrate with a fluence of \SI{30}{\milli\joule\per\square\centi\metre} \cite{Scorticati:2016aa}. It corresponds to heat diffusion in a regime where the carriers have a slightly higher temperature than the lattice, so that lattice heating is still active at the same time as heat diffuses away from the probe volume. At a fluence of $\sim\SI{68}{\milli\joule\per\square\centi\metre}$, $\tau_1$ in the EXAFS increases to $\SI{33}{\nano\second}$, which indicates that heat diffusion away from the excitation volume has a non-linear behavior with the excitation fluence. The remaining XTA transient signal is fitted with a fixed time constant of $\tau_2=\SI{10}{\micro\second}$, much longer than the maximum time delay measured experimentally, which corresponds to heat dissipation away from the ZnO layer, mainly to the substrate. Hence, no decay time constant measured experimentally is assigned to processes involving carriers such as electron-hole recombination. Time constants of radiative electron-hole recombination are in the range \SI{2}{}--\SI{20}{\nano\second} in ZnO single crystals at room temperature \cite{Skettrup:1968aa, Koida:2003kx, Teke:2004br, Chichiby:2006aa}, which compares with the time constant of heat diffusion ($\tau_1$). Hence, electron-hole recombination is most likely accidentally synchronous with heat diffusion with the time constant $\tau_1\sim\SI{15}{}$--\SI{20}{\nano\second}, which cannot be disentangled by fitting the time traces.

\begin{table}[!ht]
	\centering
	\begin{tabular}{|c|c|c|c|c|c|}
	\toprule
	Photon energy (\SI{}{\kilo\electronvolt}) & fluence (\SI{}{\milli\joule\per\square\centi\metre}) & $A_1/(A_1+A_2)$ & $\tau_1$ (\SI{}{\nano\second}) & $A_2/(A_1+A_2)$ & $\tau_2$ (\SI{}{\nano\second}) \\
	\midrule
    1.028 & 14 & $\SI{38\pm2}{}$ \% & $\SI{14\pm5}{}$ & $\SI{62\pm2}{}$ \% & 10000 (fixed) \\
	9.6610 & 11.3 & $\SI{73\pm2}{}$ \% & $\SI{15\pm3}{}$ & $\SI{27\pm2}{}$ \% & 10000 (fixed) \\
	9.6684 & 11.3 & $\SI{90\pm4}{}$ \% & $\SI{17\pm3}{}$ & $\SI{10\pm4}{}$ \% & 10000 (fixed) \\
	9.6684 & 24.3 & $\SI{88\pm1}{}$ \% & $\SI{17\pm2}{}$ & $\SI{12\pm2}{}$ \% & 10000 (fixed) \\
	9.7607 & 11.3 & $\SI{83\pm4}{}$ \% & $\SI{21\pm3}{}$ & $\SI{17\pm4}{}$ \% & 10000 (fixed) \\
	9.7607 & 24.9 & $\SI{80\pm3}{}$ \% & $\SI{21\pm3}{}$ & $\SI{20\pm3}{}$ \% & 10000 (fixed) \\
	9.7607 & 67.9 & $\SI{86\pm1}{}$ \% & $\SI{33\pm1}{}$ & $\SI{14\pm1}{}$ \% & 10000 (fixed) \\  
	\bottomrule
	\end{tabular}
	\caption{\textbf{Fitted kinetic parameters at the Zn L$_3$- and K-edge of ZnO.} Results of the biexponential fitting of the kinetics at various excitation fluences and X-ray photon energies. The standard errors of the fit parameters are indicated.}
	\label{tabS:kinetics_fitted_parameters}
\end{table}

Kinetic traces are fitted with a function given by the convolution product of a Gaussian with a FWHM given by the instrument response function (\SI{70}{\pico\second}) and the sum of two exponential decays (parallel decay model) with the decay time constant of the longest component fixed at $\tau_2=\SI{10}{\micro\second}$. The amplitude of the individual decay components ($A_1$ \& $A_2$), the time constant of the shortest component ($\tau_1$) and the position of time zero (half rise of the transient signal) are left as free parameters of the fitting. The fitting is performed in Igor Pro (version 9) with the Levenberg-Marquardt algorithm \cite{Levenberg1944:251252}. The individual points are weighted by their standard deviations during the fitting. The fitted parameters are shown in Table \ref{tabS:kinetics_fitted_parameters}. The amplitude of each decay component is given as a relative weight (for instance $A_{1,r}=A_1/(A_1+A_2)$ for the shorter-lived specie). The standard error of the amplitude A$_1$ with a relative weight $A_{1,r}$ is propagated according to the relation,
\begin{equation}
\Delta A_{1,r} = \left|\frac{A_{1,r}}{A_1+A_2}\right|\sqrt{\frac{A_2^2}{A_1^2}\sigma_1^2+\sigma_2^2-2\frac{A_2}{A_1}\sigma_{12}}
\end{equation}
where $\sigma_1$ and $\sigma_2$ are the standard deviations of $A_1$ and $A_2$, respectively, and $\sigma_{12}$ is the covariance of $A_1$ and $A_2$.

\subsection{Zn \texorpdfstring{L$_3$}{L3}-edge}

\begin{figure}[!ht]
    \centering
    \includegraphics[width=0.45\linewidth]{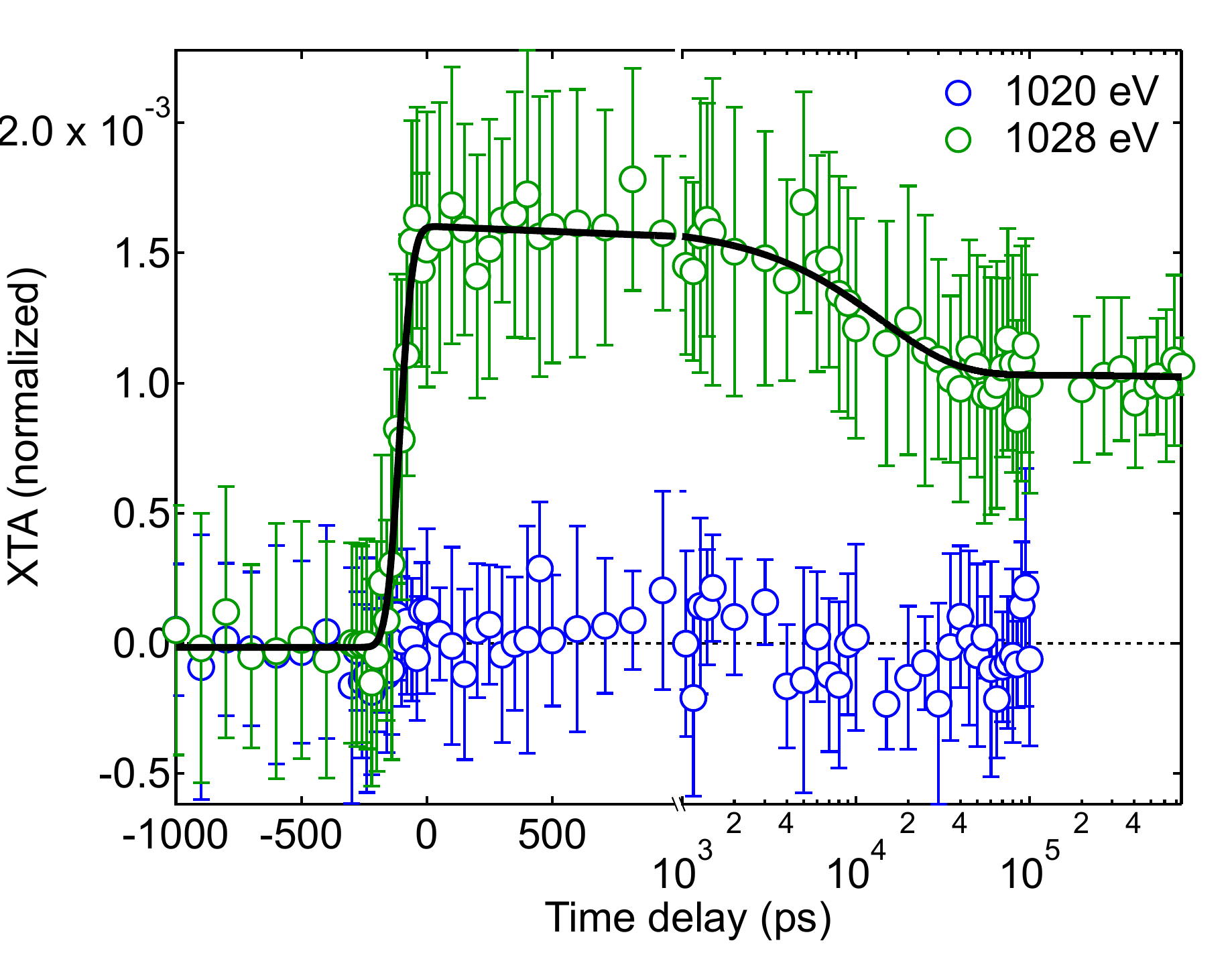}
    \caption{\textbf{XTA kinetics at the Zn L$_3$-edge.} Normalized time traces at the Zn L$_{2,3}$-edge of ZnO at \SI{1020}{\electronvolt} (blue circles) and \SI{1028}{\electronvolt} (green circles). The excitation fluence is \SI{14}{\milli\joule\per\square\centi\metre}. The black curve shows the result of a biexponential decay fitting.}
    \label{figSI:kinetics_ZnL3edge}
\end{figure}

Figure \ref{figSI:kinetics_ZnL3edge} displays time traces of the XTA at different energy points of the Zn L$_3$-edge. The time trace at \SI{1028}{\electronvolt} (green circles) is fitted with a biexponential decay (same model as at the K-edge). The decay time constants are given in Table \ref{tabS:kinetics_fitted_parameters}. The value of $\tau_1$ is the same as at the K-edge for similar excitation fluences.

\cleardoublepage


\section{Sample synthesis and characterization\label{secSI:sample_synthesis_charac}}

\subsection{ZnO thin film for measurements at the Zn K-edge}

\subsubsection{Synthesis}

The film was deposited by pulsed laser deposition (PLD) from a commercial ZnO target using a PVD Products nanoPLD 1000. The system used a KrF excimer laser (\SI{248}{\nano\metre}) with $\sim\SI{25}{\nano\second}$ pulse duration. Prior to deposition, the deposition chamber was pumped to a base pressure of $\sim\SI{2e-7}{}$ Torr. The deposition was done at \SI{600}{\degree}C in a \SI{20}{} mTorr O$_2$ atmosphere. A pulse energy of \SI{270}{\milli\joule}/pulse was used at a \SI{5}{\hertz} repetition rate. The laser was focused to a spot of $\sim\SI{1.8}{\milli\metre}\times\SI{1.0}{\milli\metre}$ (\SI{1.8}{\square\milli\metre}). Approximately \SI{55}{\percent} of the laser energy was lost in the optical train, hence the fluence at the target was $\sim\SI{6.7}{\joule\per\square\centi\metre}$. The target-substrate separation was \SI{80}{\milli\metre}, which yielded a deposition rate of $\sim\SI{0.008}{\nano\metre}$/pulse. The substrates, \textbf{c}-sapphire, were solvent cleaned with acetone and 2-propanol, then annealed ex situ at \SI{1100}{\degree}C for 2 hours, before deposition.

\subsubsection{X-ray diffraction}

\begin{figure}[!ht]
	\begin{center}
		\includegraphics[width=0.5\linewidth]{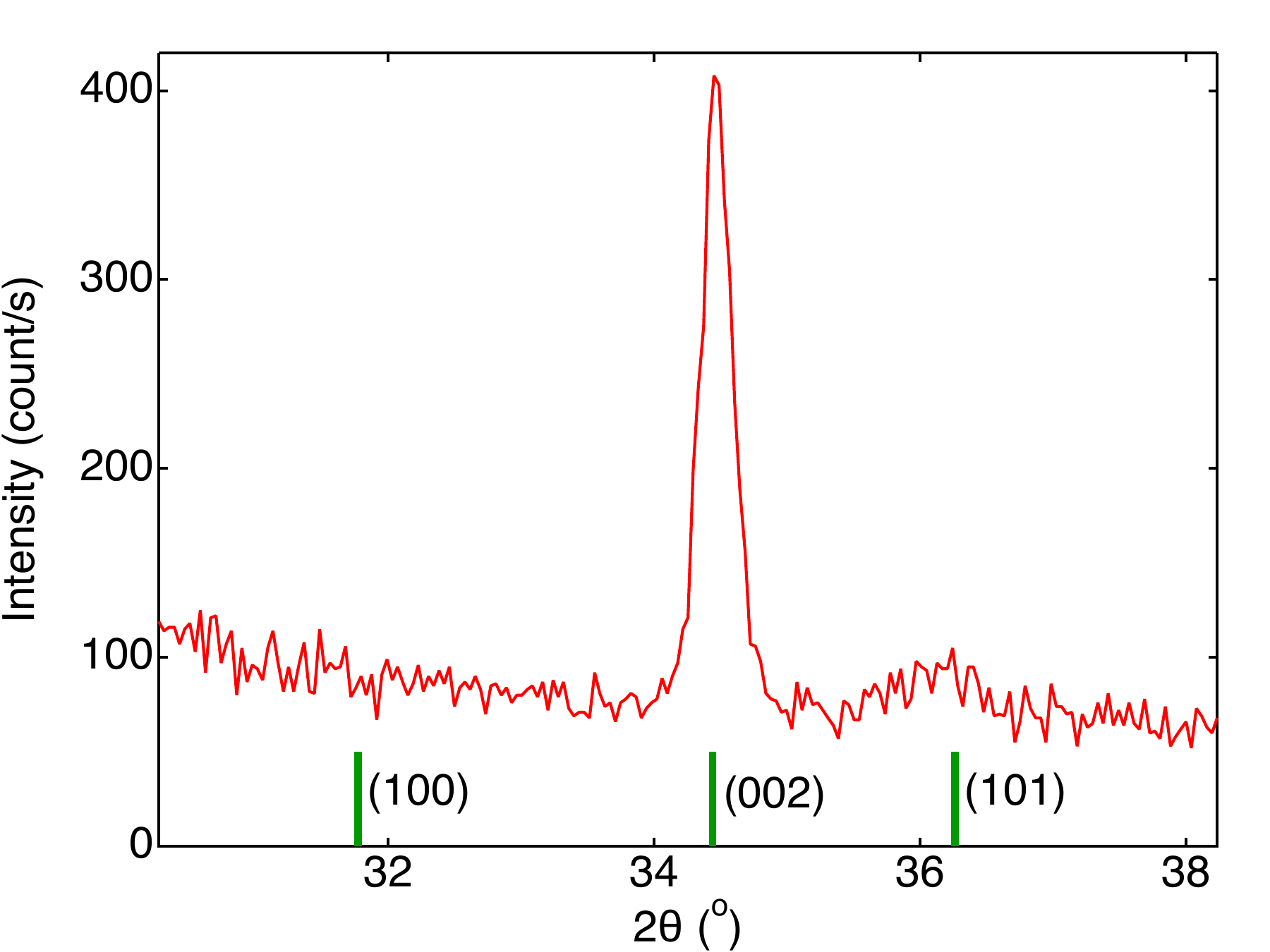}
	\end{center}
	\caption{\textbf{XRD.} X-ray diffraction pattern of ZnO thin film on \textbf{c}-sapphire substrate. Expected diffraction angles for randomly oriented ZnO from reference \cite{Kihara:1985vu} are shown with vertical green sticks.}
	\label{figSI:XRD}
\end{figure}

X-ray diffraction (XRD) patterns were measured in the Bragg-Brentano geometry with a PANanalytical instrument (at the X-ray CoreLab of the Helmholtz Zentrum Berlin) powered by a copper photocathode (\SI{40}{\kilo\volt}, \SI{40}{\milli\ampere}). The vertical divergence of the input X-ray beam is limited by a 1/16\SI{}{\degree} slit and the lateral divergence with \SI{0.04}{\degree} Soller slits. A parallel beam collimator of \SI{0.18}{\degree} is used in front of the detector. Figure \ref{figSI:XRD} shows the XRD pattern of the ZnO thin film on top of a \textbf{c}-sapphire substrate. A single (002) reflection is observed, which indicates the uniaxial orientation of the film. The expected position of Bragg reflections from isotropic ZnO are calculated based on the lattice parameters in reference \cite{Kihara:1985vu} (green sticks in Figure \ref{figSI:XRD}). The (002) peak position gives a lattice parameter $c=\SI{5.2026}{\angstrom}$, in good agreement with previous values on ZnO single crystals \cite{Kihara:1985vu, Khan:1968ig, Albertsson:1989gk}.

\subsubsection{Spectroscopic ellipsometry\label{subsec:spectroscopic_ellipsometry}}

\begin{figure}
	\begin{center}
		\includegraphics[width=0.48\linewidth]{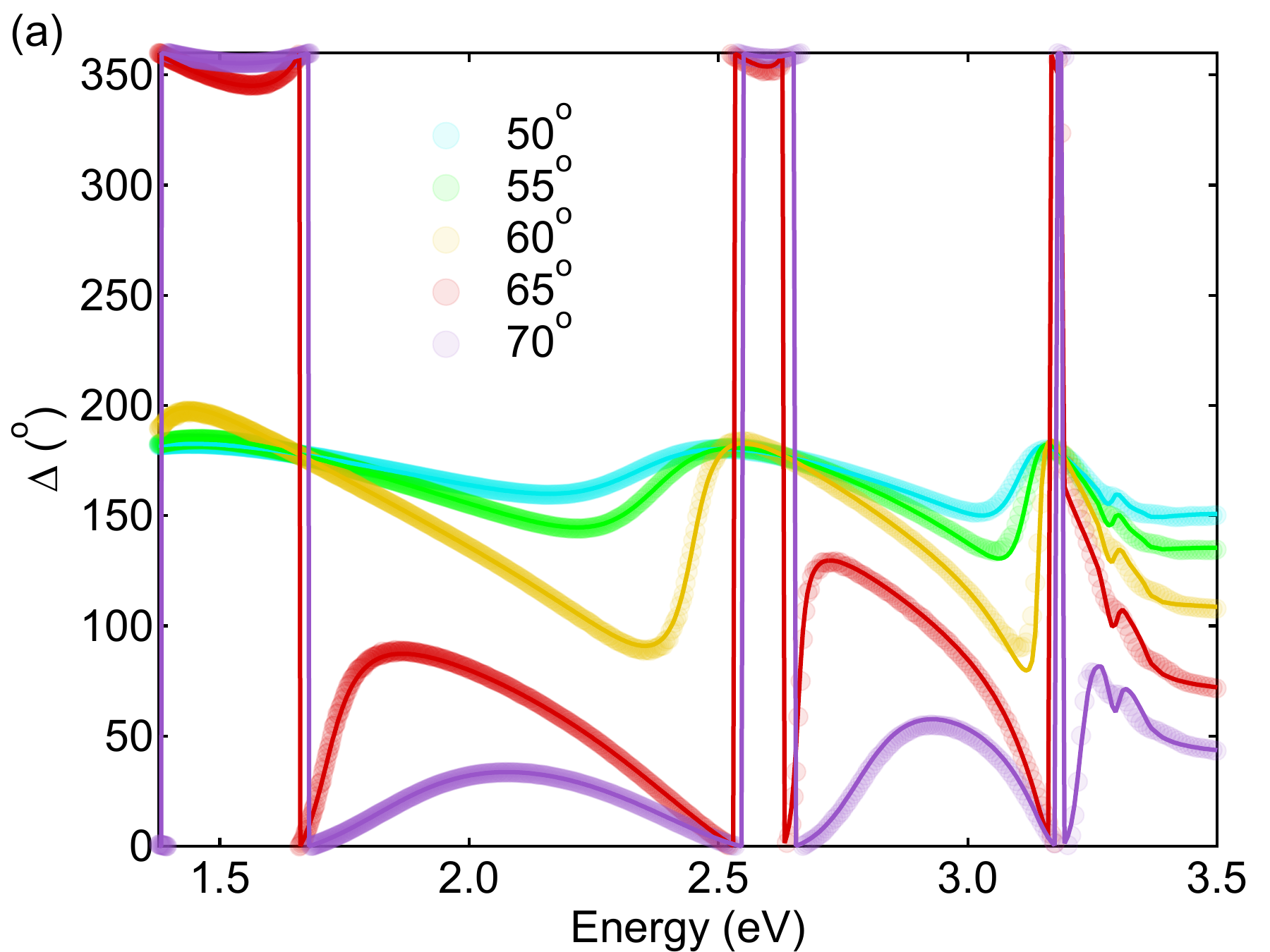}
		\includegraphics[width=0.48\linewidth]{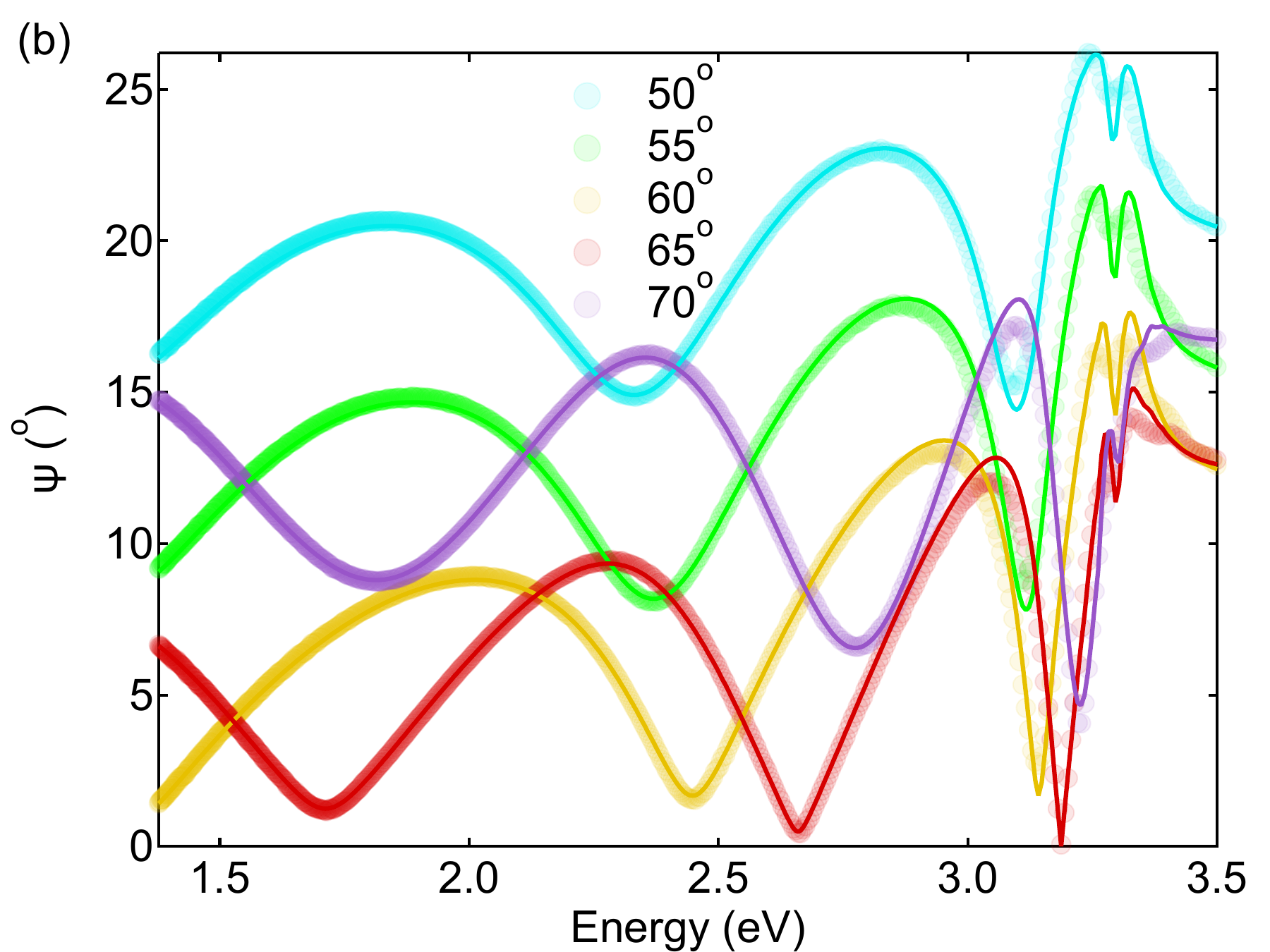}
	\end{center}
	\caption{\textbf{Angular dependence of spectroscopic ellipsometry constants.} Evolution of (a) $\Delta$ and (b) $\Psi$ measured by spectroscopic ellipsometry at various incidences angles on ZnO thin film (colored circles). Fitting results are shown with continuous curves (description of the model in the text).}
	\label{figSI:delta_psi_ellipsometry}
\end{figure}

\begin{figure}
    \centering
	\includegraphics[width=0.5\linewidth]{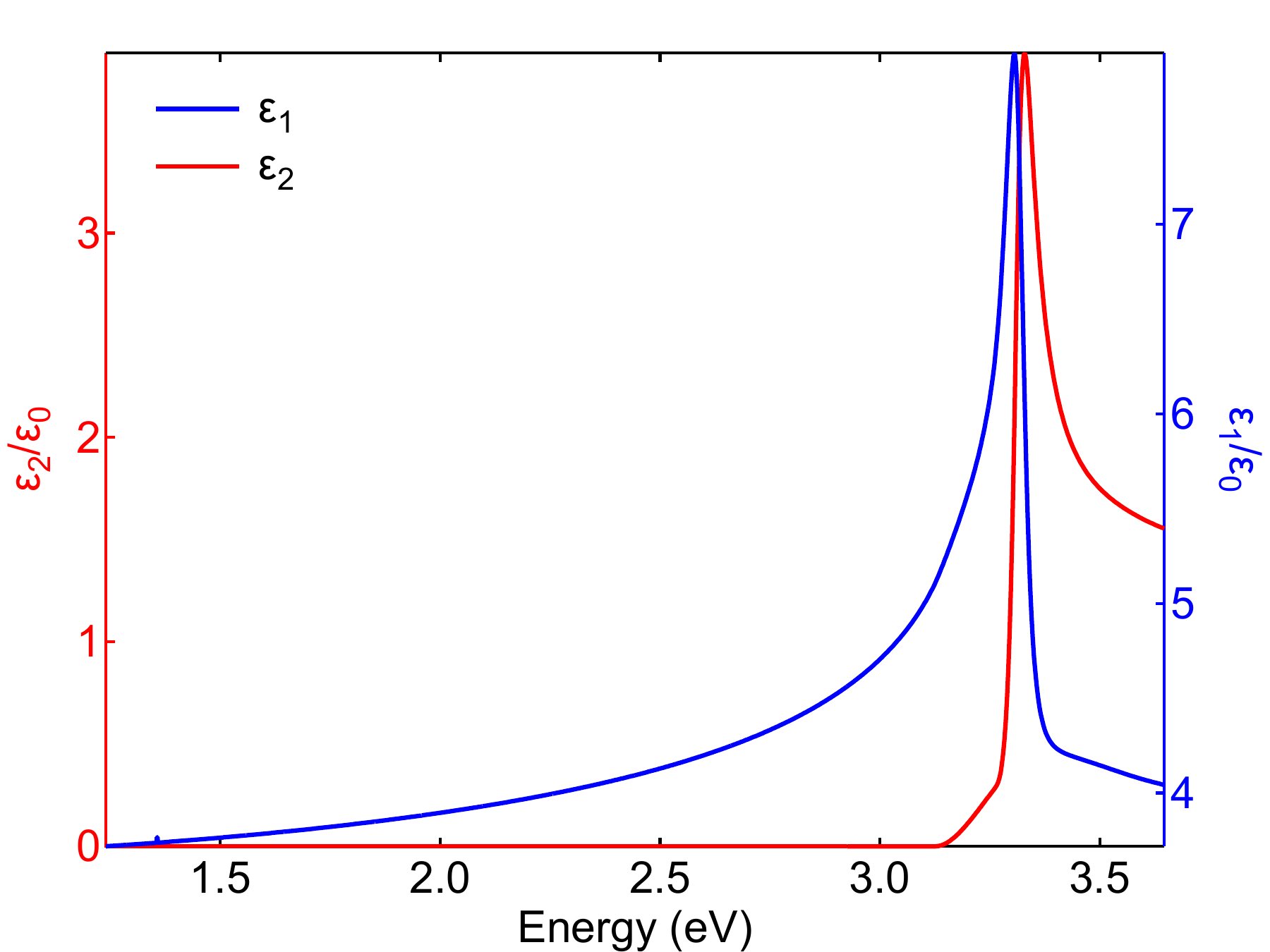}
	\caption{\textbf{Fitted permittivity of ZnO thin film.} Real (blue curve) and imaginary (red curve) part of the permittivity obtained by spectroscopic ellipsometry after deconvolution from the spectral oscillations due to the thickness of the ZnO film and contributions from the substrate.}
	\label{figSI:permittivity_ellipsometry}
\end{figure}

The thickness and optical properties of the ZnO thin film were measured by spectroscopic ellipsometry. The ellipsometer is from Sentech (SE850) and the data were analyzed with the SpectraRay 4 software. Figure \ref{figSI:delta_psi_ellipsometry} shows the evolution of the $\Delta$ and $\Psi$ ellipsometry parameters with the incidence angle (colored circles). The optical properties of the ZnO dielectric slab were modeled with a three-layer stack composed from top to bottom of: i) a semi-infinite air layer (incidence medium), ii) a ZnO thin film represented by a sum of three Tauc-Lorentz oscillators with a roughness layer at the interface with air, and iii) a \textbf{c}-sapphire substrate (semi-infinite layer) with a real refractive index (dispersion) taken from reference \cite{Palik1997}. Fittings are performed globally at all incidence angles. The results are shown with continuous curves in Figure \ref{figSI:delta_psi_ellipsometry}. The fitted parameters are given in Table \ref{tabSI:ellipsometry_fit_parameters}. The real and imaginary parts of the refractive index of the ZnO thin film after deconvolution from the thickness effect of the film are shown in Figure \ref{figSI:permittivity_ellipsometry}. The results of the fitting give for the ZnO thin film alone the permittivity $\epsilon_1/\epsilon_0=4.72$, $\epsilon_2/\epsilon_0=1.76$ at \SI{355}{\nano\metre}, which are used to calculate the reflectivity and the excitation density of the ZnO thin film (see section \ref{secSI:calculation_excitation_density}).

\begin{table}
	\centering
	\begin{tabular}{ccc}
	\toprule
	Parameter & Value & Standard error \\
	\midrule
	Roughness Air/ZnO & \SI{16.7}{\nano\metre} & \SI{0.1}{\nano\metre} \\
	Fraction of inclusion Air/ZnO & \SI{0.45}{\nano\metre} & \SI{0.01}{\nano\metre} \\
	ZnO thickness & \SI{283}{\nano\metre} & \SI{2}{\nano\metre} \\
	ZnO $E_{g,0}$ & \SI{3.13}{\electronvolt} & \SI{0.01}{\electronvolt} \\
	ZnO $A_0$ & 117 & 9 \\
	ZnO $E_{0,0}$ & \SI{3.10}{\electronvolt} & \SI{0.02}{\electronvolt} \\
	ZnO $C_0$ & \SI{0.45}{\electronvolt} & \SI{0.02}{\electronvolt} \\
	ZnO $E_{g,1}$ & \SI{3.32}{\electronvolt} & \SI{0.01}{\electronvolt} \\
	ZnO $A_1$ & 606 & 17 \\
	ZnO $E_{0,1}$ & \SI{3.26}{\electronvolt} & \SI{0.01}{\electronvolt} \\
	ZnO $C_1$ & \SI{0.055}{\electronvolt} & \SI{0.001}{\electronvolt} \\
	ZnO $E_{g,2}$ & \SI{7.26}{\electronvolt} & \SI{0.02}{\electronvolt} \\
	ZnO $A_2$ & 214 & 11 \\
	ZnO $E_{0,2}$ & \SI{3.6}{\electronvolt} & \SI{0.2}{\electronvolt} \\
	ZnO $C_2$ & \SI{30}{\electronvolt} & \SI{15}{\electronvolt} \\
	\bottomrule
	\end{tabular}
	\caption{\textbf{Spectroscopic ellipsometry fit parameters.} Optimized parameters for the fitting of the permittivity of ZnO thin film by spectroscopic ellipsometry. The Tauc-Lorentz oscillators have the following parameters: band gap ($E_g$), amplitude ($A$), resonance energy ($E_0$), and broadening ($C$). The three Tauc-Lorentz oscillators have subscripts 0, 1, and 2. The optical dielectric constant $\epsilon_{1,\infty}$=1 is kept constant for the fitting.}
    \label{tabSI:ellipsometry_fit_parameters}
\end{table}

\subsection{ZnO thin film for measurements at the Zn \texorpdfstring{L$_3$}{L3}-edge}

\subsubsection{Synthesis}

The ZnO thin films grown by plasma-assisted Molecular Beam Epitaxy (MBE) were obtained in a RIBER EPINEAT system. An effusion cell was used for Zn (\SI{99.9999}{\percent}) and a radio-frequency plasma cell was used for O (\SI{99.9999}{\percent}). The growth was conducted on \SI{100}{\nano\meter} thick SiN membranes on top of Si (\SI{500}{\micro\meter} thick) substrates (Norcada), and performed under Zn-rich conditions, mimicking optimized growth condition for \textbf{c}-plane O-polar ZnO. The growth rate, O-limited, was \SI{0.345}{\micro\meter\per\hour} (\SI{0.095}{\nano\meter\per\second}). A two-step growth was applied in order to improve the structural quality of the ZnO thin films. The first layer, about \SI{20}{\nano\meter} thick, was deposited at \SI{450}{\degree}C and was then annealed at about \SI{670}{\degree}C during \SI{30}{} minutes. The growth was then restarted using the same growth parameters. Note that the nominal temperatures were not calibrated on the membranes substrates. Such a double-step growth is needed to achieve a good ZnO structural quality, promoting in particular the growth of the (0001) orientation over the most stable (10-11) orientation. The growth rate was calibrated using both cross section measurements by Scanning Electron Microscopy (SEM) and X-ray reflectivity (XRR). The thickness of the sample studied was \SI{400\pm5}{\nano\meter}, determined by spectroscopic ellipsometry. 

ZnO thin films were also deposited by reactive magnetron sputtering in a gas mixture of Ar and O$_2$ from a ceramic ZnO target. The oxygen partial pressure was adjusted to about \SI{9}{\percent}. The total sputtering pressure was \SI{0.5}{\pascal} and the plasma was excited by RF (\SI{13.56}{\mega\hertz}) with a discharge power of \SI{50}{\watt}. The deposition rate was about \SI{1.8}{\nano\meter\per\minute}. Due to the low discharge voltages (\SI{72}{} to \SI{80}{\volt}) the energetic bombardment during the deposition is minimized. The substrates were not intentionally heated during the deposition. Due to the energy input from the plasma and the deposited atoms (Zn, O), a slight temperature increase in the samples occurred ($<\SI{100}{\degree}$C). 

\cleardoublepage


\section{Thermal effects in XAS and XTA spectra\label{secSI:XTA_thermal_effects}}

This section describes the procedure to separate thermal and non-thermal contributions to XTA spectra in two steps: i) the measurement of XAS spectra at different lattice temperatures (section \ref{secSI:XAS_temp_dep}), and ii) the simulation of XTA spectra upon lattice heating and the comparison with experimental XTA spectra in the EXAFS (section \ref{secSI:XTA_thermal_non_thermal}).

\subsection{Temperature-dependent XAS spectra\label{secSI:XAS_temp_dep}}

\begin{figure}
    \hspace{3.5mm}\includegraphics[height=0.33\linewidth]{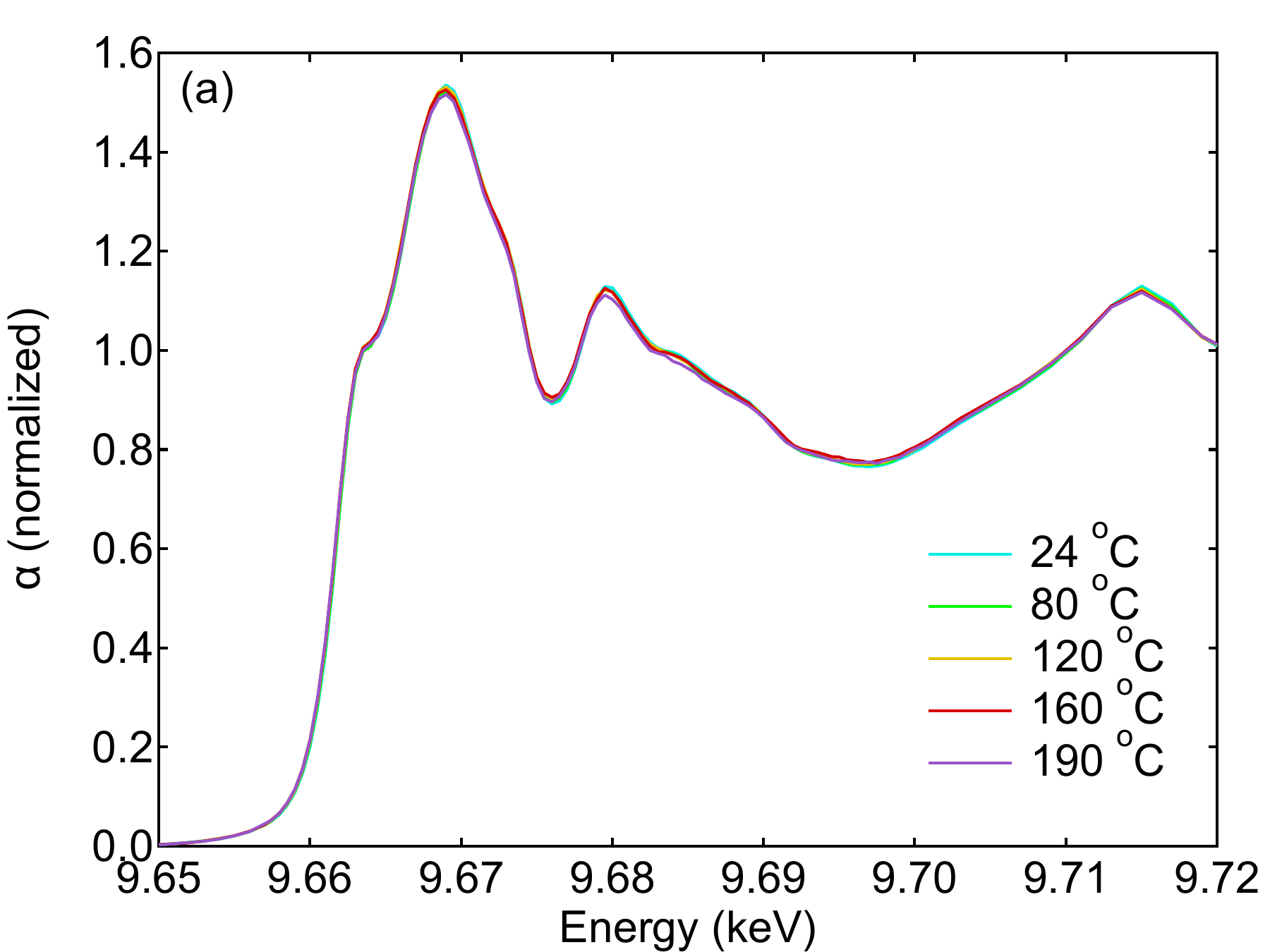}\hspace{0.95cm}
	\includegraphics[height=0.33\linewidth]{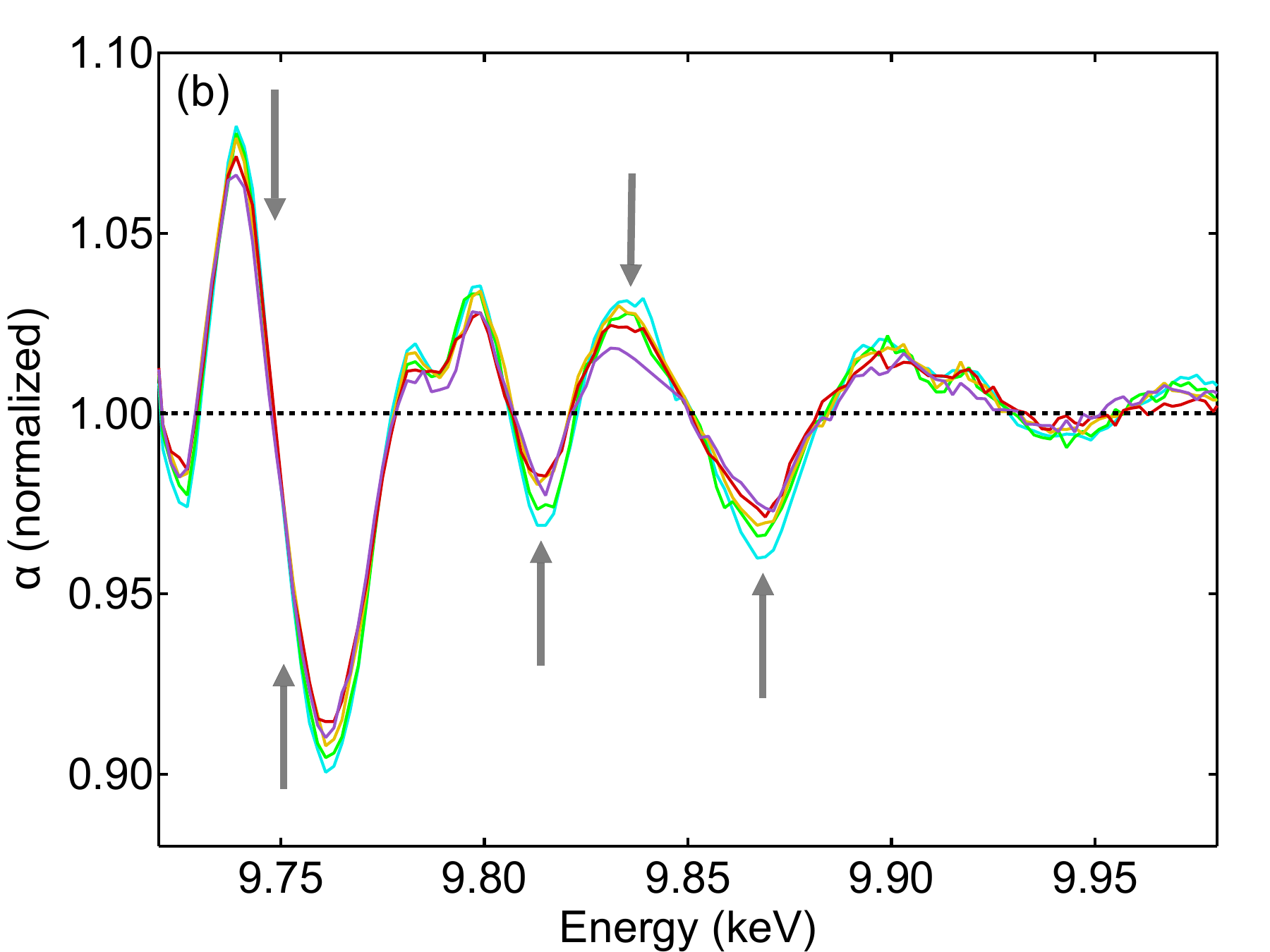}
	
    \hspace{3.5mm}\includegraphics[height=0.33\linewidth]{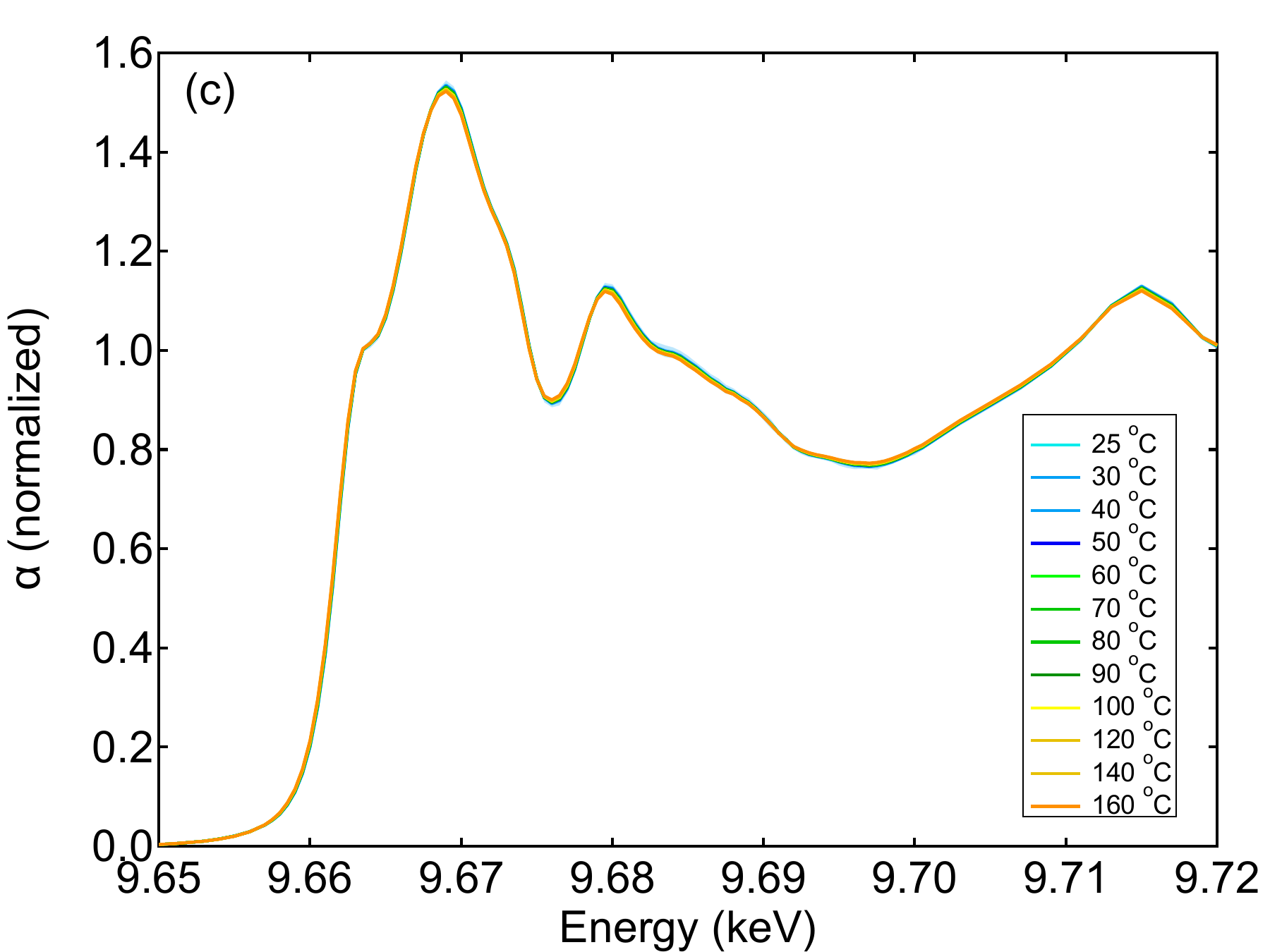}\hspace{0.95cm}
	\includegraphics[height=0.33\linewidth]{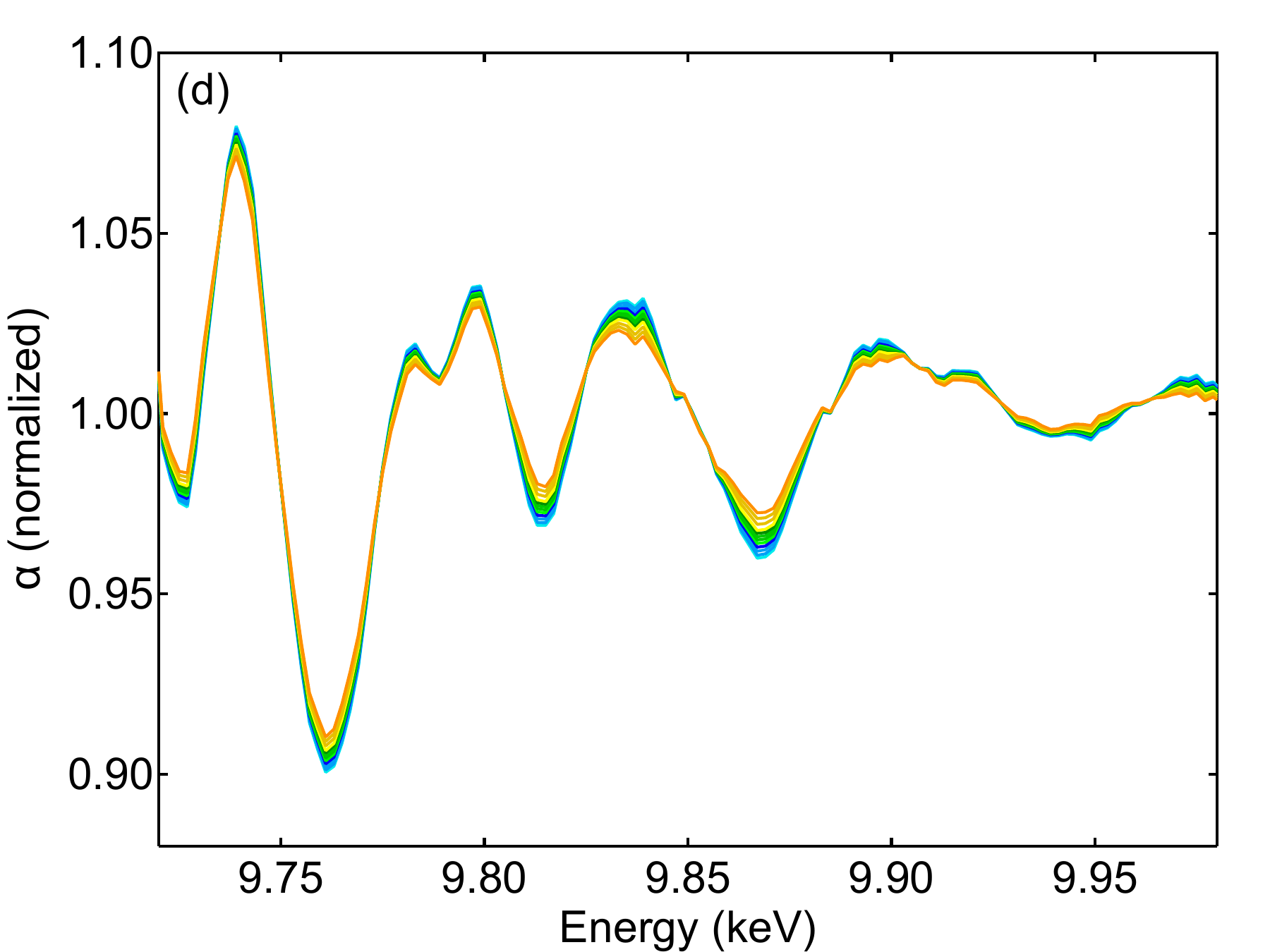}
	\begin{center}
        \includegraphics[height=0.33\linewidth]{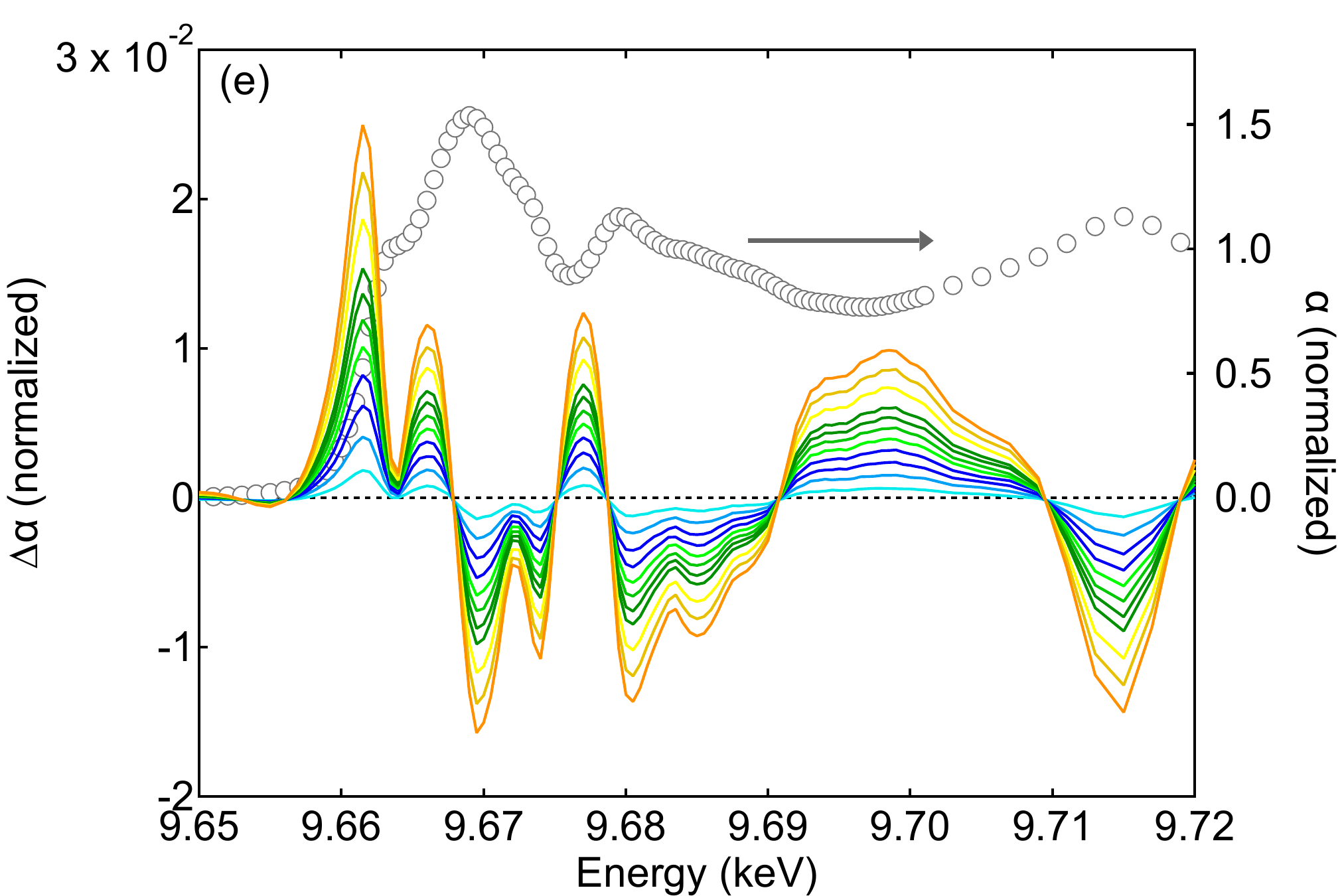}
	   \includegraphics[height=0.33\linewidth]{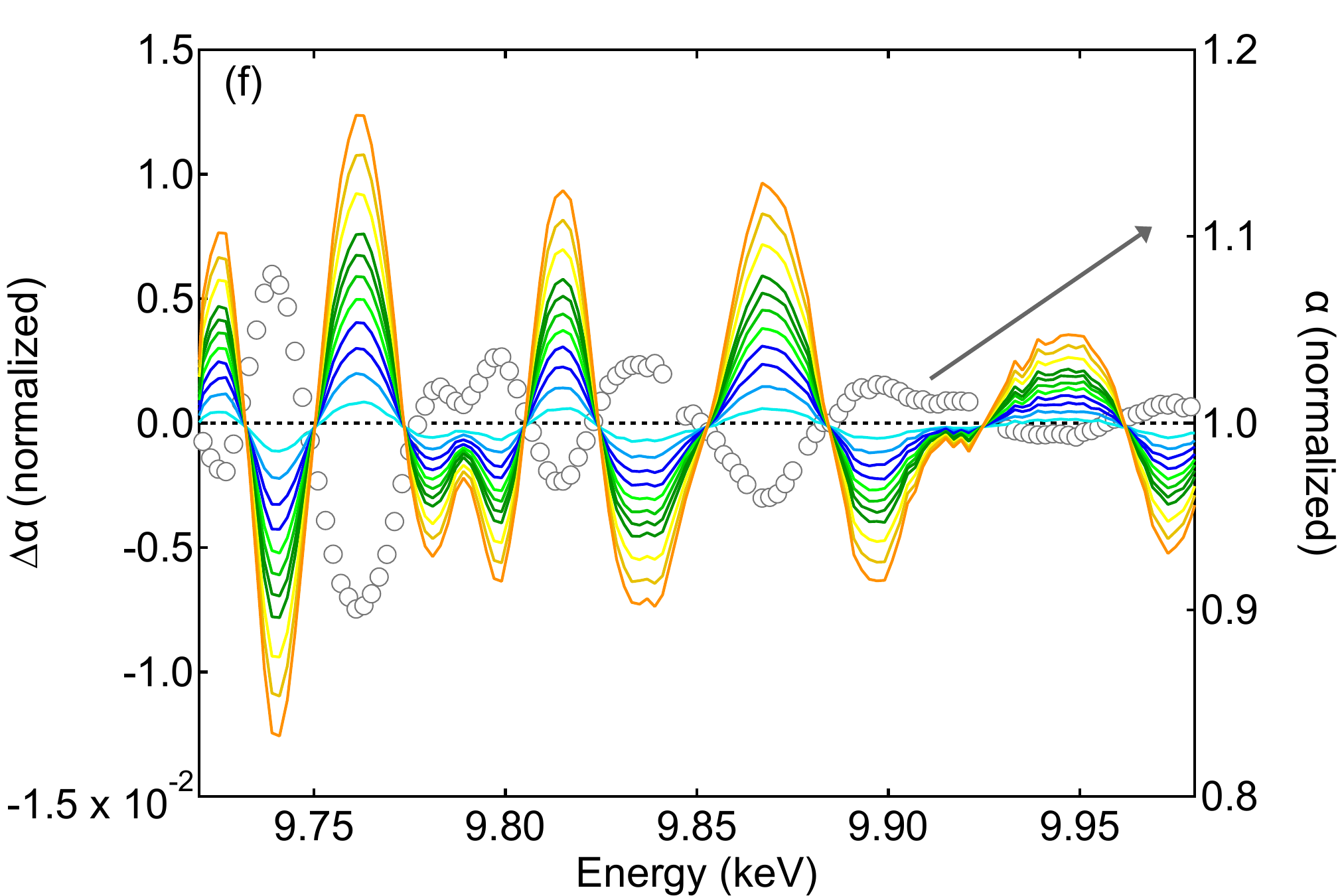}	
    \end{center}
	\caption{\textbf{Temperature-dependence of XAS spectra at the Zn K-edge.} Temperature-dependence of normalized XAS spectra at the Zn K-edge of ZnO in (a,c) the XANES, and (b,d) the EXAFS (colored curves, left axis). The spectra are (a,b) raw data and (c,d) after cubic spline interpolation along a temperature axis. Normalized temperature-difference XAS spectra with respect to the normalized XAS spectrum at \SI{24}{\degree}C are shown in (e) the XANES, and (f) the EXAFS (colored curves, left axis). The normalized room temperature XAS spectrum at \SI{24}{\degree}C is shown for reference (grey circles, right axis).}
    \label{figSI:XAS_temp_results_interpolation}
\end{figure}

\begin{figure}
	\centering
	\includegraphics[width=0.5\linewidth]{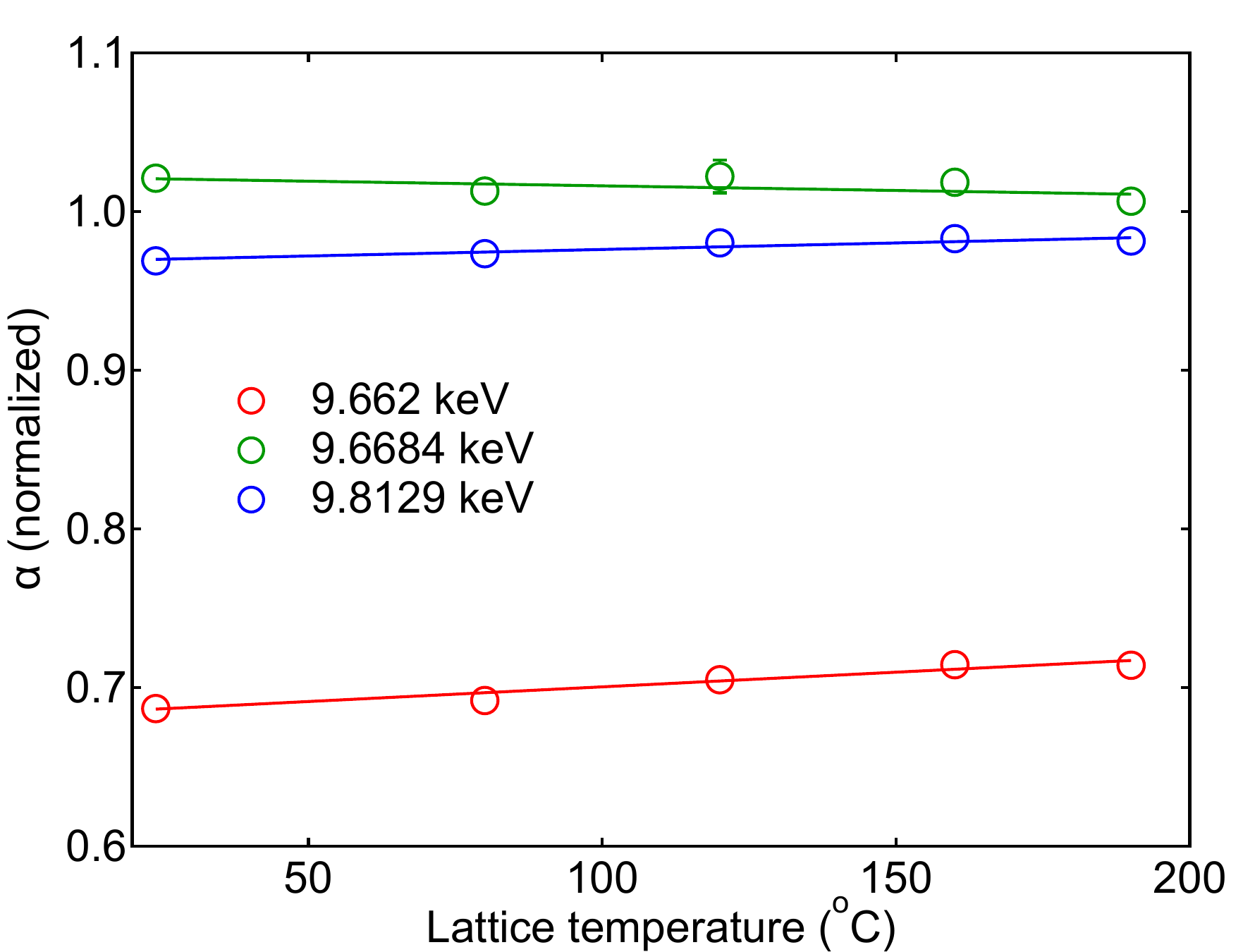}
	\caption{\textbf{XAS amplitude with lattice temperature at the Zn K-edge.} Evolution of the XAS amplitude at \SI{9.662}{\kilo\electronvolt} (red circles), \SI{9.6684}{\kilo\electronvolt} (green circles) and \SI{9.8129}{\kilo\electronvolt} (blue circles). Vertical error bars are standard deviations from the measurements. Linear fits are shown with continuous lines.}
	\label{fig:XAS_temp_dep_amplitude}
\end{figure}

Temperature-dependent XAS spectra between room temperature (\SI{24}{\degree}C) and \SI{190}{\degree}C at the Zn K-edge are shown in Figure \ref{figSI:XAS_temp_results_interpolation}a,b. A clear damping of the EXAFS oscillations around the post-edge absorption line (horizontal dashed line) is observed in the EXAFS at increasing lattice temperatures (black arrows in Figure \ref{figSI:XAS_temp_results_interpolation}b). The absorption coefficient evolves linearly with the temperature across the whole temperature range (Figure \ref{fig:XAS_temp_dep_amplitude}). For an accurate subtraction of the thermal contribution to the XTA spectra, the spectra in Figure \ref{figSI:XAS_temp_results_interpolation}a,b are interpolated using a model-free linear spline function between room temperature and \SI{190}{\degree}C, which corresponds to the temperature range covered by the experimental data. The standard deviation of the experimental data points is included in the interpolation as a statistical weight factor. The interpolated XAS spectra are shown in Figure \ref{figSI:XAS_temp_results_interpolation}c,d as well as the difference XAS spectra between a given temperature and the XAS spectrum at room temperature (\SI{24}{\degree}C) in Figure \ref{figSI:XAS_temp_results_interpolation}e,f. The standard deviations of individual energy points are conserved for the interpolated data. For the energy points that were not present in the original XAS data but were present in the XTA data, a linear interpolation was performed to obtain the corresponding XAS and its standard deviation.

\subsection{Thermal contributions to XTA spectra\label{secSI:XTA_thermal_non_thermal}}

\subsubsection{General remarks\label{subsecSI:XTA_contributions}}

\begin{figure*}
	\centering
    \includegraphics[width=0.495\linewidth]{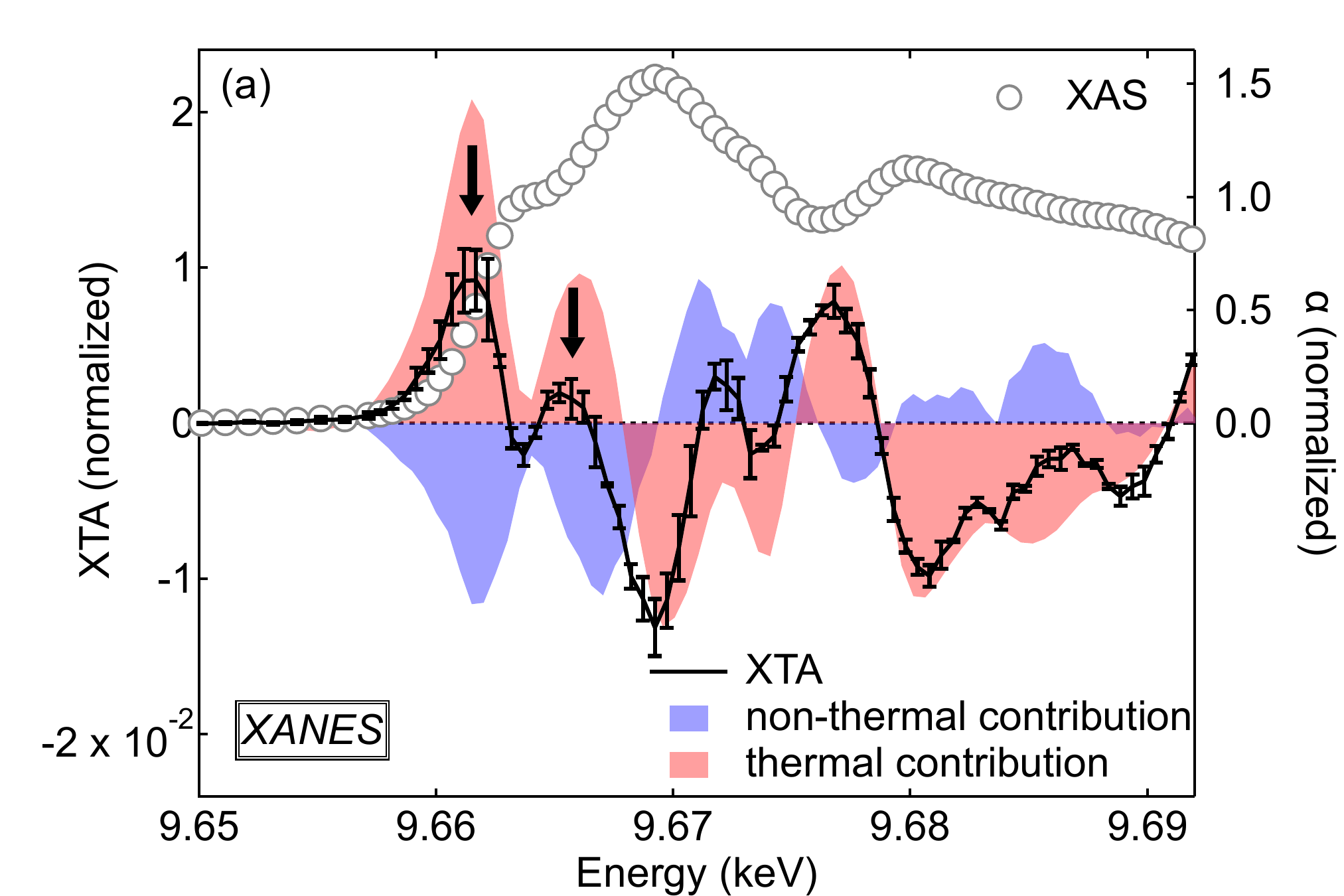}    
    \includegraphics[width=0.495\linewidth]{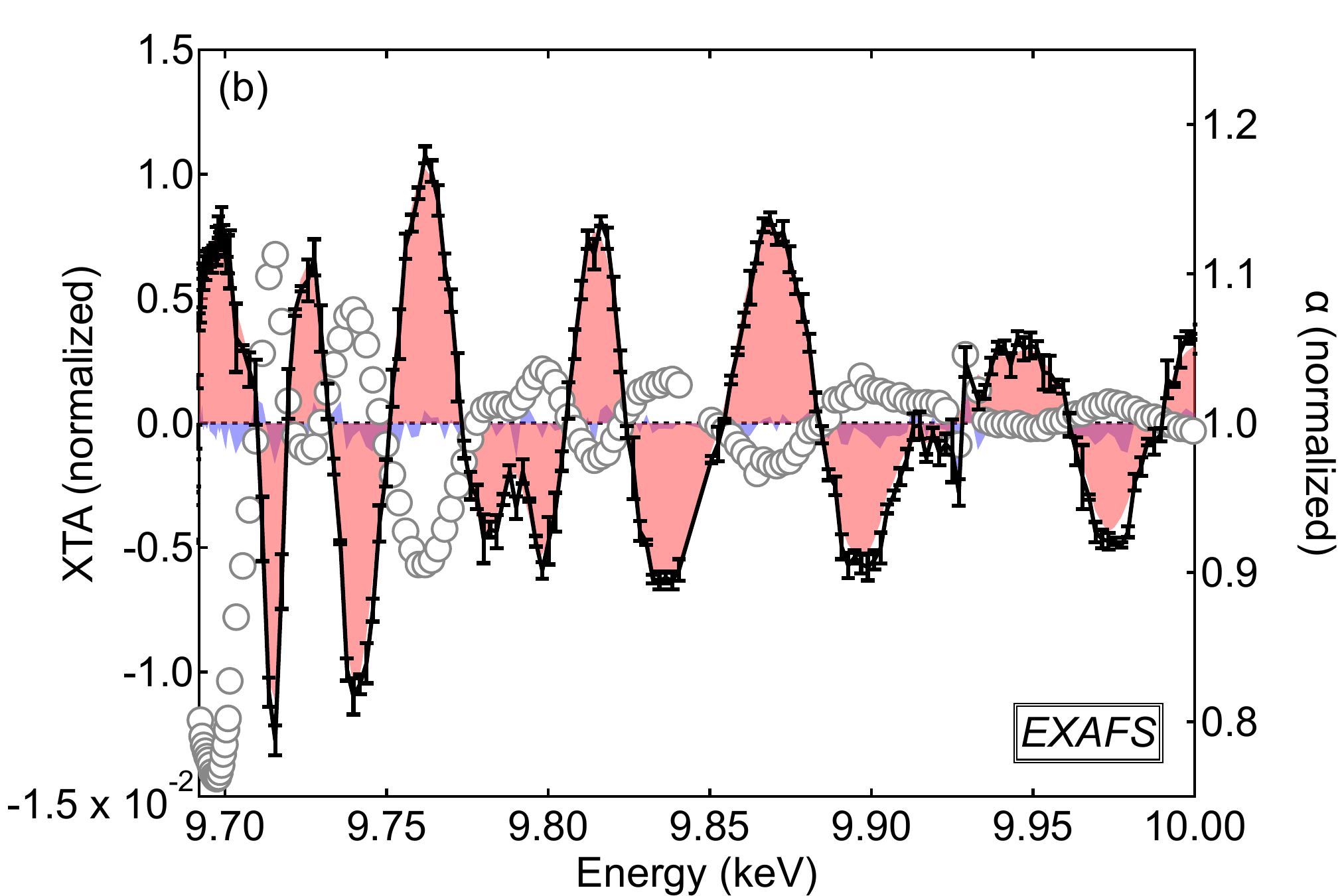}
	\caption{\textbf{Decomposition of thermal and non-thermal contributions in the X-ray transient absorption (XTA) spectrum.} Edge-jump normalized XTA spectra at \SI{67.6}{\milli\joule\per\centi\metre\squared} excitation fluence at the Zn K-edge of ZnO (0001) thin films in (a) the XANES and (b) the EXAFS (black curves with error bars, left axis). The XTA spectrum is decomposed into thermal (shaded red area) and non-thermal (shaded blue area) contributions. Time delay is \SI{100}{\pico\second}, excitation is \SI{3.49}{\electronvolt}, incidence angle is at \SI{45}{\degree} with respect to (0001). Error bars represent standard deviations between individual XTA measurements. The edge-jump normalized XAS spectrum at the Zn K-edge (absorption coefficient, $\alpha$) is shown with gray circles for reference (right axis).}
	\label{figSI:XTA_100ps_spectral_traces_decomposition}
\end{figure*}

Transient EXAFS spectra in Figure \ref{figSI:XTA_100ps_spectral_traces_decomposition}b displays a mirror image lineshape of the equilibrium EXAFS spectrum around a normalized and flattened post-edge absorption line. No phase shift or change in frequency is observed. The damping of EXAFS oscillations is related to increased thermal and/or static disorder, which is phenomenologically described by increased Debye-Waller factors (amplitude of the mean square displacement of the atoms) \cite{Timoshenko:2014el}. In the present case, the increased disorder is due to lattice heating as a consequence of energy transfer between the photoexcited carriers and the phonon bath by carrier-phonon coupling on the sub-picosecond timescale \cite{Zhukov:2010ex}. At \SI{100}{\pico\second} time delay, the excess carrier energy has been transferred to the lattice, and the electronic temperature is nearly equal to the hot lattice temperature \cite{Othonos:1998cx}. The hot lattice does not display a sizeable thermal expansion, which would increase the periodicity of the EXAFS oscillations in the excited state. We attribute this effect to the weaker changes induced by $<\SI{2}{\milli\angstrom}$ ($\sim\SI{0.02\pm0.01}{\percent}$) variations of the average Zn-O bond distance \cite{Rossi2021} and lattice parameters \cite{Albertsson:1989gk} with respect to the $\sim\SI{10}{\percent}$ change of the Debye-Waller factor \cite{Kihara:1985vu, Albertsson:1989gk} between \SI{300}{} and \SI{400}{\kelvin}, which corresponds to the typical lattice heating in the current experiment (\emph{vide infra}).

The unchanged periodicity of the EXAFS oscillations in the photoexcited state excludes the formation of polarons \cite{Sezen:2015in} or trapped holes \cite{Penfold:2018ie} as possible contributions to the transient spectra, since these localized carriers would induce a change of average bond length \footnote{Polarons can form in ZnO by direct excitation of the mid-gap states but their lifetime is only a few picoseconds \cite{Liu2021a}.}. The analysis of the XTA spectra at \SI{100}{\pico\second} time delay is based on the hypothesis that it can be decomposed into a linear combination of thermal and non-thermal signals due to the weak coupling between core excitons and lattice vibrations (further discussed in Section \ref{sec:discussion}, detail of the breakdown procedure in \ref{subsubsecSI:thermal_non_thermal_decomposition}). Figure \ref{figSI:XTA_100ps_spectral_traces_decomposition}b shows that the thermal contribution is dominant in the EXAFS (shaded red area) while the X-ray absorption near-edge structure (XANES) is made of a superposition of thermal and non-thermal (shaded blue area) contributions (Figure \ref{figSI:XTA_100ps_spectral_traces_decomposition}a). The two contributions in the XANES are distinct with different lineshapes, maxima, and zero-crossing points, in particular between \SI{9.668}{} and \SI{9.69}{\kilo\electronvolt}. 

XTA spectroscopy at the Zn K-edge is a bulk-sensitive technique (the penetration depth is $\sim\SI{8.5}{\micro\meter}$ at \SI{9.7}{\kilo\electronvolt} \cite{Hubbell:1969uu}), which is insensitive to carrier trapping at the defect-rich surface of the film. More generally, the lattice heating contribution to the XTA signal at \SI{100}{\pico\second} in the bulk of the ZnO thin film overwhelms any other signal that could originate from local electronic or structural perturbations involving only a few zinc atoms. In previous measurements on nanoparticles, however, the ratio of surface-to-bulk atoms was sufficiently large to observe carrier trapping at the defect-rich surface \cite{Penfold:2018ie} or the formation of small polarons \cite{Katz:2012ho}.

\subsubsection{Decomposition of thermal and non-thermal contributions to XTA spectra at the Zn K-edge\label{subsubsecSI:thermal_non_thermal_decomposition}}

\begin{figure}
	\centering
	\includegraphics[height=0.28\linewidth]{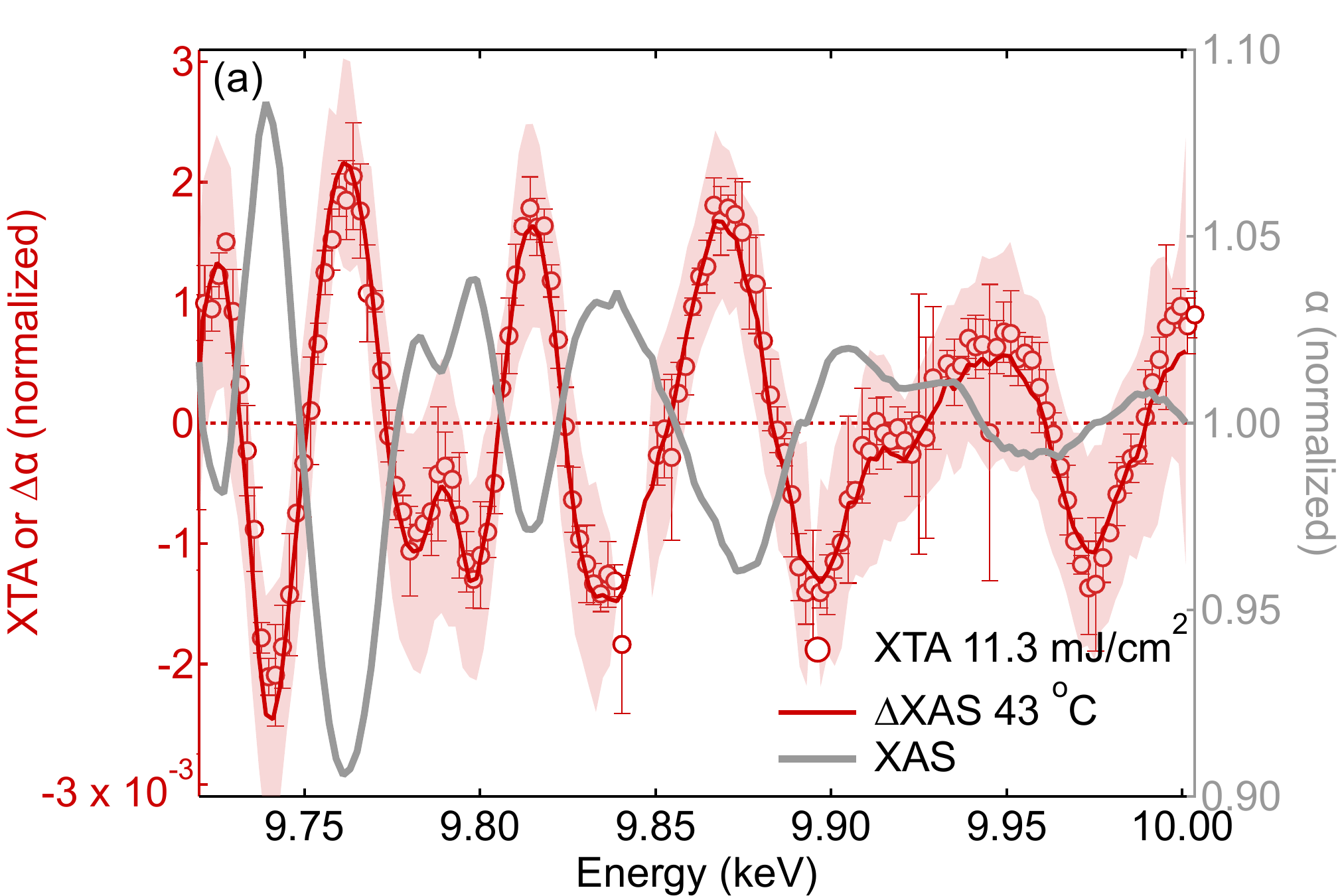}
	\includegraphics[height=0.28\linewidth]{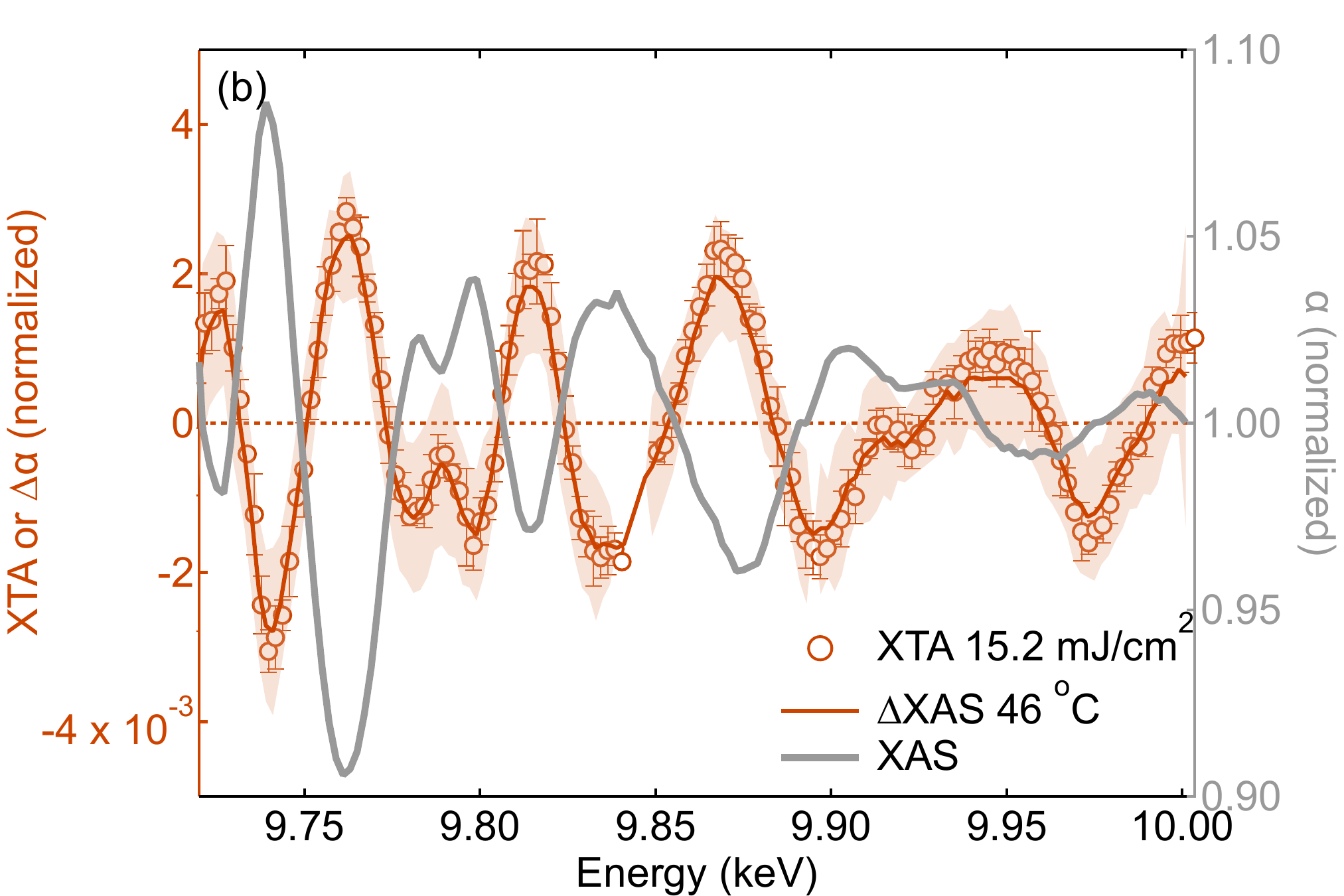}
	\includegraphics[height=0.28\linewidth]{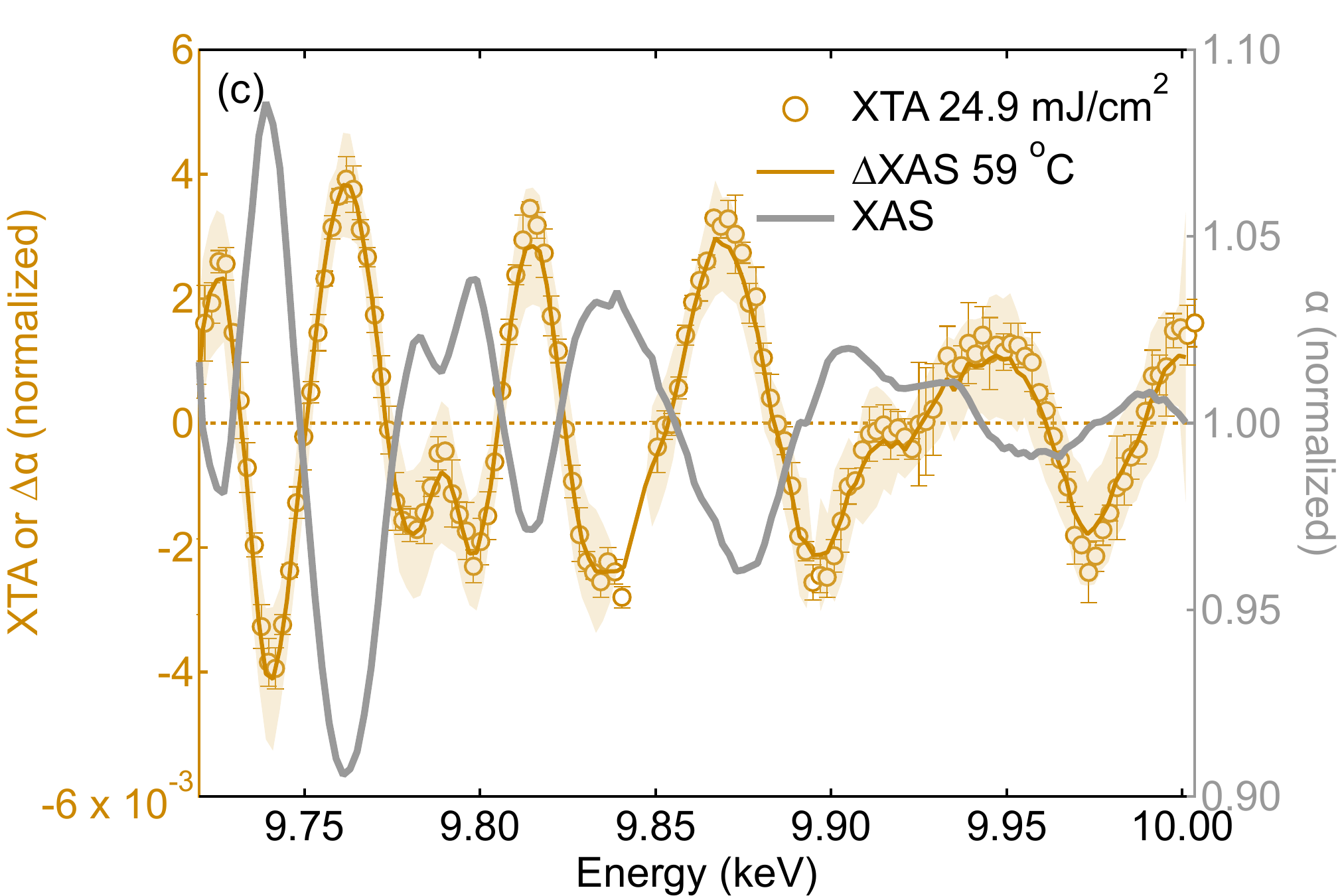}
	\includegraphics[height=0.28\linewidth]{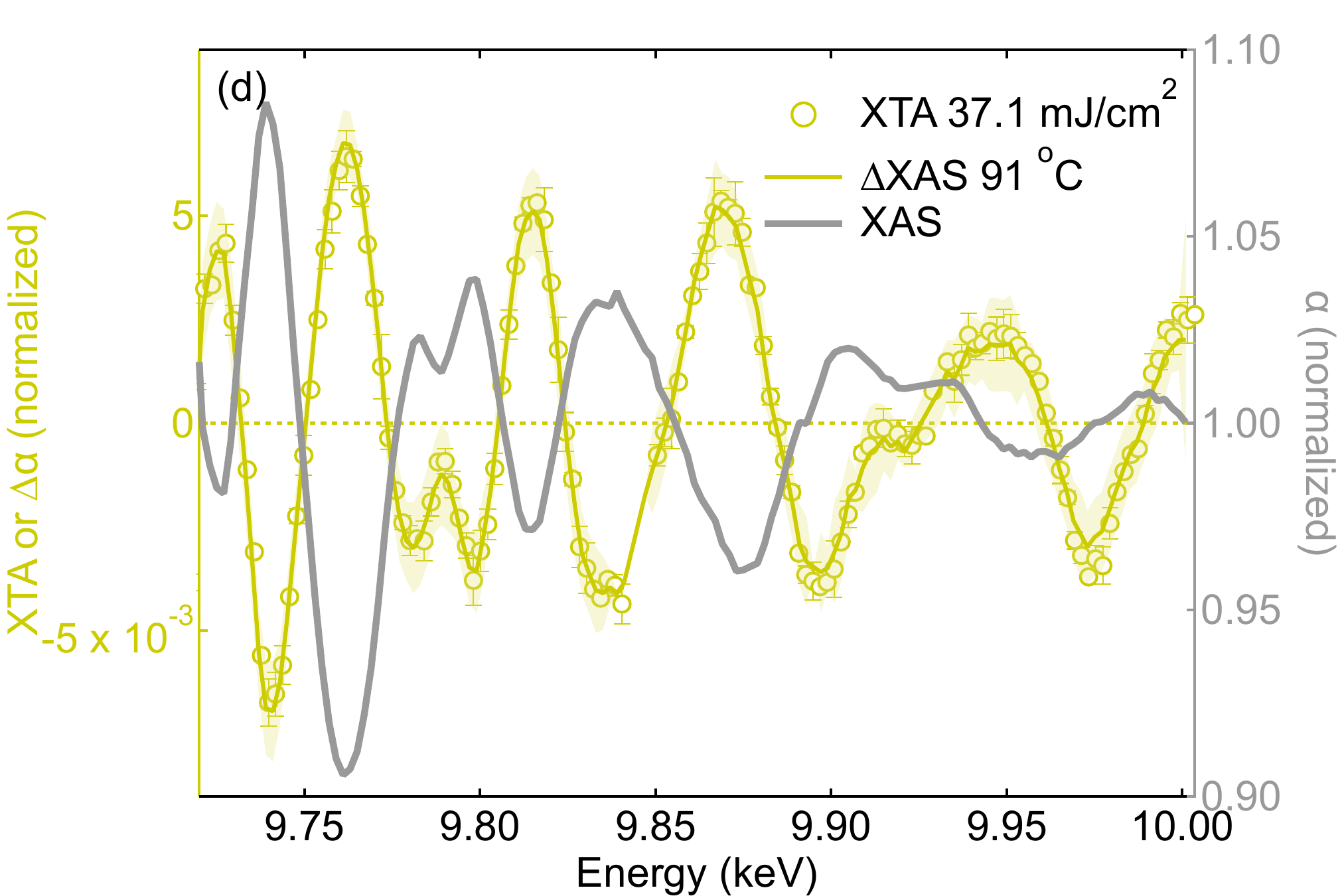}
	\includegraphics[height=0.28\linewidth]{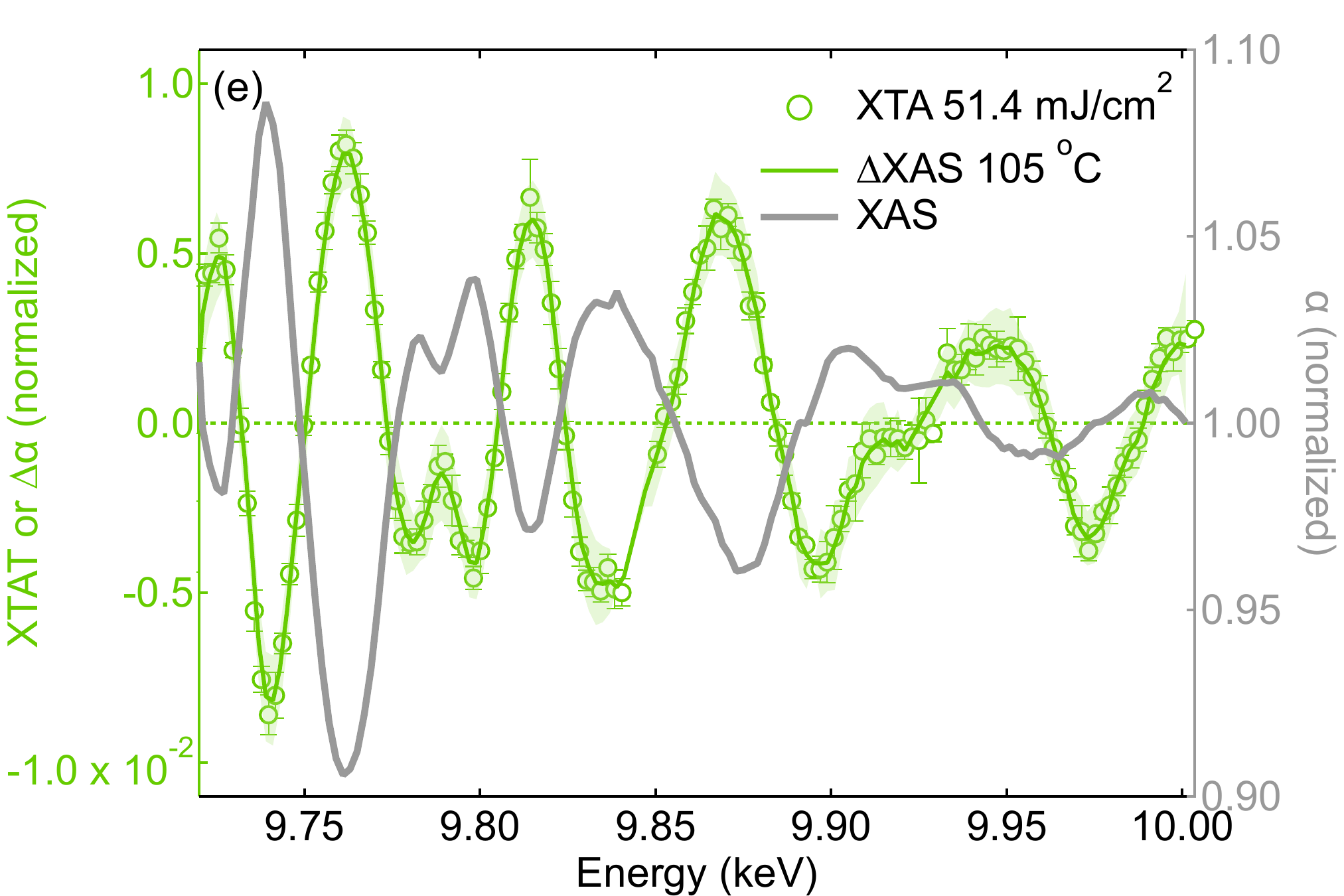}
	\includegraphics[height=0.28\linewidth]{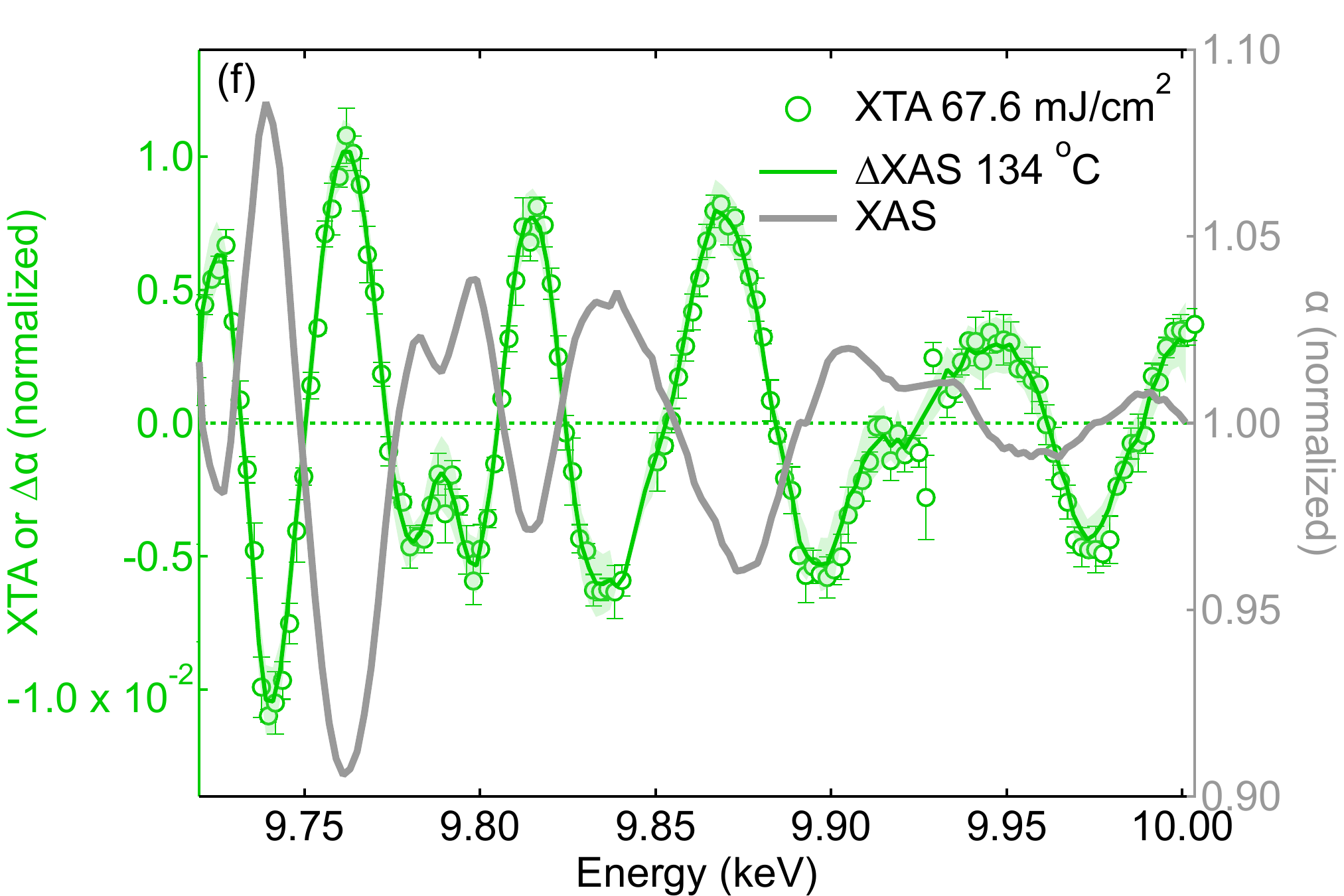}
	\includegraphics[height=0.28\linewidth]{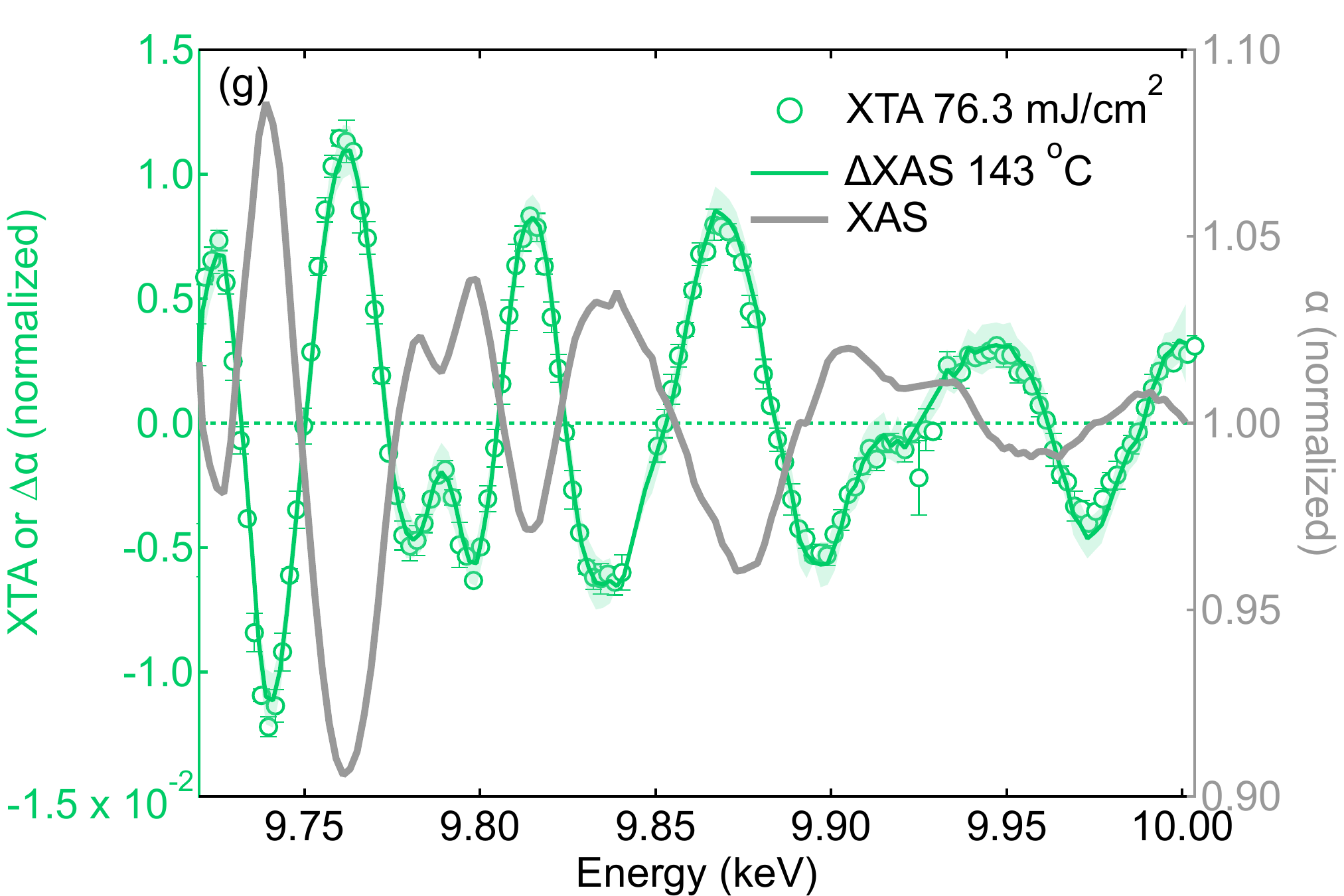}
	\includegraphics[height=0.28\linewidth]{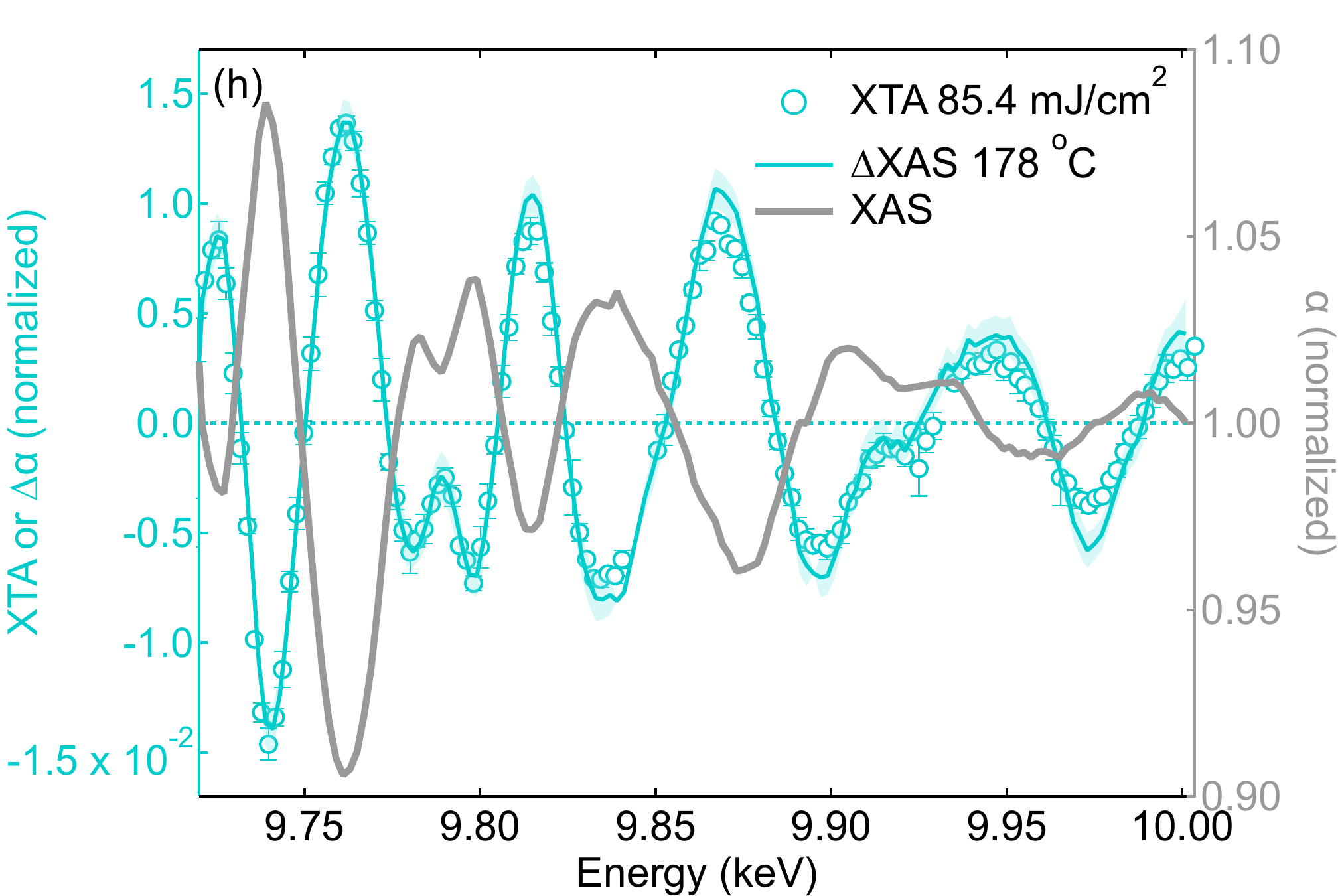}
	\caption{\textbf{Thermal contributions to XTA spectra in the EXAFS at the Zn K-edge.} Best agreement after $\chi^2$ minimization between XTA spectra in the EXAFS (colored circles, left axis) and temperature-difference XAS spectra (colored curves, left axis) at the Zn K-edge of ZnO for excitation fluences of (a) \SI{11.3}{}, (b) \SI{15.2}{}, (c) \SI{24.9}{}, (d) \SI{37.1}{}, (e) \SI{51.4}{}, (f) \SI{67.6}{}, (g) \SI{76.3}{}, and (h) \SI{85.4}{\milli\joule\per\square\centi\metre} (\SI{100}{\pico\second} time delay, excitation \SI{3.49}{\electronvolt}). Shaded areas and error bars represent standard deviations between individual measurements. The equilibrium EXAFS spectrum is shown for reference (grey curve, right axis).}
	\label{figSI:chi2_minimization_EXAFS}
\end{figure}

\begin{figure}
	\centering
	\includegraphics[height=0.35\linewidth]{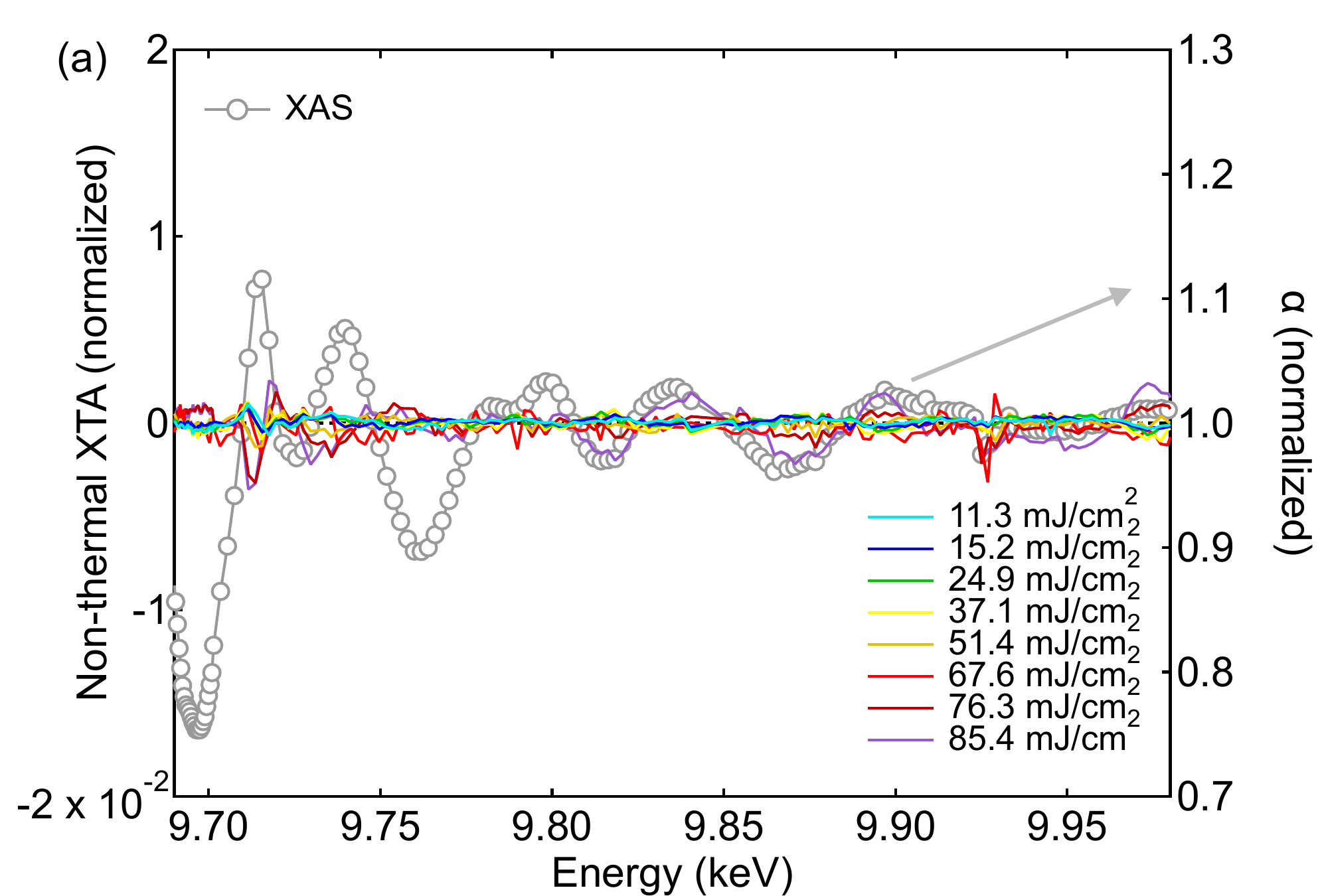}
    \includegraphics[height=0.375\linewidth]{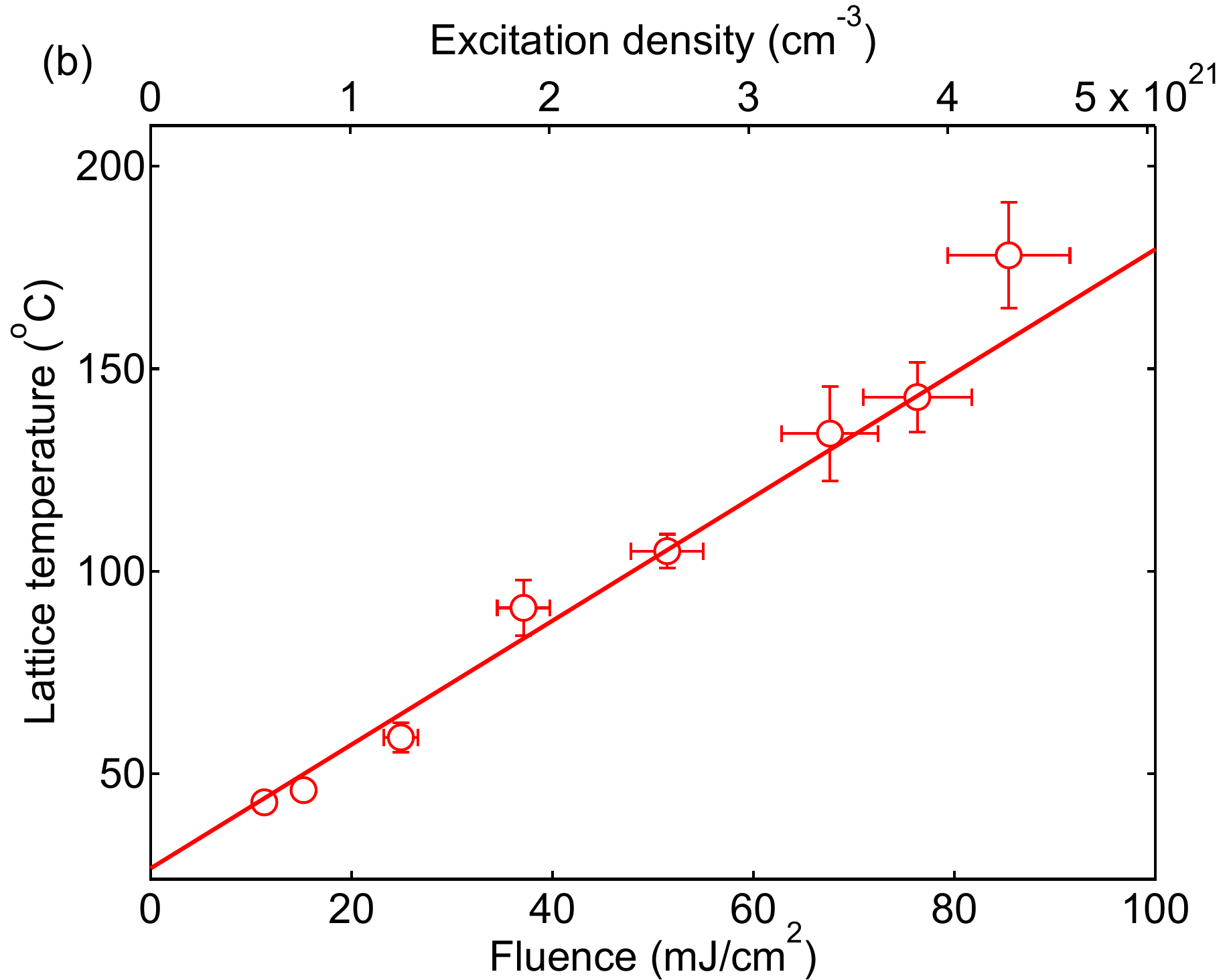}
	\caption{\textbf{$\chi^2$ residuals and optimized excited state lattice temperature.} (a) Residuals of the $\chi^2$ minimization between XTA spectra and the best-matched temperature-difference XAS spectra (colored curves, left axis). The equilibrium XAS spectrum is shown for reference (grey circles, right axis). (b) Lattice temperature obtained after $\chi^2$ optimization of the XTA spectra in the EXAFS (red circles). Vertical error bars are the variance of the fitted temperature, horizontal error bars are uncertainties in the estimate of the laser fluence. Linear fit to the data constrained to the vertical error bars (red curve). Top axis is the initial excitation density.}
	\label{figSI:chi2_minimization_residuals_lattice_temperature}
\end{figure}

XAS and XTA spectroscopy are sensitive to both electronic and structural degrees of freedom. In XTA spectra at \SI{100}{\pico\second} time delay, the effect of the electronic and structural perturbation in the excited state XAS spectrum can be considered independent, which is assigned to the fact that the incoherent thermal motion of the atoms does not perturb the formation of core excitons with a radius comparable to the interatomic bond distances (see main text). Since the EXAFS part of the spectrum is only sensitive to the local atomic displacements around the absorbing atom, it is a fingerprint of the local lattice temperature in the ZnO photoexcited state and can be used to separate thermal and non-thermal effects in XTA spectra. The procedure is in two steps: 1) perform a $\chi^2$-minimization between difference XAS spectra (in Figure \ref{figSI:XAS_temp_results_interpolation}f) and a given XTA spectrum to find the best description of the hot lattice, and 2) take the difference between the XTA spectrum and the optimized $\chi^2$ difference XAS spectrum. Steps 1) and 2) deliver the \emph{thermal} and \emph{non-thermal} contributions to the XTA spectrum, respectively. The procedure has been described in details in the Supporting Information of reference \cite{Rossi2021}. Figure \ref{figSI:chi2_minimization_EXAFS} displays the $\chi^2$-minimized difference XAS spectra (colored curves) together with XTA spectra at different excitation fluences (colored circles with error bars). Figure \ref{figSI:chi2_minimization_residuals_lattice_temperature}a displays the residuals of the $\chi^2$ minimization and \ref{figSI:chi2_minimization_residuals_lattice_temperature}b the excited state lattice temperature.

The non-thermal (electronic) part of the XTA spectrum differs from previously published spectra measured with the same polarization on ZnO nanorods \cite{Rossi2021}. The largest difference is the absence of a positive signal in the non-thermal XTA spectrum in Figure 4a of reference \cite{Rossi2021}. We expect this discrepancy may arise from 1) the distribution of nanorod orientations, which broadens the XAS spectrum, 2) larger uncertainties in the subtraction of the thermal contribution to the XTA spectrum due to a lower signal-to-noise ratio, and 3) a larger degree of structural flexibility of the ZnO nanorods, which may lead to some motion of the ZnO nanorods when the temperature increases associated with a change in the X-ray linear dichroism. The latter leads to an inaccurate subtraction of the thermal contribution to the XTA spectrum. 

\subsubsection{Decomposition of thermal and non-thermal contributions to XTA spectra at the Zn \texorpdfstring{L$_3$}{L3}-edge}

\begin{figure}[!ht]
    \centering
    \includegraphics[width=0.5\linewidth]{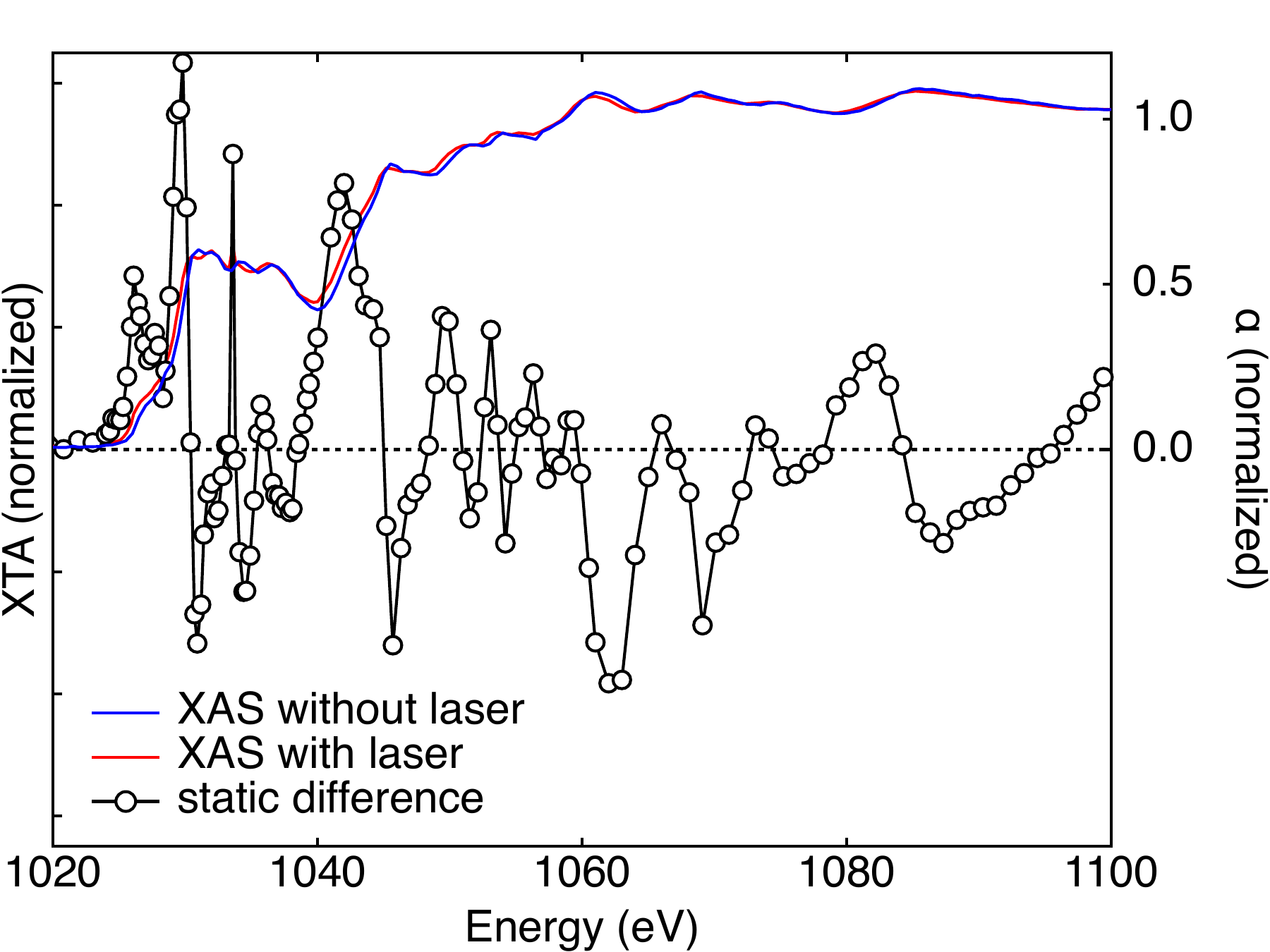}
    \caption{\textbf{Laser-induced static heating at the Zn L$_3$-edge.} Effect of static heating on the XAS spectrum at the Zn L$_3$-edge of ZnO. Consecutive XAS measurements are performed with (red curve) and without (blue curve) laser impinging on the sample. The difference of the XAS spectra with and without laser irradiation corresponds to the static heating (black circles). Normalization of the XAS spectra and the XAS difference spectra is over the edge jump. The laser fluence is \SI{14}{\milli\joule\per\square\centi\meter} at \SI{10}{\kilo\hertz}.}
    \label{figSI:ZnL23_static_heating}
\end{figure}

\begin{figure}[!ht]
    \centering
    \includegraphics[width=0.45\linewidth]{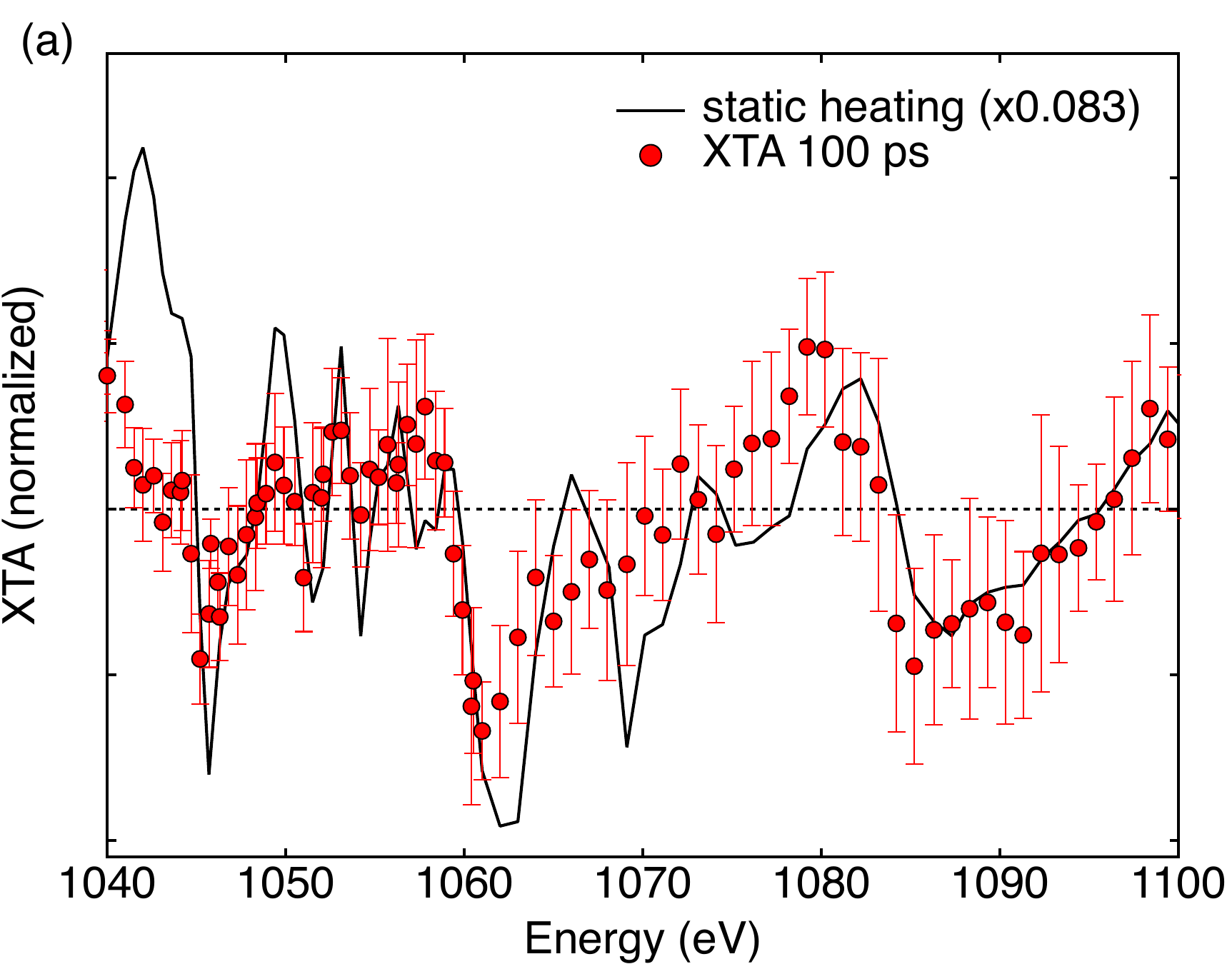}
    \includegraphics[width=0.45\linewidth]{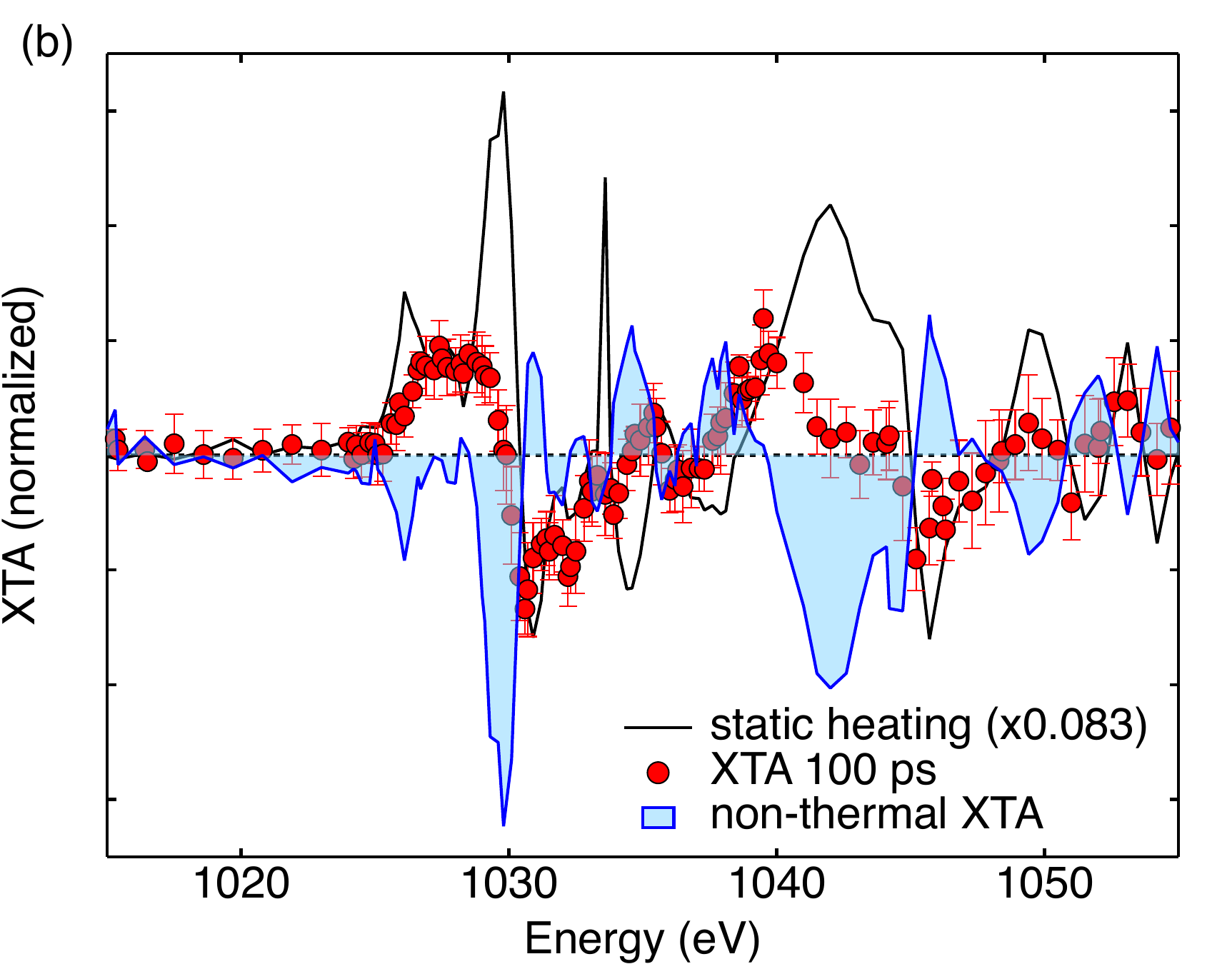}
    \caption{\textbf{Thermal and non-thermal contributions to XTA spectra at the Zn L$_3$-edge.} (a) Best agreement between a scaled XAS difference spectrum corresponding to static heating (black curve, vertical scaling factor of 0.083) and the XTA spectrum at the Zn L$_3$-edge of ZnO at \SI{100}{\pico\second} (red circles with error bars). (b) Difference spectrum between XTA at \SI{100}{\pico\second} (red circles) and the lattice heating contribution (black curve) resulting in a simulation of the non-thermal XTA spectrum (shaded blue curve). Error bars correspond to standard deviations between individual measurements. The laser fluence for the XTA measurement is \SI{14}{\milli\joule\per\centi\metre\squared}.}
    \label{figSI:ZnL23_heating_contribution}
\end{figure}

The subtraction of the lattice heating contribution to the XTA spectrum at the Zn L$_3$-edge is achieved by performing separate measurements of static heating. The effect of static heating on the XAS is measured with the full X-ray intensity by the difference between two XAS spectra measured with and without laser impinging on the sample (black markers in Figure \ref{figSI:ZnL23_static_heating}). The laser fluence is the same as in the pump-probe measurements (\SI{14}{\milli\joule\per\square\centi\meter}). Since lattice heating is an overwhelming contribution to XTA spectra at \SI{100}{\pico\second} far above the absorption edge \cite{Rossi2021}, we make the hypothesis that the XAS difference spectrum obtained from the static heating measurement is a good model of the dynamic heating contribution to the XTA spectrum at \SI{100}{\pico\second}, with only a constant multiplicative factor correction to apply to get an agreement between the two spectra. This approximation is valid when the dynamic structural fluctuations induced by lattice heating (quantified by the Debye-Waller factor) lead to a linear change of the absorption cross-section with the temperature, which occurs when the lattice temperature is larger than the Debye temperature \cite{AlsNielsen:2011vn}. In ZnO, the Debye temperature ($\Theta_D$) is in the range \SI{370}{}--\SI{383}{\kelvin} \cite{Kihara:1985vu, Albertsson:1989gk} and since the excited state temperature is $\gg\SI{500}{\kelvin}$ at \SI{100}{\pico\second} (because of the vacuum environment), the linear scaling of the XTA amplitude above the edge with the lattice temperature change is a valid approximation. A multiplicative factor is then applied to the difference XAS spectrum induced by static heating (black circles in Figure \ref{figSI:ZnL23_static_heating}) to best-match the XTA spectrum between \SI{1045}{} and \SI{1100}{\electronvolt} (red circles in Figure \ref{figSI:ZnL23_heating_contribution}a) by $\chi^2$ minimization. The results are shown in Figure \ref{figSI:ZnL23_heating_contribution}a for a best-matched scaling factor of 0.083 for the XAS difference spectrum coming from the static heating measurement. Following the same procedure as at the Zn K-edge, the obtained model of the lattice heating contribution to the XTA spectrum is then subtracted from the XTA spectrum in the near-edge region of the Zn L$_3$-edge to simulate the non-thermal contribution to the XTA spectrum, shown with a shaded blue curve in Figure \ref{figSI:ZnL23_heating_contribution}b and in Figure 7 in the main text.

\subsection{Chemical shift and broadening simulations of the excited state XAS spectrum at the Zn K-edge\label{secSI:XAS_chemical_broadening}}

\begin{figure}[!ht]
    \centering
    \includegraphics[width=0.49\linewidth]{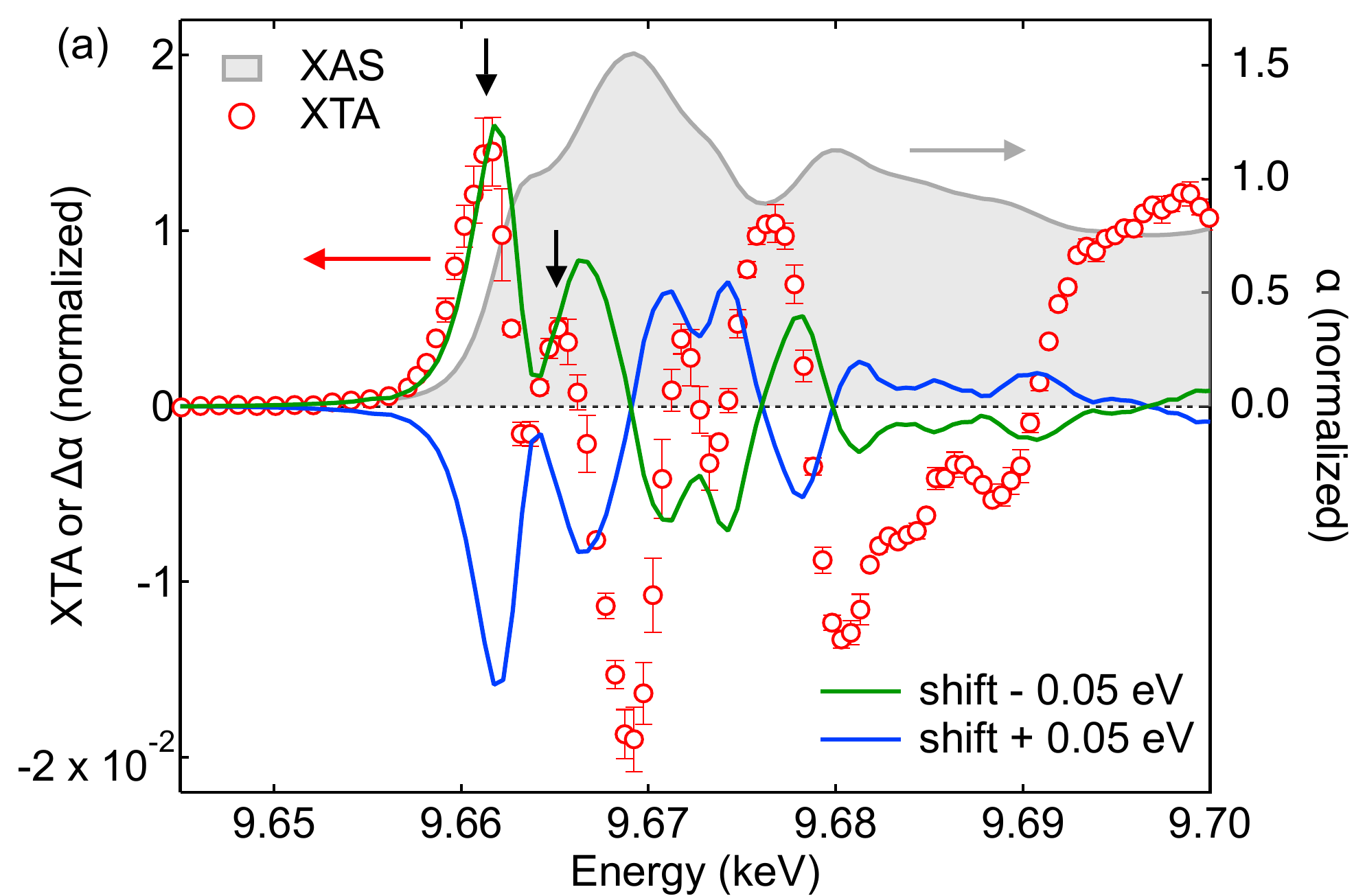}
    \includegraphics[width=0.49\linewidth]{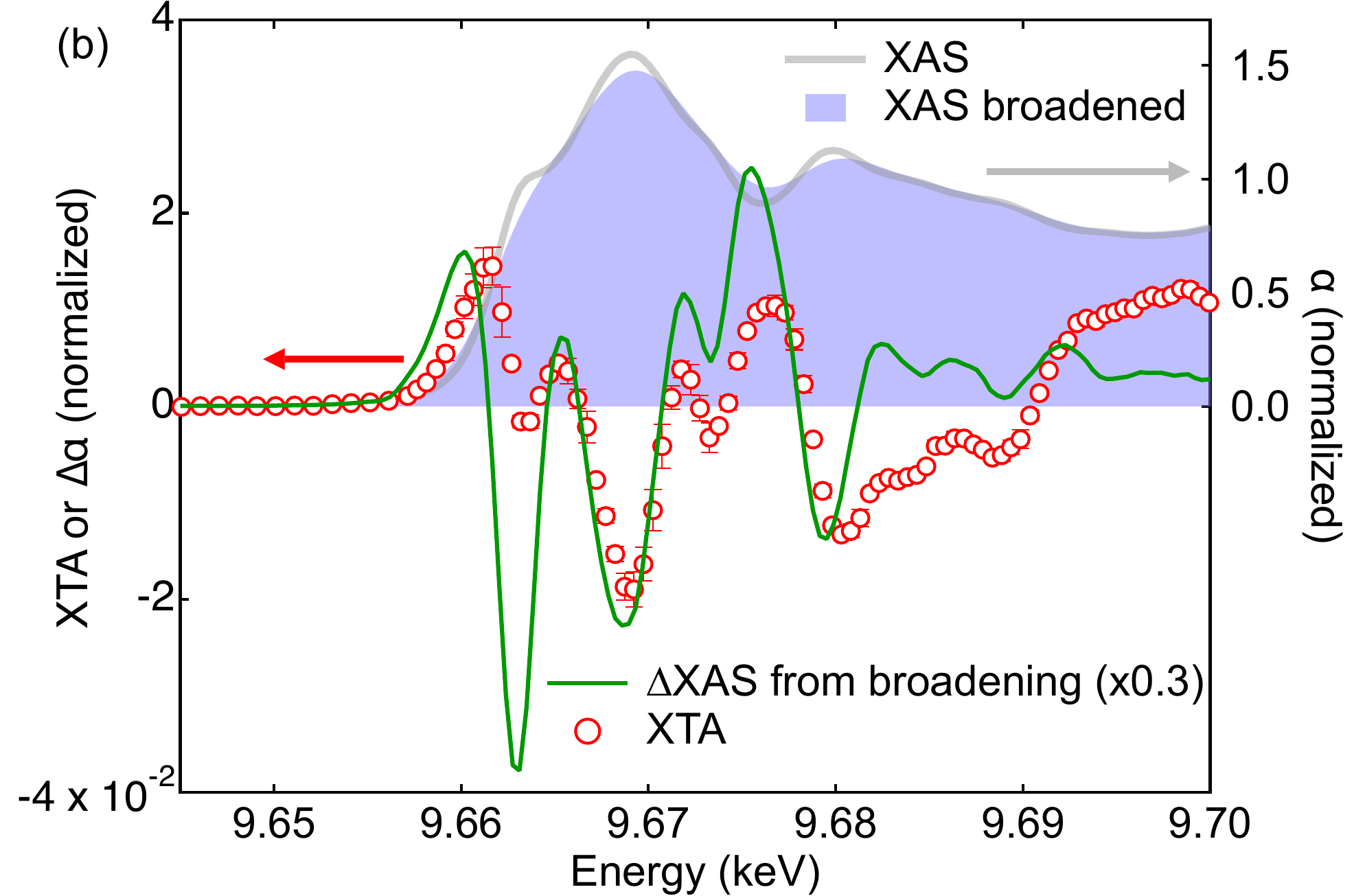}
    \caption{Simulation of the normalized difference XAS spectrum of ZnO at the Zn K-edge upon (a) chemical shift, and (b) spectral broadening. The experimental XTA spectrum (fluence \SI{85.4}{\milli\joule\per\square\centi\metre}, \SI{3.49}{\electronvolt} excitation, \SI{100}{\pico\second} time delay) is shown with red circles with error bars corresponding to standard deviations between individual measurements. In (a), the normalized room temperature XAS spectrum (shaded grey) is energy shifted to simulate XTA spectra upon chemical shifts of \SI{-50}{\milli\electronvolt} (green curve) and +\SI{50}{\milli\electronvolt} (blue curve). Black arrows indicate two positive features of the XTA spectrum at the absorption edge originating from lattice heating. In (b), the normalized room temperature XAS spectrum is shown with a shaded grey curve and the broadened spectrum by a gaussian of \SI{1}{\electronvolt} width (FWHM) is shown with a shaded blue curve. The difference between the broadened and normalized room temperature XAS spectra is shown with a green curve (rescaled by a constant amplitude factor of 0.3).}
    \label{figSI:spectral_shift_broadening_simulation}
\end{figure}

Figure \ref{figSI:spectral_shift_broadening_simulation} displays simulated XTA spectra based on a chemical shift (panel a) or spectral broadening (panel b) of the equilibrium XAS spectrum to model the excited (pumped) state XAS spectrum.

\cleardoublepage


\section{Fluence dependence of the XTA spectra at the Zn K-edge\label{secSI:XTA_fluence_dependence}}

\begin{figure}[!ht]
	\centering
	\includegraphics[width=0.45\linewidth]{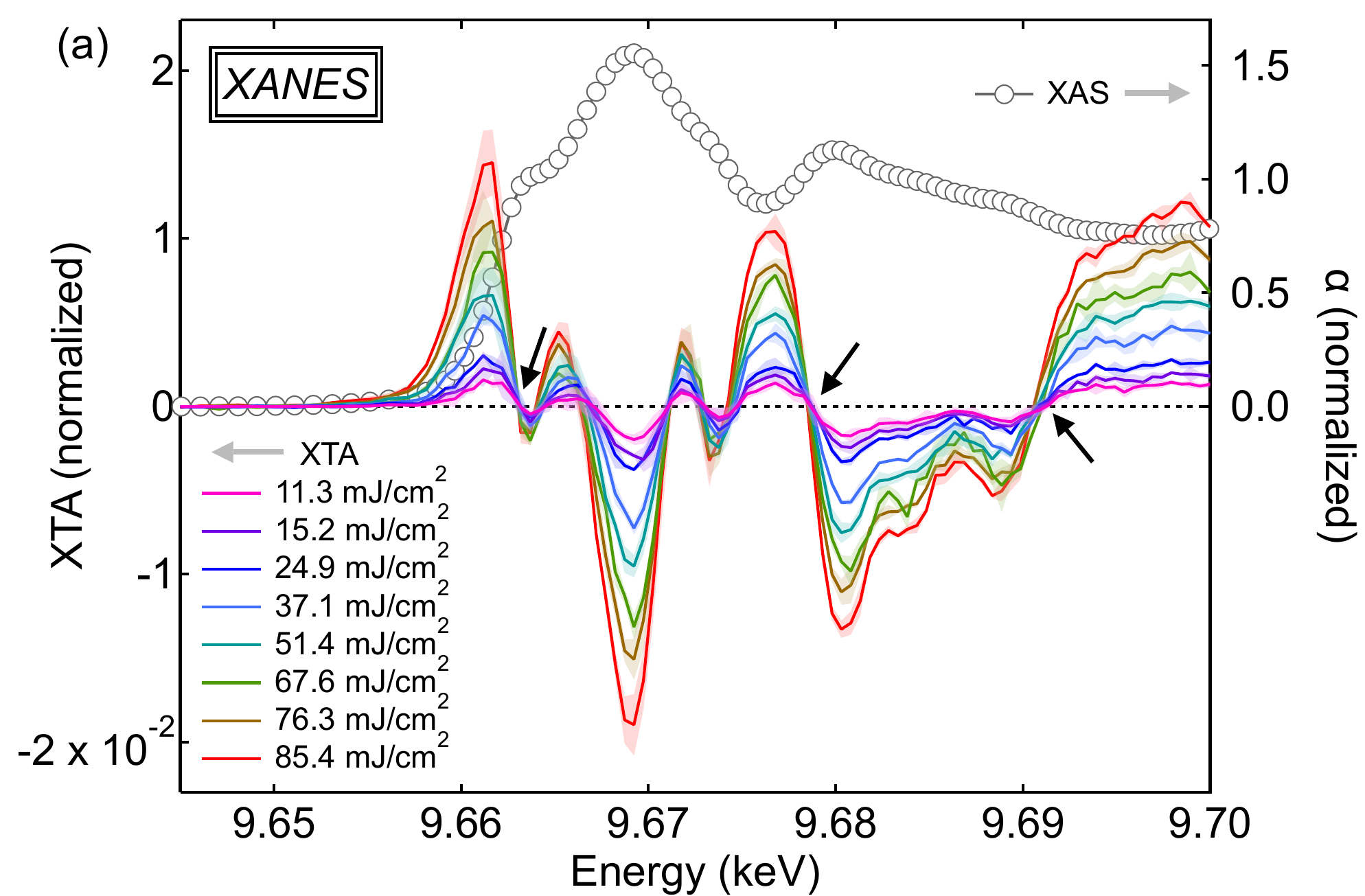}
	\includegraphics[width=0.45\linewidth]{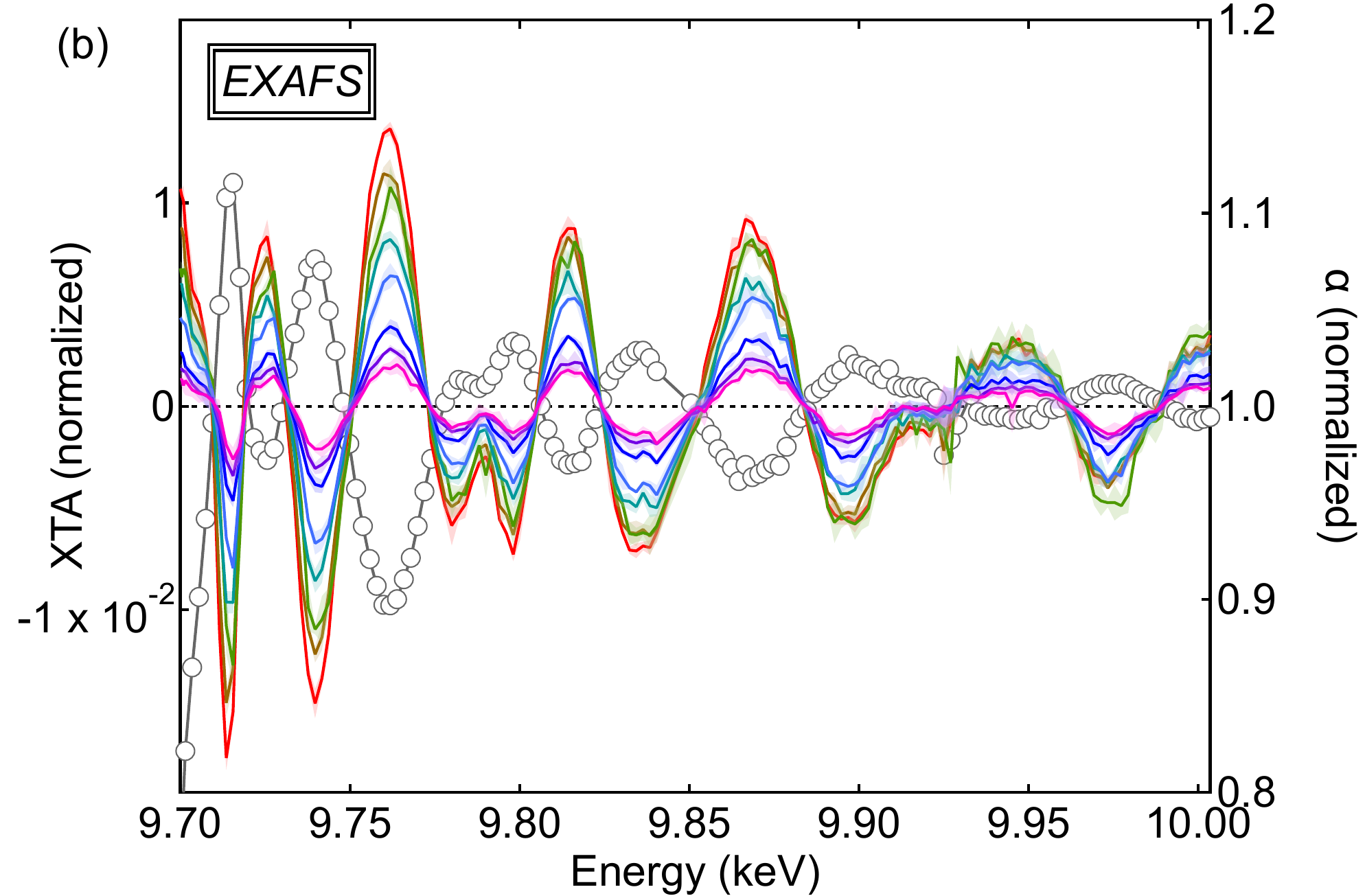}
    \vspace{-3mm}
    \caption{Edge-jump normalized XTA spectra at various excitation fluences at the Zn K-edge of ZnO (0001) thin films in (a) the XANES and (b) the EXAFS regions (colored curves, left axis) for a time delay of \SI{100}{\pico\second}, an excitation energy of \SI{3.49}{\electronvolt},  and an incidence angle of \SI{45}{\degree} with respect to (0001). Shaded areas represent standard deviations between individual measurements, black arrows indicate isosbestic points in the XANES. The edge-jump normalized XAS spectrum at the Zn K-edge (absorption coefficient, $\alpha$) is shown with gray circles for reference (right axis).}
    \label{figSI:XTA_100ps_spectral_traces}
\end{figure}

\begin{figure}[!ht]
    \centering
    \includegraphics[width=0.44\linewidth]{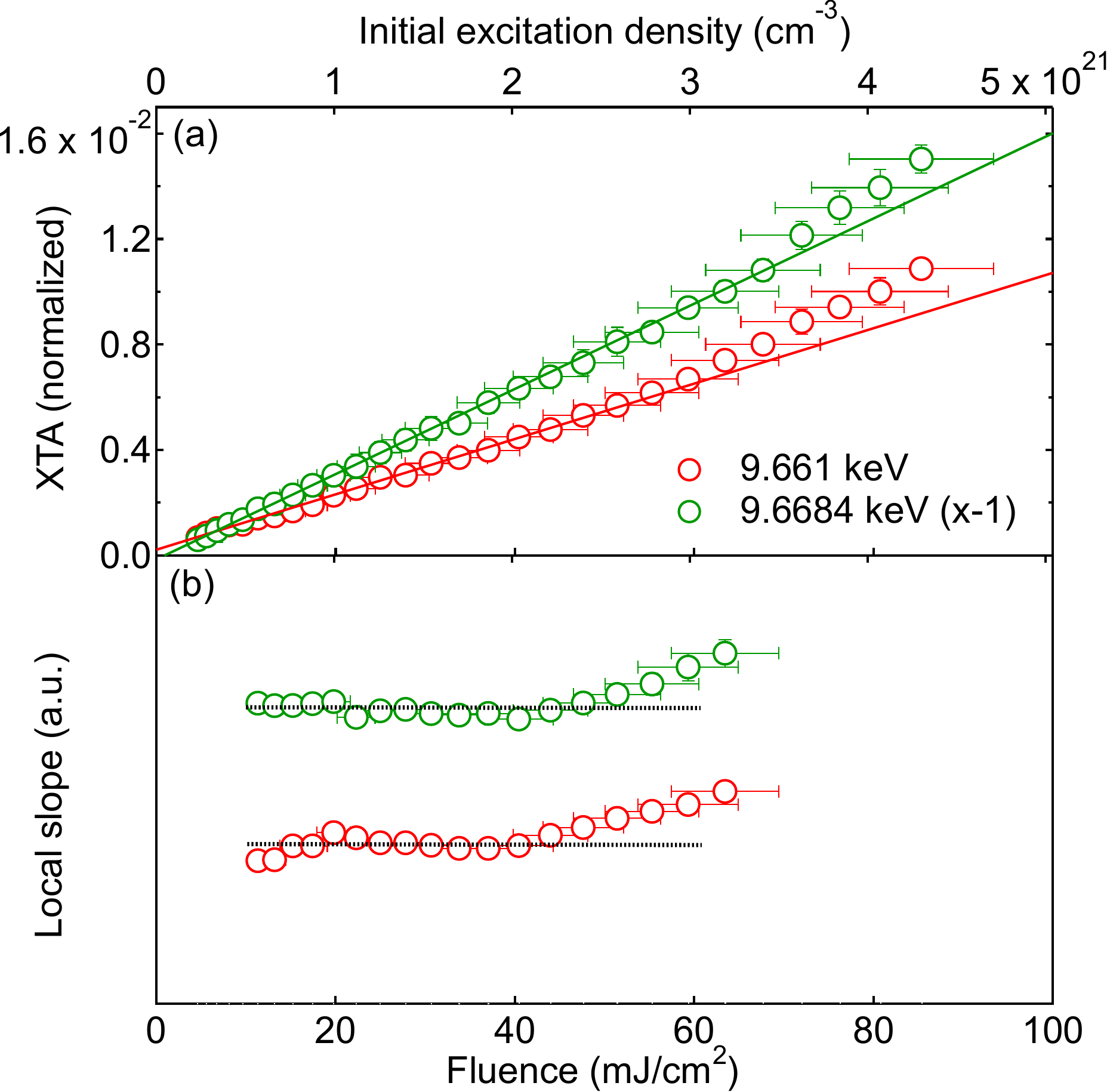}
    \vspace{-3mm}
    \caption{Evolution of (a) the XTA amplitude and (b) the local slope (derivative) of the XTA amplitude with the excitation fluence at \SI{100}{\pico\second} at \SI{9.661}{\kilo\electronvolt} (red circles) and \SI{9.6684}{\kilo\electronvolt} (green circles). Vertical error bars represent standard deviations between individual measurements, horizontal error bars are uncertainties in the calculation of the laser fluence. Linear fits in (a) include data points at fluences  $<\SI{50}{\milli\joule\per\square\centi\metre}$ and are weighted by the standard deviations. The local slope in (b) is based on a linear fit of 11 consecutive data points in panel (a) around a mean fluence. Horizontal black dashed lines are guides to the eye.}
    \label{figSI:XTA_amplitude_fluence_dependence}
\end{figure}

Figure \ref{figSI:XTA_100ps_spectral_traces} shows XTA spectra measured at increasing excitation fluences focusing on the XANES (panel a) and the EXAFS (panel b). Figure \ref{figSI:XTA_amplitude_fluence_dependence}a shows the evolution of the XTA amplitude at \SI{9.661}{\kilo\electronvolt} (red circles) and \SI{9.6684}{\kilo\electronvolt} (green circles). Figure \ref{figSI:XTA_amplitude_fluence_dependence}b shows the evolution of the local slope (derivative) of the evolution of XTA amplitude with the excitation fluence in panel a.

\vspace{-5mm}

\cleardoublepage


\section{Calculation of the excitation density\label{secSI:calculation_excitation_density}}


\subsection{Initial excitation density}

The procedure for the calculation of the initial excitation density has been discussed extensively in the Supporting Information of reference \cite{Rossi2021}. We neglect the birefringence of ZnO which leads to variations of the permittivity between \SI{7}{\percent} and \SI{11}{\percent} for the real and imaginary parts between the extraordinary and ordinary polarizations at the pump wavelength, respectively \cite{Shokhovets:2010by}. Hence, the relative permittivity of ZnO is obtained from spectroscopic ellipsometry measurements (see section \ref{subsec:spectroscopic_ellipsometry}), which yield $\epsilon_1=4.72$ and $\epsilon_2=1.76$ at the pump wavelength of \SI{355}{\nano\meter} (\SI{3.49}{\electronvolt}) and at room temperature. The permittivity converts into a complex refractive index with the real part $n=2.21$ and the imaginary part $\kappa=0.40$. The reflectance for the $p$-polarized pump at \SI{45}{\degree} incidence angle is \SI{0.2}{} and the absorption coefficient is $\alpha=\SI{1.41e7}{\per\metre}$ corresponding to a penetration depth of \SI{71}{\nano\metre}. In these conditions, the laser pulse energy is fully absorbed by the sample. We have shown earlier how the excitation density gets homogeneous in the probe volume at \SI{100}{\pico\second} \cite{Rossi2021}. Finally, the initial excitation density $n_0$ delivered by the pump pulse is calculated with the relation,
\begin{equation}
n_0=F\frac{(1-R)}{h\nu t}
\label{eq:excitation_density_initial}
\end{equation}
where $R$ is the reflectivity, $F$ is the fluence (corrected for the incidence angle), $h\nu$ is the pump photon energy (\SI{3.49}{\electronvolt}) and $t$ is the sample thickness (obtained by spectroscopic ellipsometry, see section \ref{subsec:spectroscopic_ellipsometry}). Equation \ref{eq:excitation_density_initial} accounts for reflectance losses as well as the film thickness at an oblique incidence angle. The conversion between the incident fluence and the initial excitation density is given in Table \ref{tabS:conversion_fluence_excitation_density}.

\begin{table}[!ht]
	\centering
	\begin{tabular}{ccc}
		\toprule
		$F$ (\SI{}{\milli\joule\per\square\centi\metre}) & $n_0$ (\SI{}{\per\cubic\centi\metre}) & $n(\SI{100}{\pico\second})$ (\SI{}{\per\cubic\centi\metre}) \\
		\midrule
		11.3 & $5.7\times10^{20}$ & $2.4\times10^{19}$ \\
		15.2 & $7.7\times10^{20}$ & $3.2\times10^{19}$ \\
		24.9 & $1.3\times10^{21}$ & $5.3\times10^{19}$ \\
		37.1 & $1.9\times10^{21}$ & $7.9\times10^{19}$ \\
		51.4 & $2.6\times10^{21}$ & $1.1\times10^{20}$ \\
		67.6 & $3.4\times10^{21}$ & $1.4\times10^{20}$ \\
		76.3 & $3.9\times10^{21}$ & $1.6\times10^{20}$ \\
		85.4 & $4.3\times10^{21}$ & $1.8\times10^{20}$ \\
		\bottomrule
	\end{tabular}
	\caption{\textbf{Fluence and excitation densities.} Equivalence between the incident laser fluence ($F$), the initial excitation density from the absorbed pump pulse energy ($n_0$), and the excitation density at \SI{100}{\pico\second} ($n(\SI{100}{\pico\second})$).}
	\label{tabS:conversion_fluence_excitation_density}
\end{table}

\subsection{Decay of the excitation density at \SI{100}{\pico\second}}

\begin{figure}
    \centering
    \includegraphics[width=0.48\linewidth]{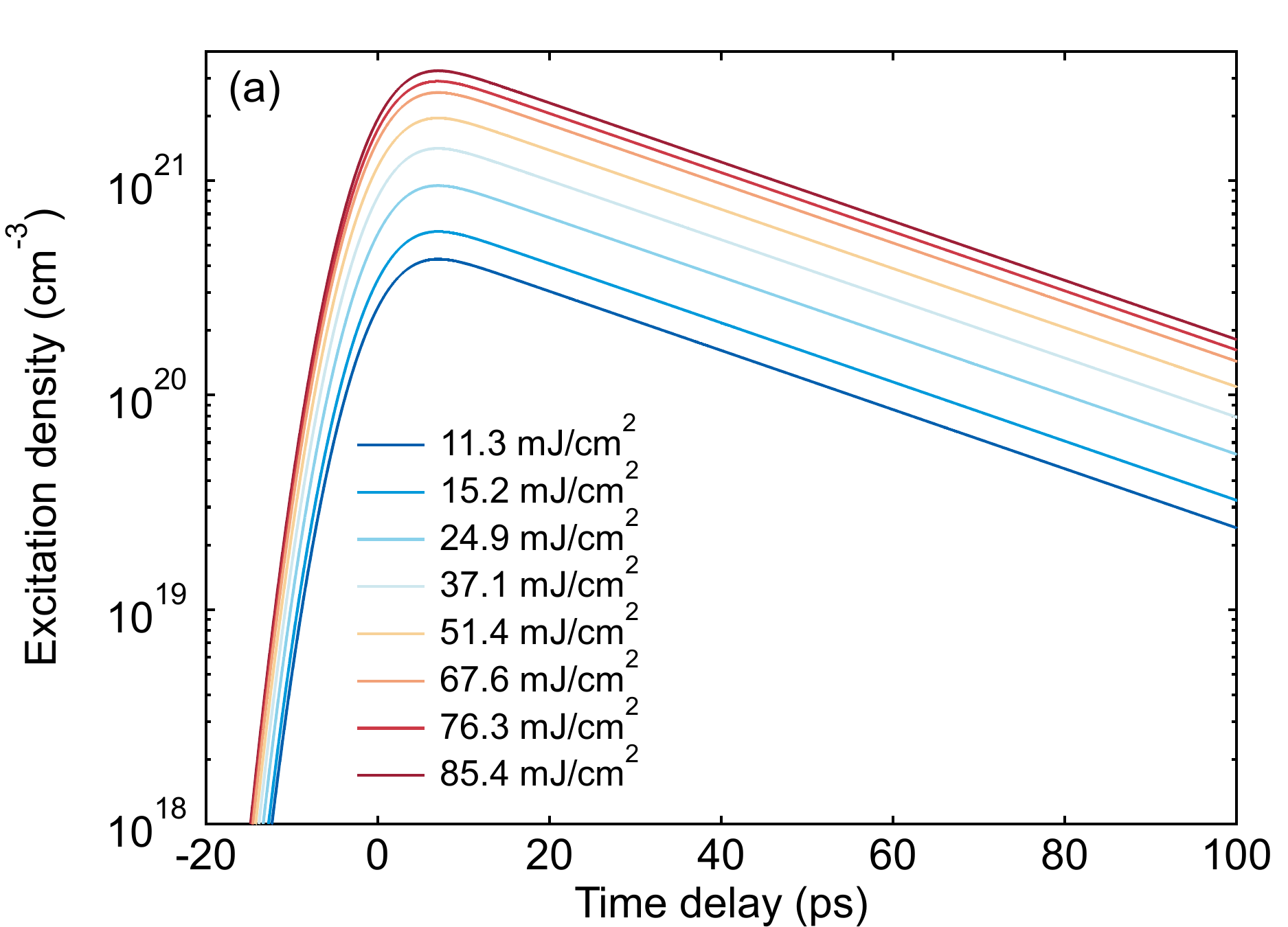}
    \includegraphics[width=0.48\linewidth]{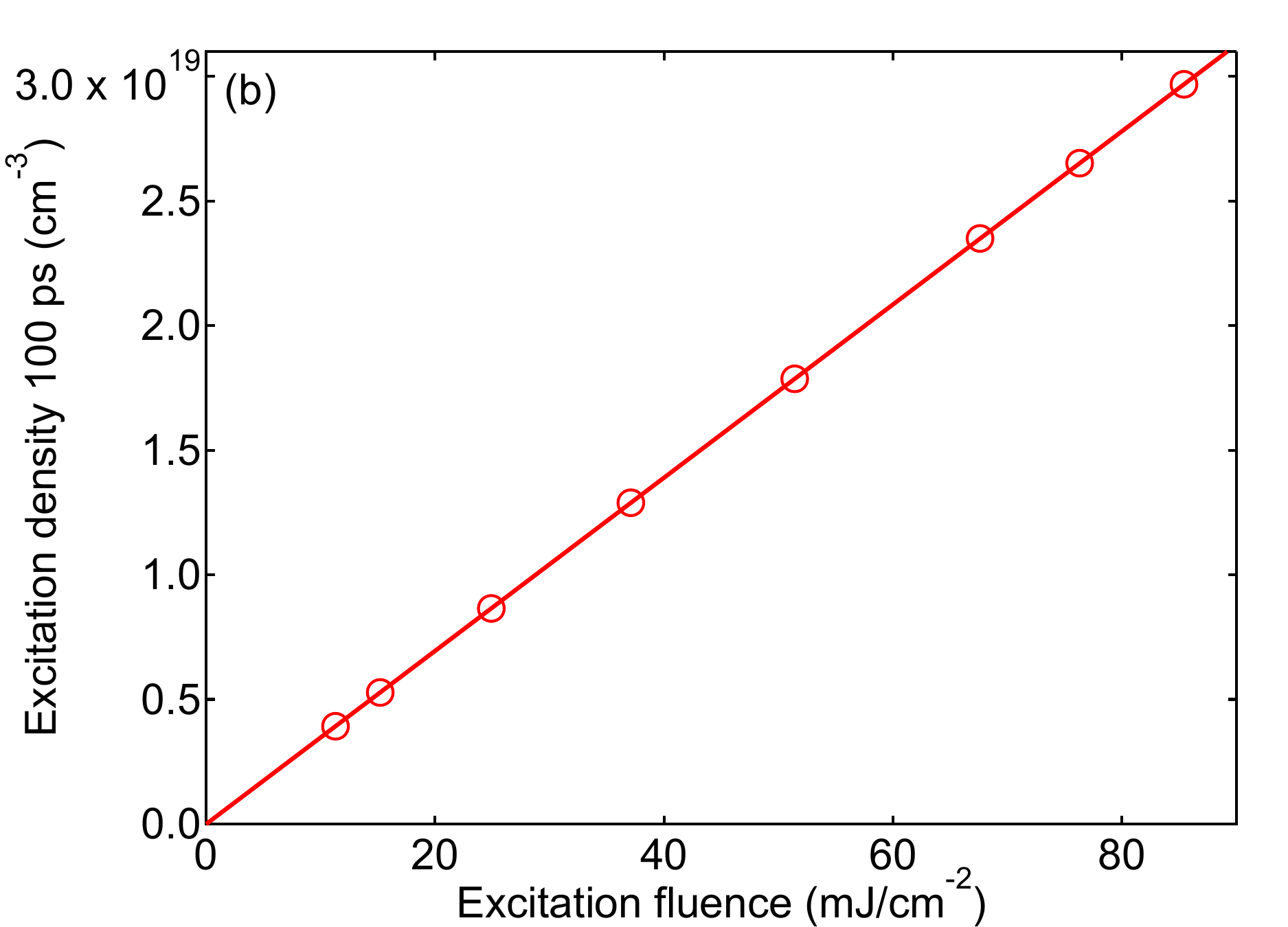}
    \caption{\textbf{Time evolution of the excitation density.} (a) Numerical evaluation of the excitation density with time. (b) Excitation density at \SI{100}{\pico\second} (circles) with linear fit.}
    \label{figSI:evolution_excitation_density}
\end{figure}

\begin{figure}[!ht]
	\centering
	\includegraphics[width=0.5\linewidth]{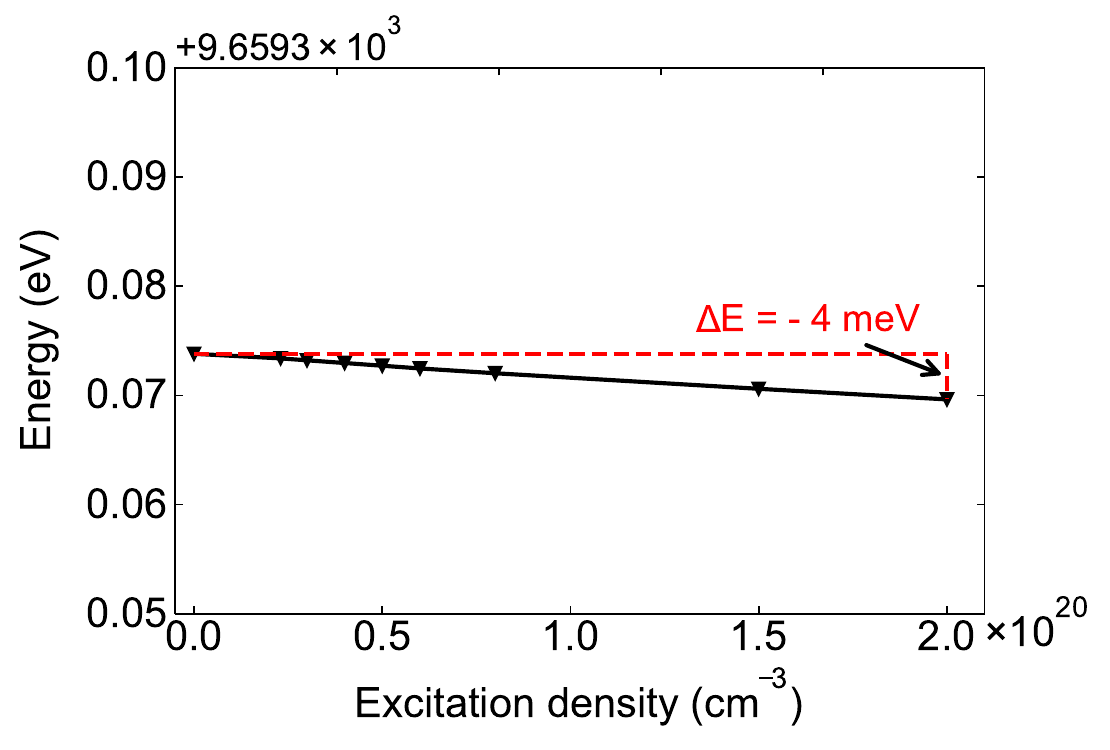}
	\caption{Change of the Kohn-Sham gap between the Zn 1s orbital and CBM+1 state at the $\Gamma$-point with increasing excitation density.}
	\label{figS:IP_renormalization}
\end{figure}

\begin{figure}[H]
        {\includegraphics[width=0.5\linewidth]{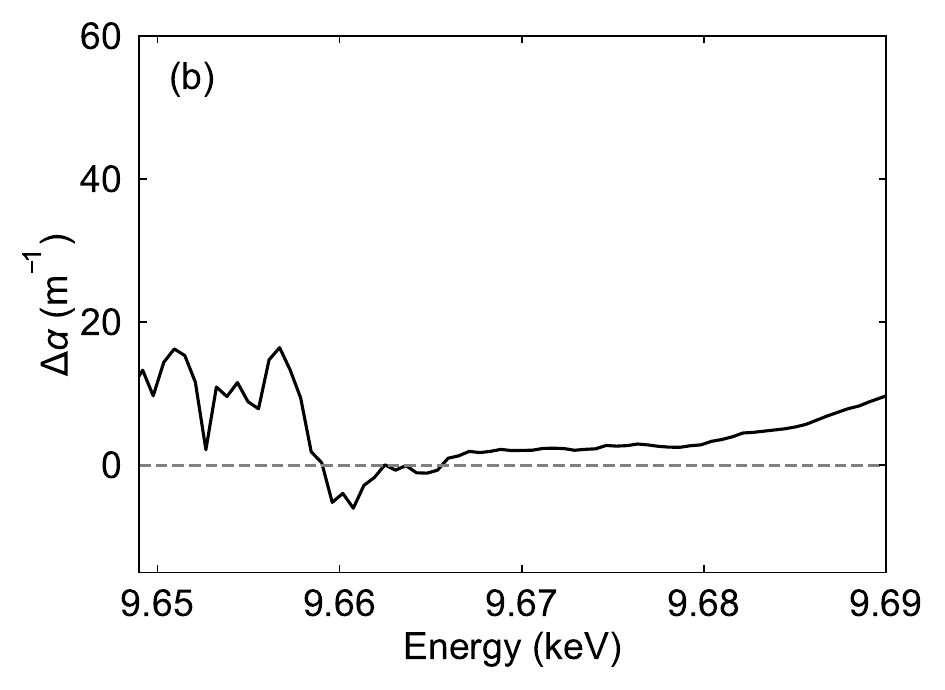}}
        {\includegraphics[width=0.5\linewidth]{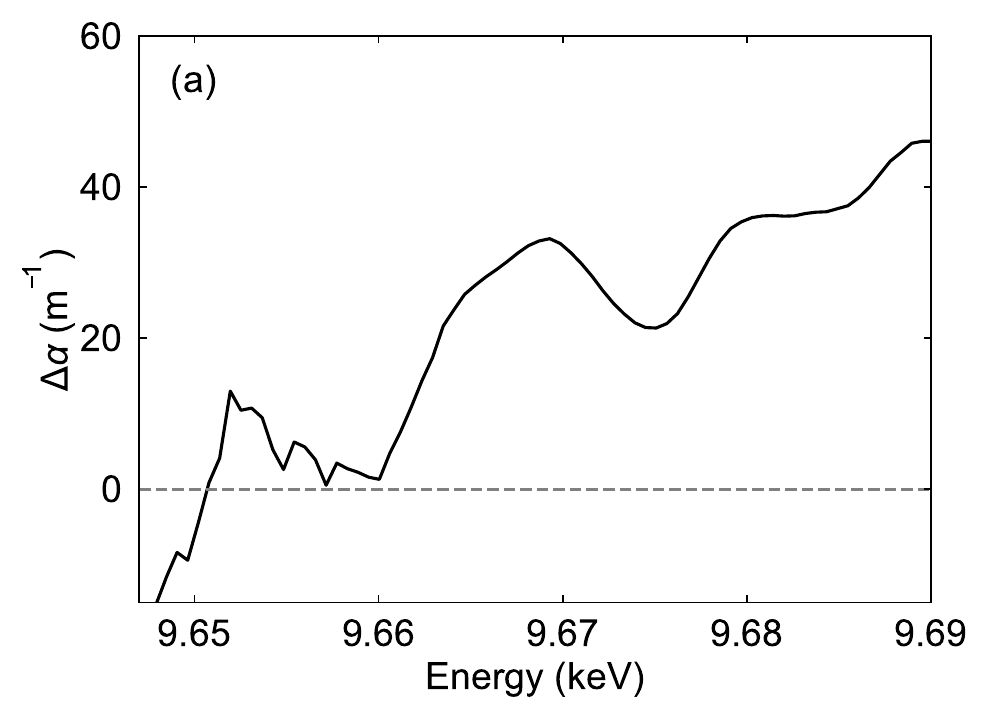}}    \caption{(a) XTA computed in the IPA. (b) Change in the XTA coefficient by adding electron-hole exchange to the IPA at an excitation density or of \SI{5.0e19}{\per\cubic\centi\metre}.}
    \label{figS:BSE_IP}
\end{figure}

\begin{figure}[!ht]
    \centering
    \includegraphics[width=0.5\linewidth]{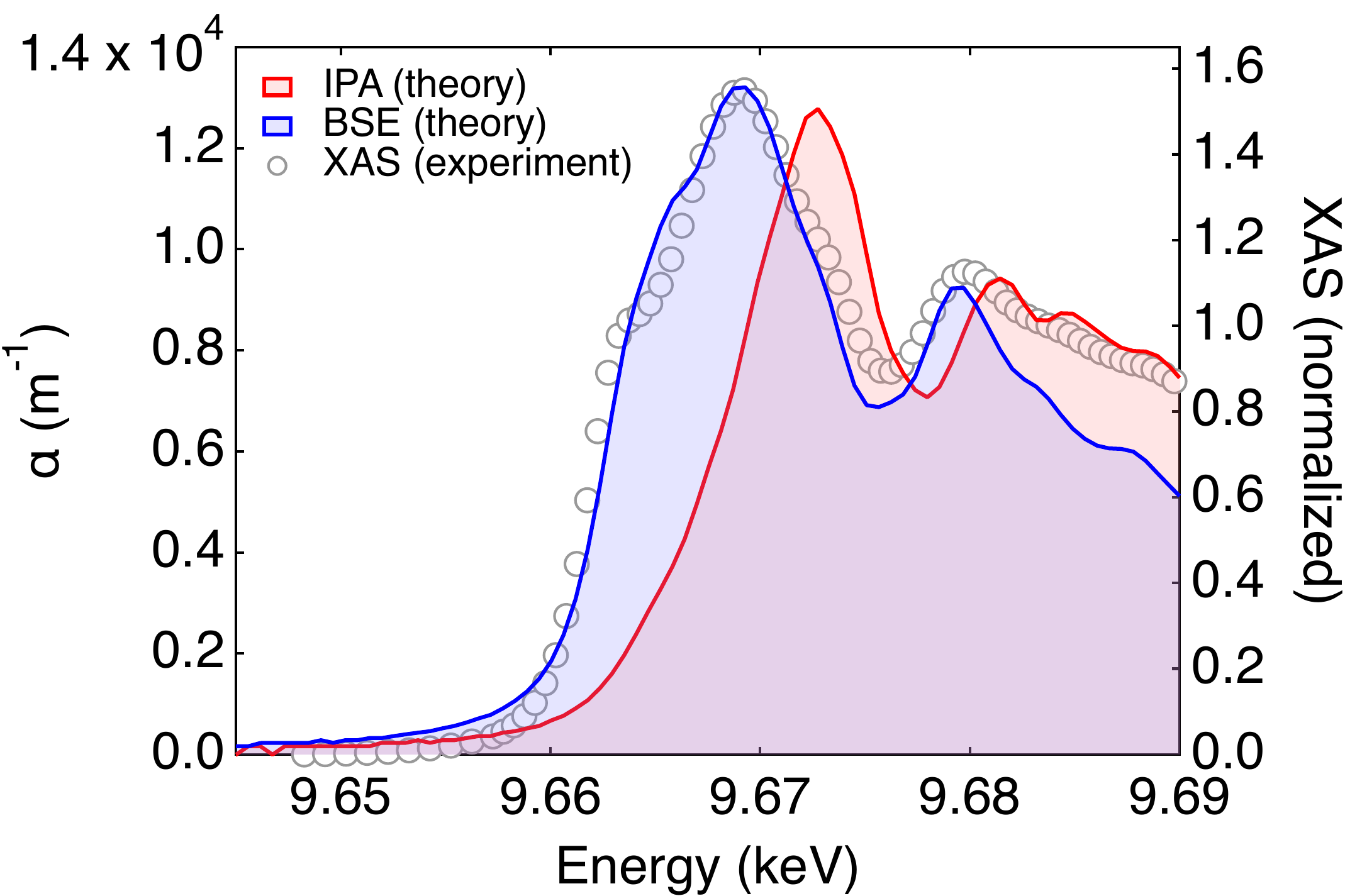}
    \caption{Comparison between calculated XAS spectra within the IPA (shaded red curve, left axis) and from the BSE (shaded blue curve, left axis). The normalized equilibrium XAS spectrum at the Zn K-edge in the experiment is shown for reference (gray circles, right axis).}
    \label{figSI:XAS_BSE_VS_IPA}
\end{figure}

\begin{figure}[!ht]
	\centering
    \includegraphics[width=0.28\linewidth]{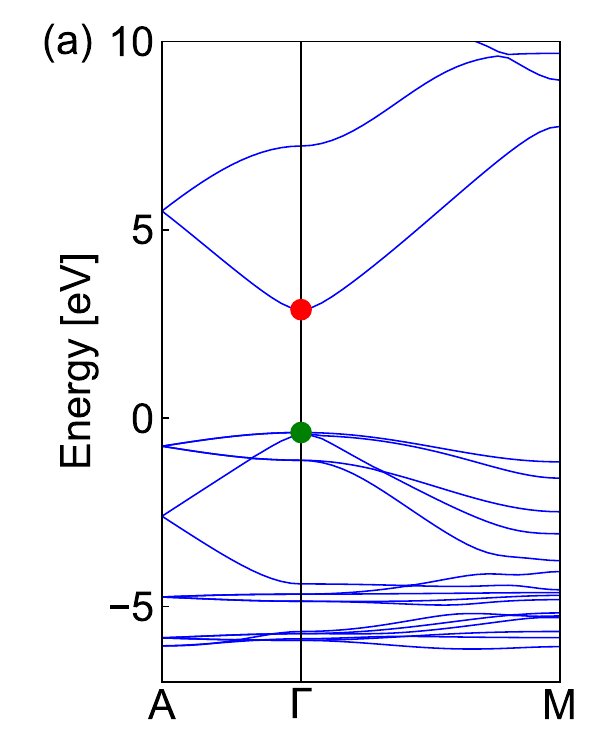}
    \includegraphics[width=0.28\linewidth]{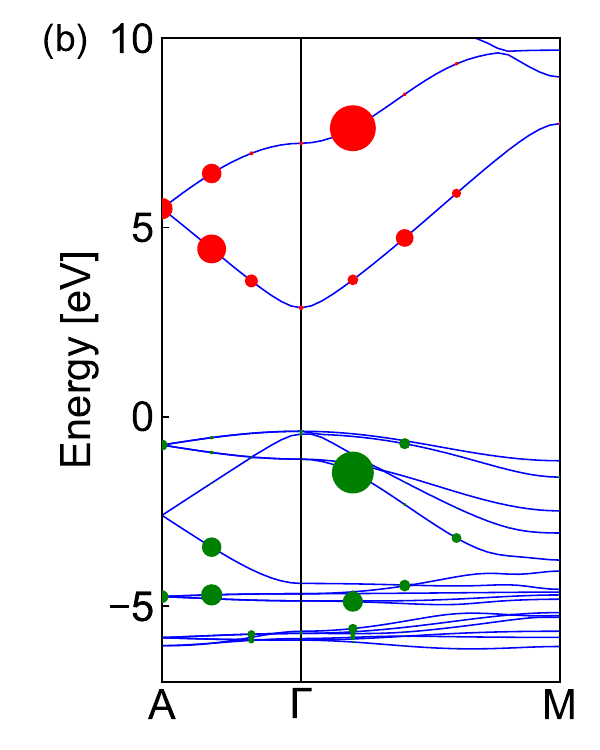}
	\caption{Distribution of excited electrons (red) and holes (green) in the BZ at time delays of (a) \SI{100}{\pico\second} and (b) \SI{20}{\femto\second}. Panel (a) shows the distribution obtained from constrained DFT with an excitation density of $\SI{2.3e19}{\per\cubic\centi\metre}$, while panel (b) is a distribution obtained by RT-TDDFT with a pump fluence of \SI{74.3}{\milli\joule\per\square\centi\metre}. The area of the circles is proportional to the occupation numbers at the given \textbf{k}-points.}
	\label{figS:ps_fs_occupation}
\end{figure}

\begin{figure}[H]
    \centering
    \includegraphics[width=0.5\linewidth]{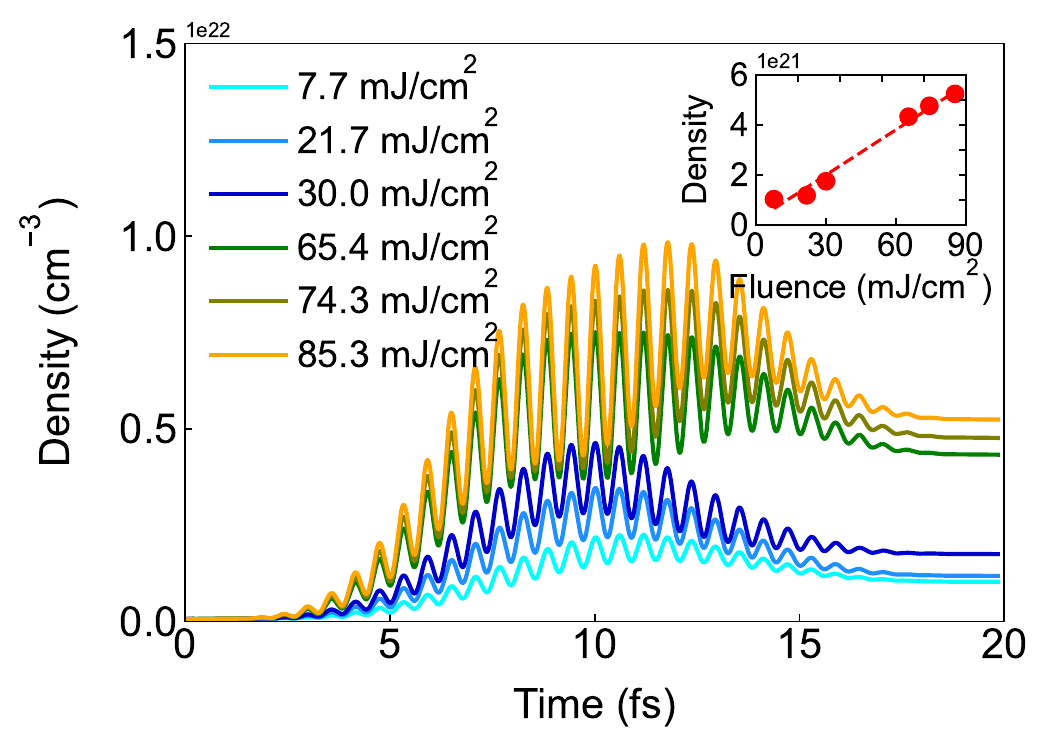}
    \caption{Evolution of the excitation density following the pump excitation at different fluences (\SI{10}{\femto\second} pulse duration FWHM, \SI{3.49}{\electronvolt} photon energy, polarization along the (100) crystal axis). Inset: Excitation density at \SI{20}{\femto\second} with increasing pump fluence.}
    \label{figSI:density_vs_time}
\end{figure}

\begin{figure}[H]
    \centering
    \includegraphics[width=0.5\linewidth]{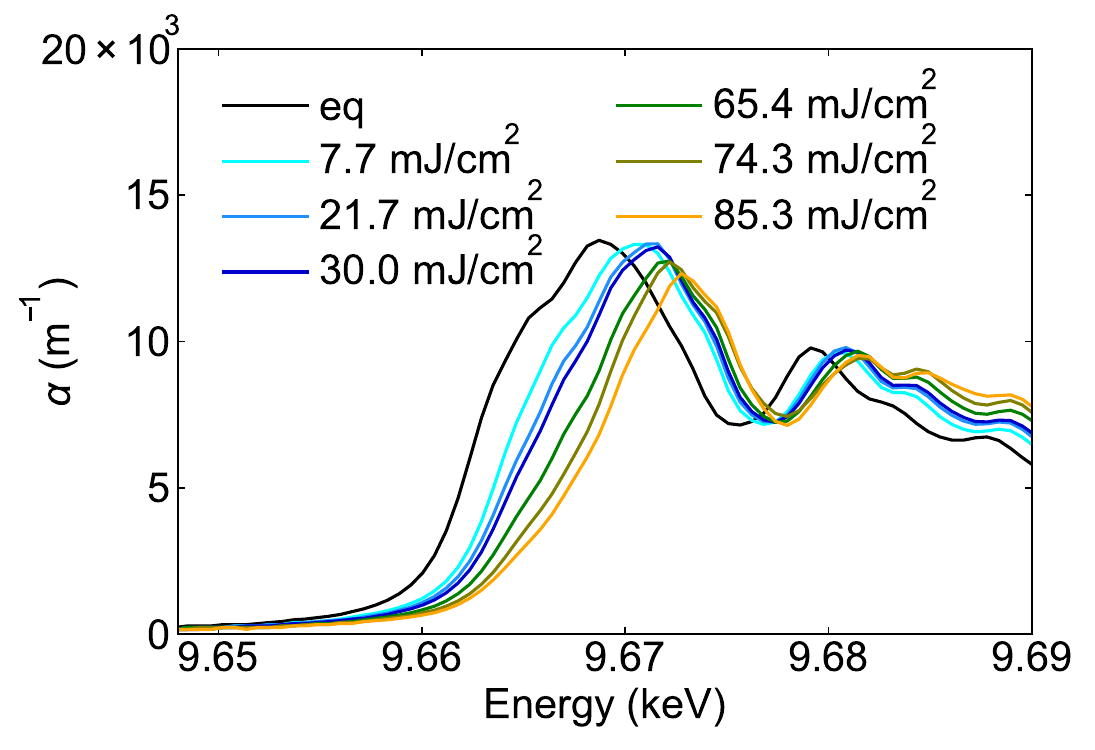}
    \caption{XAS spectra on the femtosecond timescale. Computed XAS spectra (colored curves) at the Zn K-edge for different pump fluences at \SI{20}{\femto\second}. The equilibrium XAS spectrum (black curve) is shown for reference.}
    \label{figSI:XAS_fs}
\end{figure}

\begin{figure}[ht!]
    \centering
    \includegraphics[width=0.5\linewidth]{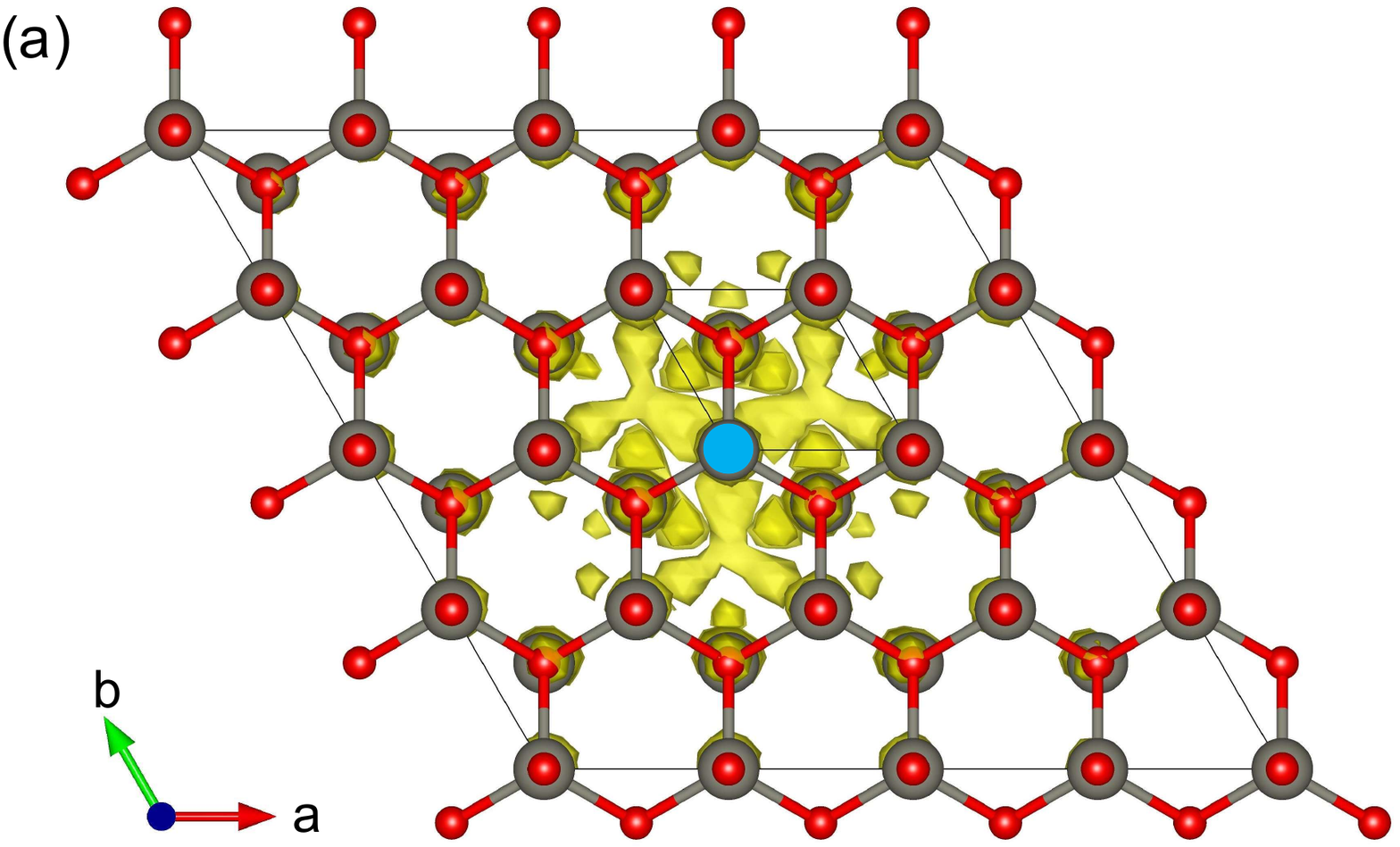}\\ \includegraphics[width=0.5\linewidth]{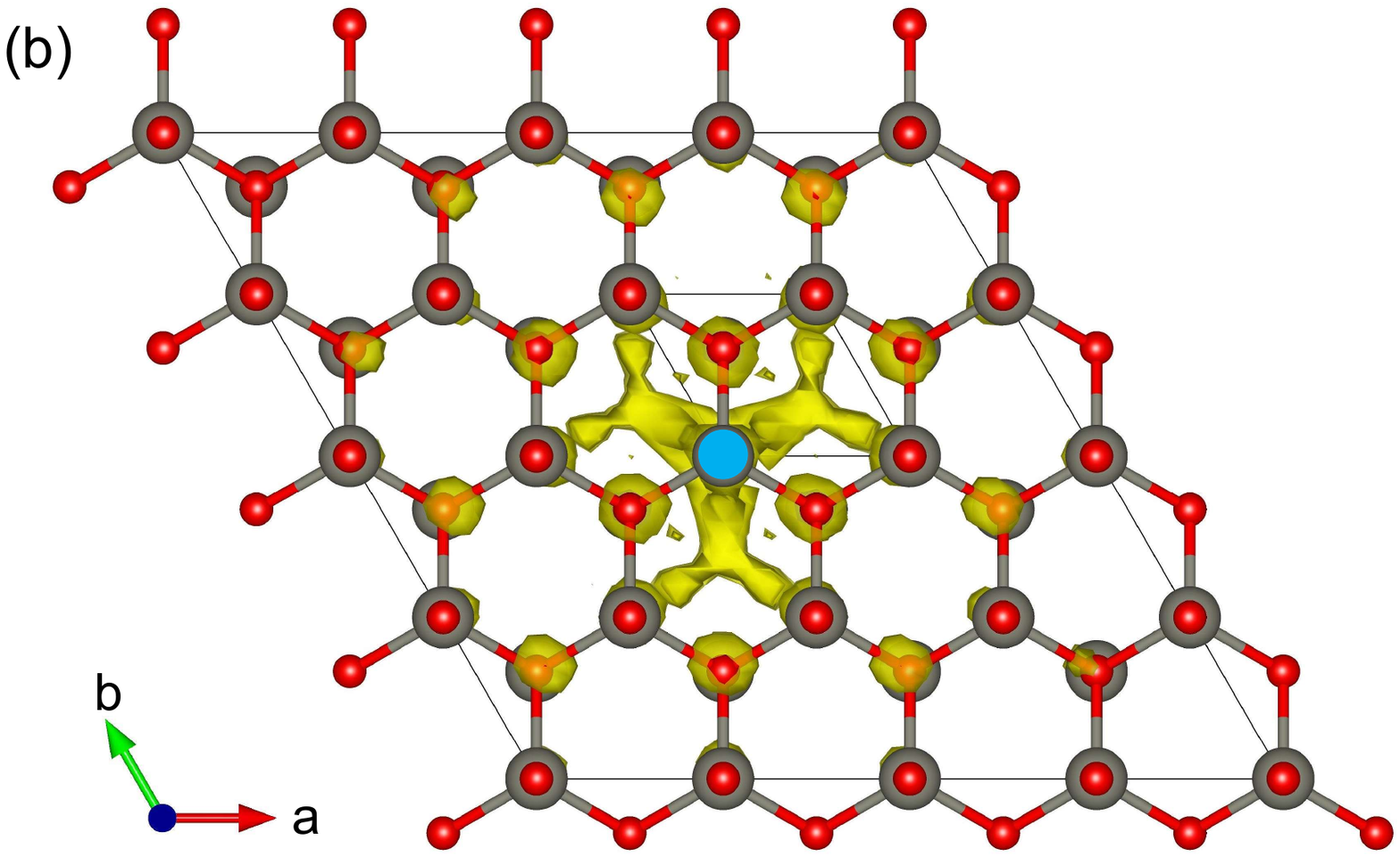}
    \caption{Electron distribution of the lowest-energy core exciton at equilibrium at the (a) K-edge, and (b) L$_3$-edge. The core-hole positions are marked by the blue circles. The zinc and oxygen atoms are gray and red, respectively. The solid gray line indicates the unit cell. Representations generated with \texttt{VESTA} \cite{Momma2011}.} 
    \label{figSI:electron_distribution}
\end{figure}

The excitation density over time is calculated numerically in order to account for the similar timescales of excitation density increase upon absorption of the pump pulse, and decrease due to electron-hole population decay. The time evolution of the excitation density is obtained from a finite difference method evaluation of the differential equation:
\begin{equation}
\frac{dn}{dt}=-kn+\dot{S}
\end{equation}
where $k$ is the effective rate of free carrier decay and $\dot{S}$ the source term derived from a gaussian source $S(t)=n_0e^{-t^2/(2\sigma^2)}$ with $n_0$ the peak excitation density of the pump pulse (such that $n_{exc}=\int_{-\infty}^\infty n_0e^{-t^2/(2\sigma^2)}dt$ with $n_{exc}$ the total excitation density delivered by the pump pulse and calculated from the absorbed fluence), and $\sigma$ the duration of the pump pulse related to the full width at half maximum (FWHM, \SI{10}{\pico\second}) with $\sigma=\mathrm{FWHM}/(2\sqrt{2\ln2})$. The effective decay rate is taken from previous time-resolved photoluminescence measurements at the band gap using a similar pulse duration and photon energy, which shows that the overwhelming contribution to the decay is an exponential decay with time constant in the range $\SI{15}{}-\SI{40}{\pico\second}$ \cite{Guo:2003hr, Lee:2011em, Chernikov2011}, which is set to \SI{31}{\pico\second} in this work corresponding to the average decay time of the free exciton emission at room temperature in reference \cite{Lee:2011em}. The results of the simulation are displayed in Figure \ref{figSI:evolution_excitation_density}a. The evolution of the excitation density at \SI{100}{\pico\second} displays a linear evolution with the excitation fluence (Figure \ref{figSI:evolution_excitation_density}b). The values of the excitation density at \SI{100}{\pico\second} ($n_0(\SI{100}{\pico\second})$) are reported in Table \ref{tabS:conversion_fluence_excitation_density}.

\cleardoublepage

\section{Theoretical Background}

\subsection{Constrained density functional theory}
\label{secSI:theory_background}

\noindent \hspace{1em} The Kohn-Sham (KS) equations provide a practical method for determining the ground state density of a many-electron system in the external potential of the atomic nuclei. Every KS state $\psi_{i\mathbf{k}}(\mathbf{r})$ is an eigenfunction of the KS Hamiltonian with the corresponding eigenenergy $\epsilon_{i\mathbf{k}}$:
\begin{equation}
\left[-\frac{\nabla^2}{2}+v_{\text{KS}}(\mathbf{r})\right]\psi_{i\mathbf{k}}(\mathbf{r})=\epsilon_{i\mathbf{k}}\psi_{i\mathbf{k}}(\mathbf{r}),
\end{equation}
where $\mathbf{k}$ denotes the wavevector, and $v_{\text{KS}}(\mathbf{r})$, the KS potential. The ground state electron density $n_{0}(\mathbf{r})$ is expressed in terms of $\psi_{i\mathbf{k}}(\mathbf{r})$ as:
\begin{equation}
n_{0}(\mathbf{r})=\sum_\mathbf{k}w_\mathbf{k}\sum_if_{i\mathbf{k}}^0|\psi_{i\mathbf{k}}(\mathbf{r})|^2,
\end{equation}
where $w_\mathbf{k}$ stands for the weight of $\mathbf{k}$, and $f_{i\mathbf{k}}^0$ is the ground state occupation of the KS state labeled with $i\mathbf{k}$. 

Excited states can be modeled by considering that electrons have been promoted to empty states, leaving holes behind in the previously occupied states. In this approach, the resulting electron density $n(\mathbf{r})$ is given by:
\begin{equation}
n(\mathbf{r})=n_0(\mathbf{r})+n_{e}(\mathbf{r})-n_{h}(\mathbf{r}),
\label{eq: excited_e_density}
\end{equation}
where $n_{e}(\mathbf{r})$ and $n_{h}(\mathbf{r})$ are the density of excited electrons and holes, respectively.

In cDFT as implemented in \exciting, we express $n(\mathbf{r})$ as:
\begin{equation}
n(\mathbf{r})=\sum_\mathbf{k}w_\mathbf{k}\sum_if_{i\mathbf{k}}^0|\psi_{i\mathbf{k}}(\mathbf{r})|^2+\sum_{c\mathbf{k}}f_{c\mathbf{k}}|\psi_{c\mathbf{k}}(\mathbf{r})|^2-\sum_{v\mathbf{k}}f_{v\mathbf{k}}|\psi_{v\mathbf{k}}(\mathbf{r})|^2,
\end{equation}
where $v$ and $c$ represent the indices of valence and conduction states, respectively. The distribution of excited electrons ($f_{c\mathbf{k}}$) and holes ($f_{v\mathbf{k}}$) is constrained based on some physical insight that fulfills the following conditions:
\begin{equation}
\sum_{c}f_{c\mathbf{k}} = \sum_{v}f_{v\mathbf{k}},
\end{equation}
\begin{equation}
\sum_{c\mathbf{k}}f_{c\mathbf{k}} = N_\mathrm{exc},
\end{equation}
where $N_\mathrm{exc}$ represents the number of excitations in a unit cell.

\subsection{Real-time time-dependent density functional theory}

\noindent \hspace{1em} Real-time time-dependent density functional theory (RT-TDDFT) can be employed to investigate the time evolution of electrons submitted to a time-dependent perturbation, such as electric or magnetic fields. In RT-TDDFT, a KS state $\psi_{j\mathbf{k}}(\mathbf{r},t)$ evolves under the equation
\begin{equation}
    \frac\partial{\partial t}\psi_{j\mathbf{k}}(\mathbf{r},t) = -\mathrm{i}\hat{H}(\mathbf{r},t)\psi_{j\mathbf{k}}(\mathbf{r},t),
\end{equation}
where $\hat{H}(\mathbf{r},t)$ is the time-dependent KS Hamiltonian. Considering that a laser pulse, with an electric field described by the vector potential $\mathbf{A}(t)$, excites the electron system, the corresponding KS Hamiltonian can be expressed as \cite{Yabana2012:19771}:
\begin{equation}
    \hat{H}(\mathbf{r},t)=\frac12\left(-\mathrm{i}\nabla+\frac1c\mathbf{A}(t)\right)^2+v_{\text{KS}}(\mathbf{r},t), 
\end{equation}
\noindent where $v_{\text{KS}}(\mathbf{r},t)$ is the time-dependent KS potential.

Tracking the creation of electron-hole pairs over time is crucial to understand the excitation dynamics. In RT-TDDFT, this is determined by projecting $\psi_{i\mathbf{k}}(\mathbf{r},t)$ onto the KS states at the initial time, $\psi_{i\mathbf{k}}(\mathbf{r}, 0)$. The number of excited electrons at a specific \textbf{k}-point within a conduction state $u$ \cite{RodriguesPela2021:86963} is: 
\begin{equation}
    f_{c\mathbf{k}}(t)=\sum_i f^0_{i\mathbf{k}}|\langle\psi_{c\mathbf{k}}(0)|\psi_{i\mathbf{k}}(t)\rangle|^{2} .
\end{equation}
\noindent The number of holes left in a valence state $v$ is:
\begin{equation}
    f_{v\mathbf{k}}(t)=f_{v\mathbf{k}}^0-\sum_{i\mathbf{k}} f^0_{i\mathbf{k}} |\langle\psi_{v\mathbf{k}}(0)|\psi_{i\mathbf{k}}(t)\rangle|^2,
\end{equation}
and the number of excited electrons per unit cell is:
\begin{equation}
    N_{\mathrm{exc}}(t)=\sum_{c\mathbf{k}} w_\mathbf{k}f_{c\mathbf{k}}(t)=\sum_{v\mathbf{k}}w_\mathbf{k}f_{v\mathbf{k}}(t).
\end{equation}


\subsection{Bethe-Salpeter equation}

\noindent \hspace{1em} X-ray absorption spectra are computed by solving the Bethe-Salpeter equation (BSE) for the two-particle Green's function. In matrix form, the BSE is mapped onto an effective eigenvalue problem \cite{Rohlfing_2000}:
\begin{equation}  \sum_{o^{\prime}u^{\prime}\mathbf{k}^{\prime}}H_{ou\mathbf{k},o^{\prime}u^{\prime}\mathbf{k}^{\prime}}^{BSE}A_{o^{\prime}u^{\prime}\mathbf{k}^{\prime}}^{\lambda}=E_{\lambda}A_{ou\mathbf{k}}^{\lambda},
\label{Schrödinger equation}
\end{equation}
with the eigenenergies $E_{\lambda}$ and eigenvectors $A^{\lambda}_{ou\mathbf{k}}$, where $o$ and $u$ label the occupied and unoccupied states, respectively. The $H^{BSE}$ Hamiltonian consists of three contributions:
\begin{equation}
    H^{BSE}=H^{diag}+2H^{x}+H^{dir}.
\end{equation}
$H^{diag}$ represents the diagonal term, accounting for the single-particle transitions; $H^{x}$ is the exchange term, which includes the repulsive bare Coulomb interaction $\nu(\mathbf{r},\mathbf{r}^{\prime})$; and $H^{dir}$ is the direct term containing the attractive screened Coulomb interaction $W(\mathbf{r},\mathbf{r^{\prime}})$:
\begin{equation}    H_{ou\mathbf{k},o^{\prime}u^{\prime}\mathbf{k^{\prime}}}^{diag}=\left(\epsilon_{u\mathbf{k}}-\epsilon_{o\mathbf{k}}\right)\delta_{oo^{\prime}}\delta_{uu^{\prime}}\delta_{\mathbf{kk^{\prime}}},
\end{equation}

\begin{equation}
H_{vo\mathbf{k},v^{\prime}o^{\prime}\mathbf{k}^{\prime}}^{x}  \\  =\int d^3\mathbf{r}d^3\mathbf{r^\prime}\psi_{o\mathbf{k}}(\mathbf{r})\psi_{u\mathbf{k}}^{*}(\mathbf{r})\nu(\mathbf{r},\mathbf{r}^{\prime})\psi_{o^{\prime}\mathbf{k}^{\prime}}^{*}(\mathbf{r}^{\prime})\psi_{u^{\prime}\mathbf{k}^{\prime}}(\mathbf{r}^{\prime}),
\end{equation}

\begin{equation}
H_{ou\mathbf{k},o^{\prime}u^{\prime}\mathbf{k^{\prime}}}^{dir}  =-\int d^3\mathbf{r}d^3\mathbf{r^\prime}\psi_{o\mathbf{k}}(\mathbf{r})\psi_{u\mathbf{k}}^{*}(\mathbf{r^{\prime}})W(\mathbf{r},\mathbf{r^{\prime}})\psi_{o^{\prime}\mathbf{k^{\prime}}}^{*}(\mathbf{r})\psi_{u^{\prime}\mathbf{k^{\prime}}}(\mathbf{r^{\prime}}).
\label{H_direct_term}
\end{equation}

\noindent $\nu(\mathbf{r},\mathbf{r}^{\prime})$ and $W(\mathbf{r},\mathbf{r}^{\prime})$ stand for the short-range bare and statically screened Coulomb potentials, respectively. Taking the Fourier transform gives $W(\mathbf{r},\mathbf{r}^{\prime})$ as:
\begin{equation}
W_{\mathbf{GG}^{\prime}}(\mathbf{q})=4\pi\frac{\varepsilon_{\mathbf{GG}^{\prime}}^{-1}(\mathbf{q},\omega=0)}{|\mathbf{q}+\mathbf{G}||\mathbf{q}+\mathbf{G^{\prime}}|},
\end{equation}
where $\varepsilon$ is the microscopic dielectric function given, in the random-phase approximation, by
\begin{equation}
\varepsilon_{\mathbf{GG}^{\prime}}(\mathbf{q},\omega)=\delta_{\mathbf{GG}^{\prime}}-\frac1V\nu_{\mathbf{G}^{\prime}}(\mathbf{q})\sum_{ou\mathbf{k}}\frac{f_{u\mathbf{k}+\mathbf{q}}-f_{o\mathbf{k}}}{\epsilon_{u\mathbf{k}+\mathbf{q}}-\epsilon_{o\mathbf{k}}-\omega}\left[M_{ou}^{\mathbf{G}}(\mathbf{k},\mathbf{q})\right]^*M_{ou}^{\mathbf{G}^{\prime}}(\mathbf{k},\mathbf{q}).
\end{equation}
\noindent $V$ is the unit cell volume, $f_{o\mathbf{k}}$ is the occupation factor of the single-particle state $o$ at $\mathbf{k}$, and $M_{ou}^{\mathbf{G}}(\mathbf{k},\mathbf{q})$ is the plane-wave matrix element between the single-particle states $o$ and $u$ \cite{Vorwerk:2017gs, Vorwerk_2019}.

The occupations of excited states, obtained by using constrained DFT and RT-TDDFT, are incorporated here to calculate the Coulomb screening contribution to the spectra. Then, the occupations of excited states are used to calculate the dipole moment to obtain the Pauli blocking contribution to the spectra. The dipole moment matrix ($\mathrm{D}^*$) is given by:
\begin{equation}
\mathrm{D}^* =\mathrm{i}\sqrt{|f_{o\mathbf{k}}-f_{u\mathbf{k}+\mathbf{q}}|} \times \frac{\langle c\mathbf{k}|\hat{\mathbf{p}}|u\mathbf{k}\rangle}{\epsilon_{u\mathbf{k}+\mathbf{q}} - \epsilon_{o\mathbf{k}}},
\end{equation}
which enters the transition coefficient $t_\lambda$:
\begin{equation}
t_\lambda(0,\mathbf{q})=-\mathrm{i}\frac{\hat{\mathbf{q}}}{|\mathbf{q}|}\mathbf{X}_\lambda^\dagger\mathrm{D}^*,
\end{equation}
where $\mathbf{X}_\lambda$ is the resonant part of eigenvectors. The $\beta\beta$ tensor components of the macroscopic dielectric function can be obtained from the transition weights:
\begin{equation}
\begin{aligned}
\varepsilon_M^{\beta\beta}(\omega)& =1-\frac{8\pi}{V}\sum_\lambda|t_{\lambda}^\beta|^2\delta(\omega-E_\lambda).
\end{aligned}
\end{equation}
Averaging over the Cartesian components of the macroscopic dielectric function, and decomposing it into real and imaginary parts, 
\begin{equation}
\begin{aligned}
\frac{\varepsilon_M^{xx}(\omega)+\varepsilon_M^{yy}(\omega)+\varepsilon_M^{zz}(\omega)}{3} = \varepsilon_1(\omega) + i\varepsilon_2(\omega), 
\end{aligned}
\end{equation}
allows one to express the absorption coefficient as:
\begin{equation}
   \alpha(\omega) = \frac{\omega}{v_{\mathrm{light}}} \cdot \sqrt{2} \cdot \sqrt{\sqrt{\varepsilon_1^2(\omega) + \varepsilon_2^2(\omega)} - \varepsilon_1(\omega)},
\end{equation}
where $v_{\mathrm{light}}$ is the light speed in vacuum.

\section{Computational details \label{sec:computational_details}}

\subsection{Methods}

In the ground-state calculations, the Perdew-Burke-Ernzerhof (PBE) functional \cite{perdew:1996PBE} was used to treat exchange-correlation effects. A $10\times10\times6$ mesh of \textbf{k}-points was utilized to sample the first Brillouin zone. The muffin-tin radii $R_{\mathrm{MT}}$ were set to 2.0 bohr for Zn and 1.45 bohr for O. The basis-set cut-off was set to $R_{\mathrm{MT}}|\mathbf{G}+\mathbf{k}|_{\mathrm{max}}= 8.0$. Structural relaxation of the wurtzite unit cell yielded lattice parameters of $\mathbf{a} = \SI{3.289}{\angstrom}$ and $\mathbf{c} = \SI{5.307}{\angstrom}$.

RT-TDDFT calculations were carried out to simulate the dynamics of the photoexcited carriers generated by a laser pump pulse with a FWHM pulse duration of \SI{10}{\femto\second}, a pump photon energy of \SI{3.49}{\electronvolt}, and with the electric field parallel to the (100) crystal axis. 

To obtain the XAS spectra within the BSE formalism, we computed the electron-hole ($e$-$h$) interaction matrix elements using a $10\times10\times6$ mesh of \textbf{k}-points with an offset of $(0.097, \thinspace 0.173, \thinspace0.193)$. Such a non-symmetric shift leads to an effective finer sampling by avoiding symmetrically redundant contributions to the spectrum. A scissors shift of \SI{141.6}{\electronvolt} was employed to align the computed spectra to experiment. Local-field effects were included, setting ($|\mathbf{G}+\mathbf{q}|_\mathrm{max}$) to $\mathrm{3.0\thinspace a.u.^{\scriptscriptstyle{-1}}}$. The screened Coulomb interaction was calculated within the random phase approximation. A Lorentzian lineshape with a FWHM broadening of $\delta = \SI{1.36}{\electronvolt}$ is employed to simulate the core-hole lifetime, which corresponds to the tabulated bare core-hole broadening of Zn \cite{Krause1979}. The energy dependence of the spectral broadening induced by the inelastic mean free path of the photoelectron is neglected because it follows the core-hole lifetime at K-edges \cite{Nishihata1998:121653}. 

\subsection{Decomposition of non-thermal XTA spectra\label{secSI:supp_decomposition_XTA}}

\subsection{Carrier distribution at femtosecond time delays\label{secSI:occ_fs}}


\subsection{Electron distribution of the core exciton\label{secSI:electron distribution}}

\end{document}